# Understanding Space in Proof Complexity: Separations and Trade-offs via Substitutions

Eli Ben-Sasson*
Computer Science Department
Technion — Israel Institute of Technology
Haifa, 32000, Israel
eli@cs.technion.ac.il

Jakob Nordström†
Computer Science and Artificial Intelligence Laboratory
Massachusetts Institute of Technology
Cambridge, MA 02139, USA
jakobn@mit.edu

August 31, 2018

**Abstract**

For current state-of-the-art satisfiability algorithms based on the DPLL procedure and clause learning, the two main bottlenecks are the amounts of time and memory used. In the field of proof complexity, these resources correspond to the length and space of resolution proofs for formulas in conjunctive normal form (CNF). There has been a long line of research investigating these proof complexity measures, but while strong results have been established for length, our understanding of space and how it relates to length has remained quite poor. In particular, the question whether resolution proofs can be optimized for length and space simultaneously, or whether there are trade-offs between these two measures, has remained essentially open apart from a few results in restricted settings.

In this paper, we remedy this situation by proving a host of length-space trade-off results for resolution in a completely general setting. Our collection of trade-offs cover almost the whole range of values for the space complexity of formulas, and most of the trade-offs are superpolynomial or even exponential and essentially tight. Using similar techniques, we show that these trade-offs in fact extend (albeit with worse parameters) to the exponentially stronger $k$-DNF resolution proof systems, which operate with formulas in disjunctive normal form with terms of bounded arity $k$. We also answer the open question whether the $k$-DNF resolution systems form a strict hierarchy with respect to space in the affirmative.

Our key technical contribution is the following, somewhat surprising, theorem: Any CNF formula $F$ can be transformed by simple variable substitution into a new formula $F'$ such that if $F$ has the right properties, $F'$ can be proven in essentially the same length as $F$, whereas on the other hand the minimal number of *lines* one needs to keep in memory simultaneously in any proof of $F'$ is lower-bounded by the minimal number of *variables* needed simultaneously in any proof of $F$. Applying this theorem to so-called pebbling formulas defined in terms of pebble games on directed acyclic graphs, we obtain our results.

*The research leading to these results has received funding from the European Community's Seventh Framework Programme (FP7/2007-2013) under grant agreement number 240258 and was supported by the Israeli Science Foundation and by the US-Israel Binational Science Foundation under grant number 2006104.

†Research supported by the Royal Swedish Academy of Sciences, the Ericsson Research Foundation, the Sweden-America Foundation, the Foundation Olle Engkvist Byggmästare, the Sven and Dagmar Salén Foundation, and the Foundation Blancheflor Boncompagni-Ludovisi, née Bildt.

# 1 Introduction

A central theme in the field of propositional proof complexity is the study of limitations of natural proof systems. Such a study is typically conducted by considering a *complexity measure* of propositional proofs and investigating under which circumstances this measure is large. The most thoroughly examined complexity measure is that of *proof size/length*. The interest in this measure is motivated by its connections to the NP vs. co-NP problem (since by [CR79], proving superpolynomial lower bounds for arbitrary proof systems would separate NP and co-NP), by methods for proving independence results in first order theories of bounded arithmetic (for an example, see [Ajt88]), and because lower bounds on proof length imply lower bounds on the running time of algorithms for solving NP-complete problems such as SATISFIABILITY (such algorithms are usually referred to as *SAT solvers*).

**Proof space** This paper focuses on a more recently suggested complexity measure known as *space*. The space measure was first defined and studied by Esteban and Torán [ET01] in the context of the famous *resolution* proof system introduced by Blake [Bla37], and was generalized to other proof systems by Alekhnovich et al. in [ABRW02]. Roughly speaking, the space of a proof corresponds to the minimal size of a blackboard needed to give a self-contained presentation of the proof, where the correctness of each step is verifiable from what is currently on the blackboard. The interest in space complexity stems from two main sources that we survey next.

First, there are intricate and often surprising connections between space and length. For resolution, it follows from the elegant results of Atserias and Dalmau [AD08] that upper bounds on space imply upper bounds on length. Esteban and Torán [ET01] showed the converse that length upper bounds imply space upper bounds for the restricted case of *tree-like resolution*. Recall that the *tree-like* version of a *sequential*[1] proof system has the added constraint that every line in the proof can be used at most once to derive a subsequent line. In terms of space, a proof is tree-like if any claim appearing on the blackboard must be erased immediately after it has been used to derive a new claim. Another question which has attracted interest is whether space and length can display *trade-offs*, that is whether there are formulas having proofs in both short length and small space, but for which there are no proofs in short length and small space *simultaneously*. Such length-space trade-offs have been established in restricted settings in [Ben09, Nor09b][2] but nothing has been known for refutations of explicit formulas in general, unrestricted resolution.

A second motivation to study space is its connection to the memory consumption of SAT solvers. For instance, the family of backtracking heuristics suggested by [DP60, DLL62] and known as *Davis-Putnam-Logemann-Loveland (DPLL)* SAT solvers have the following property. When given as input an unsatisfiable formula $F$ in conjunctive normal form—called henceforth a *CNF formula*—the description of the execution of a DPLL SAT solver corresponds to a tree-like resolution proof refuting $F$. Thus, lower bounds on tree-like refutation space imply lower bounds on the *memory consumption* of DPLL SAT solvers, much like lower bounds on tree-like refutation length imply lower bounds on the *running time* of DPLL heuristics. During the last 10-15 years, a family of SAT solvers known as *DPLL with clause learning* [SS96, BS97] (denoted DPLL+) has been put to practical use with impressive success. For instance, an overwhelming majority of the best algorithms in recent rounds of the international SAT competitions (see [SAT]) belong to this class. These SAT solvers have the property that an execution trace corresponds to a (non-tree-like) resolution refutation. Hence, space lower bounds in general resolution translate into memory lower bounds for these algorithms, and length-space trade-offs could have implications for trade-offs between time efficiency and memory consumption.

---

[1] A proof system is said to be *sequential* if a proof $\pi$ in the system is a *sequence* of lines $\pi = \{L_1, \ldots, L_\tau\}$ where each line is derived from previous lines by one of a finite set of allowed *inference rules* (See Section 3 for formal definitions).

[2] A related result, claimed in [HP07], has been later retracted by the authors in [HP10].

We end this discussion by pointing out that there is still much left to explore regarding the connection between space lower bounds in proof complexity and memory consumption of SAT solvers. On the one hand, the memory consumption of a "typical" DPLL+ SAT solver can be far greater than the theoretical upper bounds on refutation space. On the other hand, the theoretical lower bounds on refutation space are *worst-case* bounds for *non-deterministic algorithms*, and hence apply even to the most memory-efficient proofs theoretically possible, which is not remotely close to the kind of proofs produced by a typical SAT solver. Understanding what kind of practical implications one can get on the memory consumption of SAT solvers from refutation space lower bounds remains an interesting open problem.

**$k$-DNF resolution** The family of sequential proof systems known as $k$-*DNF resolution* was introduced by Krajíček [Kra01] as a intermediate step between resolution and depth-2 Frege. Roughly speaking, for integers $k > 0$ the $k$th member of this family, denoted henceforth by $\mathfrak{R}(k)$, is only allowed to reason in terms of formulas in disjunctive normal form (*DNF formulas*) with the added restriction that any conjunction in any formula is over at most $k$ literals. For $k = 1$, the lines in the proof are hence disjunctions of literals, and the system $\mathfrak{R}(1) = \mathfrak{R}$ is standard resolution. At the other extreme, $\mathfrak{R}(\infty)$ is equivalent to depth-2 Frege.

The original motivation to study this family of proof systems, as stated in [Kra01], was to better understand the complexity of counting in weak models of bounded arithmetic, and it was later observed that these systems are also related to SAT solvers that reason using multi-valued logic (see [JN02] for a discussion of this point). By now a number of works have shown superpolynomial lower bounds on the length of $\mathfrak{R}(k)$-refutations, most notably for (various formulations of) the pigeonhole principle and for random CNF formulas [AB04, ABE02, Ale05, JN02, Raz03, SBI04, Seg05]. Of particular relevance to our current work are the results of Segerlind et al. [SBI04] and of Segerlind [Seg05] showing that the family of $\mathfrak{R}(k)$-systems form a *strict hierarchy* with respect to proof length. More precisely, they prove that there exists a constant $\epsilon > 0$ such that for every integer $k > 0$ there exists a family of formulas $\{F_n\}_{n=1}^{\infty}$ of arbitrarily large size $n$ such that $F_n$ has a $\mathfrak{R}(k+1)$-refutation of polynomial length $n^{O(1)}$ but all $\mathfrak{R}(k)$-refutations of $F_n$ require exponential length $2^{n^{\epsilon}}$.

Just as in the case for standard resolution, the understanding of space complexity in $k$-DNF resolution has remained more limited. We are aware of only one prior work by Esteban et al. [EGM04] shedding light on this question. Their paper establishes essentially optimal linear space lower bounds for $\mathfrak{R}(k)$ and also prove that the family of *tree-like* $\mathfrak{R}(k)$ systems form a strict hierarchy with respect to space. What they show is that there exist arbitrarily large formulas $F_n$ of size $n$ that can be refuted in tree-like $\mathfrak{R}(k+1)$ in constant space but require space $\Omega(n/\log^2 n)$ to be refuted in tree-like $\mathfrak{R}(k)$. It should be pointed out, however, that as observed in [Kra01, EGM04] the family of tree-like $\mathfrak{R}(k)$ systems for all $k > 0$ are strictly weaker than standard resolution. As was noted above, the family of general, unrestricted $\mathfrak{R}(k)$ systems are strictly stronger than resolution, so the results in [EGM04] leave completely open the question of whether there is a strict space hierarchy for (non-tree-like) $\mathfrak{R}(k)$ or not.

**Definition of length and space** As a final point before turning to our results, we briefly and informally recall what is meant by "length" and "space" (formal definitions will follow in Section 3). We view a refutation of an unsatisfiable CNF formula $F$ as being presented on a blackboard. The refutation is represented as a sequence of sets of $k$-DNF formulas $\pi = \{\mathbb{D}_0, \ldots, \mathbb{D}_\tau\}$, where $\mathbb{D}_t$ is a snapshot of the blackboard at time $t$. In particular, $\mathbb{D}_0$ should be the empty set, $\mathbb{D}_\tau$ should contain the contradictory empty formula, and at time $t$ we can go from from $\mathbb{D}_{t-1}$ to $\mathbb{D}_t$ by *(i)* writing a clause of $F$ (an *axiom*) on the blackboard, *(ii)* erasing a line from the board, or *(iii)* inferring a new line from those lines present on the board according to the inference rules of $k$-DNF resolution. We do not discuss the details of these rules here, since the exact definitions in fact do not matter—the lower bounds we prove hold for any *arbitrarily strong* (but sound) rules. What is important is that the only new formulas that can be derived are those implied by the set of formulas that are currently on the blackboard, and that these formulas are all $k$-DNFs.

The length of a refutation is the number of formulas appearing in the refutation counted with repetitions, or equivalently (within a factor of 2) the number of derivation steps. There are several different ways to measure the space of a set $\mathbb{D}_t$ in our refutation. The crudest way is to count the number of $k$-DNF formulas on the board, i.e., to measure the size of $\mathbb{D}_t$. We call this the *formula space*, or simply, *space* of $\mathbb{D}_t$. (For resolution, i.e., when $k = 1$, this is the well-studied measure of *clause space*.) A finer granulation is to measure the *total space*—the number of appearances of literals in $\mathbb{D}_t$, counted with repetitions. Formula space and total space are the two space measures that have received the most attention in previous research, and they are also the focus of the current paper. A third, closely related, measure that will also be of interest to us is *variable space*, defined as the number of distinct variables appearing on the board. It is easily seen that variable space is a lower bound on total space. For all of these space measures, the space of a refutation $\pi = \{\mathbb{D}_0, \ldots, \mathbb{D}_\tau\}$ is the maximal space of any $k$-DNF set $\mathbb{D}_t$ in it.

## 1.1 Our Results in Brief

**Space Hierarchy for $k$-DNF resolution**  Our first main result is that Krajíček's family of $k$-DNF resolution proof systems form a strict hierarchy with respect to space. More precisely, we separate $k$-DNF resolution from $(k + 1)$-DNF resolution in the sense that we exhibit for every $k$ a family of *explicitly constructible*[3] CNF formulas of size $n$ that can be refuted in constant formula space and linear length simultaneously in $(k + 1)$-DNF resolution (i.e., they are very easy with respect to both measures) but have the property that any $k$-DNF resolution, no matter how long or short, must by necessity use at least order of $\sqrt[k+1]{n}$ formula space.

**Length-space trade-offs**  Our second main result is a collection of strong length-space trade-offs for $k$-DNF resolution. For $k = 1$, i.e., standard resolution, these are the first trade-off results for resolution refutations of explicit formulas in the general, unrestricted resolution proof system, thus eliminating all technical restrictions in the previous works [Ben09, Nor09b]. For $k > 1$, to the best of our knowledge no trade-offs have been known even in restricted settings.

We also want to emphasize two other novel aspects of our results. First, as was discussed above there are several different ways of measuring space, and previous papers have focused on one particular measure and proven results specifically for that measure. Our techniques, however, allow us to obtain results that hold for both total space and formula space (i.e., the largest and smallest space measure) simultaneously. Second, our upper bounds hold for the standard *syntactic* version of the proof systems, where new formulas can be derived from two existing formulas by a limited set of structural rules, whereas the lower bounds hold for *semantic* versions where new formulas can be derived from an unlimited set of formulas by arbitrary sound rules. The reason this is worth pointing out is that in general, semantic $k$-DNF resolution proof systems are known to be exponentially stronger than syntactic $k$-DNF resolution.

We give the formal statements of our trade-offs in Section 2, but a general template for the kind of trade-off theorems we are able to prove is as follows.

**Theorem 1.1 (Trade-off theorem template (informal)).** *Let $K$ be a fixed positive integer and let $s_{\mathrm{lo}}(n)$ and $s_{\mathrm{hi}}(n)$ be suitable function such that $s_{\mathrm{lo}}(n) \ll s_{\mathrm{hi}}(n) = \mathrm{O}(n/\log\log n)$. Then there are explicitly constructible CNF formulas $\{F_n\}_{n=1}^{\infty}$ of size $\mathrm{O}(n)$ and width $\mathrm{O}(1)$ (with constants depending on $K$) such that the following holds:*

- *The formulas $F_n$ are refutable in syntactic resolution in (small) total space $\mathrm{O}(s_{\mathrm{lo}}(n))$.*

---

[3] A family of formulas is *explicitly constructible* if there exists a polynomial time algorithm that on input $1^n$ produces the $n$th member of the family.

- *There are also syntactic resolution refutations $\pi_n$ of $F_n$ in simultaneous length $\mathrm{O}(n)$ and (much larger) total space $\mathrm{O}(s_{\mathrm{hi}}(n))$.*

- *However, any resolution refutation, even semantic, in formula space $\mathrm{o}(s_{\mathrm{hi}}(n))$ must have superpolynomial or sometimes even exponential length.*

- *Even for the much stronger semantic $k$-DNF resolution proof systems, $k \leq K$, it holds that any $\mathfrak{R}(k)$-refutation of $F_n$ in formula space $\mathrm{o}\bigl(\sqrt[k+1]{s_{\mathrm{hi}}(n)}\bigr)$ must have superpolynomial length (or exponential length, correspondingly).*

We instantiate this theorem template for a wide range of space functions $s_{\mathrm{lo}}(n)$ and $s_{\mathrm{hi}}(n)$ from constant space all the way up to nearly linear space. This is in contrast to [Nor09b], where the trade-off results are obtained only for a very carefully selected ratio of space to formula size. Moreover, our trade-offs are robust in that they are not sensitive to small variations in either length or space (as in [Nor09b]). That is, intuitively speaking they will not show up only for a SAT solver being unlucky and picking just the wrong threshold when trying to hold down the memory consumption. Instead, any refutation having length or space in the same general vicinity will be subject to the same qualitative trade-off behaviour.

## 1.2 Overview of Technical Contributions

We want to highlight three technical contributions underpinning the results discussed informally above.

**Substitution space theorems** Our first key technical contribution is a general way to derive strong space lower bounds in resolution from weak lower bounds on the number of variables that occur simultaneously in a proof. Very loosely, we show the following: Suppose we have a formula that has short refutations but where any such short refutation must mention many variables at some point. Then by making variable substitutions in this formula and expanding the result into a CNF formula in the natural way, this new formula will still have short refutations, but now any such refutation must use lots of space, in the sense that lower bounds on *variable space* will translate into lower bounds on *formula space*.

We believe that this generic procedure of transforming weak space lower bounds into stronger ones is an interesting result in its own right that sheds new light on space measures in proof complexity. To support this point, we strengthen the theorem by showing that not only can we obtain strong resolution lower bounds in this way, but it is also possible to lift weak resolution lower bounds to strong lower bounds in other more powerful proof systems, namely $k$-DNF resolution systems. We remark that this general idea of "hardness amplification" in proof complexity has also been used in the recent work of Beame et al. [BHP09], although the actual techniques there appear somewhat orthogonal to ours (and, in particular, incomparable in the sense that it seems neither paper can be used to derive the results in the other).

**Minimally unsatisfiable $k$-DNF sets** One crucial ingredient in the proof of the substitution space theorem for resolution is analyzing the structure of sets of disjunctive clauses that imply many other clauses. Intuitively, it seems reasonable that if the set of implied clauses is sufficiently large and disjoint, the set of clauses implying all these clauses cannot itself be too small. One important special case of this is for clause sets containing many variables but being *minimally unsatifiable*—that is, every clause places a necessary constraint on the variables to enforce unsatifiability and if just one arbitrary clause is removed from the set, then the rest can be satisfied. It is well known that such a clause set must contain strictly less variables than clauses, and we can use similar proof techniques to derive the more general result that we need.

When we want to extend our theorem to $k$-DNF resolution, it becomes essential to understand instead the structure of sets of $k$-DNF formulas that imply many other $k$-DNF formulas. Here there are no previous results to build on, as the proof techniques that yield tight results for disjunctive clauses can be shown to break down fundamentally. Instead, we have to develop new methods. One important step along the way is to understand the structure of minimally unsatisfiable sets of $k$-DNF formulas, which appears to be a natural combinatorial problems of independent interest. We prove that a minimally unsatisfiable $k$-DNF set of size $m$ can contain at most $\lesssim m^{k+1}$ variables, and this bound turns out to be tight up to an additive one in the exponent in view of recent joint work [NR09] of Razborov and the second author.

**Reductions between resolution and pebbling**   Using the substitution space theorems, we can construct reductions between ($k$-DNF) resolution on the one hand and so-called pebble games played on directed acyclic graphs (DAGs) on the other. In one direction, this reduction is easy, but the other direction is nontrivial. Moreover, our reductions are time- and space-preserving. This allows us (modulo some technical complications which we ignore for the moment) to translate known trade-off results for pebbling into corresponding trade-offs for resolution. This is done by transforming the pebble game played on a DAG $G$ into a CNF formula that encodes this particular problem instance of the game, and showing that this formula has similar trade-off properties in resolution as the DAG $G$ has for the pebble game.

With hindsight, such a correspondence might seem more or less obvious, so let us stress that this is not the case. Pebble games on graphs and $\mathfrak{R}(k)$-refutations of CNF formulas are very different objects. Once we have translated a pebbling instance into a CNF formula, it is not at all clear why a $\mathfrak{R}(k)$-prover refuting this formula would have to care about how it was constructed. There might be shortcuts in the proof complexity world that do not correspond to anything meaningful in the pebbling world. And indeed, reading previous literature on pebbling formulas in proof complexity reveals a few such surprising shortcuts, and there has been no consensus on what properties these formulas are likely to have in general.

What we show is that for the right flavour of pebbling formulas, any prover refuting such formulas must in effect reason in terms of pebblings. More precisely, we show that given any $\mathfrak{R}(k)$-refutation, no matter how it is structured, we can extract from it a pebbling of the underlying DAG, and this pebbling has at least as good time and space properties as the refutation from which it was extracted. In other words, the pebbling formula inherits the time-space trade-off properties of the DAG in terms of which it is defined. This allows us to draw on the rich literature on pebbling trade-offs from the 1970s and 1980s, as well as on newer results by the second author in [Nor10b], to obtain strong trade-offs in proof complexity.

## 1.3   Organization of the Rest of This Paper

In Section 2, we present formal statements of our main results and—in order not to let all the notation and terminology obscure what is in essence a clean and simple proof construction—briefly outline some of the key ingredients in the proofs. Having presented our results and sketched the main ideas in the proof constructions, in Section 3 we then give the necessary preliminaries for the formal proofs that will follow. Our first main technical contribution, the substitution space theorem for resolution, is proven in Section 4. In Section 5, we extend this theorem to $k$-DNF resolution, along the way establishing upper and lower bounds on the size of minimally unsatisfiable sets of $k$-DNF formulas. Our final technical tool, namely the method for converting strong pebbling trade-offs into length-space trade-offs for resolution, is described in Section 6. We apply this tool in Section 7 to prove our collection of length-space trade-off results. Concluding the paper, in Section 8 we briefly discuss some open questions.

## 2   Formal Statements of Results and Outline of Proofs

In what follows, let us write $L(\pi)$ to denote the length of a resolution refutation and $Sp(\pi)$, $TotSp(\pi)$, and $VarSp(\pi)$ to denote the formula space, total space and variable space, respectively. Taking the minimum over all resolution refutations of $F$, we let $L_{\mathfrak{R}}(F \vdash 0)$ denote the length of a shortest refutation, and

$Sp_{\mathfrak{R}}(F \vdash 0)$, $TotSp_{\mathfrak{R}}(F \vdash 0)$, and $VarSp_{\mathfrak{R}}(F \vdash 0)$ are defined completely analogously. These definitions are also generalized to $\mathfrak{R}(k)$ for general $k$. To state our results it will also convenient to use the notation $W(\pi)$ for the *width* of a standard resolution refutation, i.e., the size of a largest clause in it, and $W_{\mathfrak{R}}(F \vdash 0)$ for the minimal width of any standard resolution refutation of $F$. Again, more formal definitions are given in Section 3.

## 2.1 Substitution Space Theorems

If $F$ is a CNF formula over variables $x, y, z, \ldots$ and $f : \{0,1\}^d \mapsto \{0,1\}$ is a Boolean function over $d$ variables, we can obtain a new CNF formula by substituting $f(x_1, \ldots, x_d)$ for every variable $x$ and then expand to conjunctive normal form. We will write $F[f]$ to denote the resulting *substitution formula*. For example, for the disjunctive clause $C = x \vee \overline{y}$ and the exclusive or function $f = x_1 \oplus x_2$ we have

$$C[f] = (x_1 \vee x_2 \vee y_1 \vee \overline{y}_2) \wedge (x_1 \vee x_2 \vee \overline{y}_1 \vee y_2) \\ \wedge (\overline{x}_1 \vee \overline{x}_2 \vee y_1 \vee \overline{y}_2) \wedge (\overline{x}_1 \vee \overline{x}_2 \vee \overline{y}_1 \vee y_2) \ . \tag{2.1}$$

We say that $f$ is *k-non-authoritarian* if no partial assignment to any subset of $k$ variables can fix the value of $f$ to true or false and that $f$ is *k-authoritarian* otherwise. For instance, the XOR function $\oplus$ on $d+1$ variables is $d$-non-authoritarian, as is the majority function on $2d+1$ variables. If $f$ is 1-non-authoritarian (1-authoritarian) we say that the function is simply *non-authoritarian* (*authoritarian*). For example, non-exclusive or $\vee$ of any arity is always authoritarian.

Loosely put, the substitution space theorem for resolution says that if a CNF formula $F$ can be refuted in resolution in small length and width simultaneously, then so can the substitution formula $F[f_d]$. There is an analogous result in the other direction as well in the sense that we can translate any refutation $\pi_f$ of $F[f_d]$ into a refutation $\pi$ of the original formula $F$ where the length of $\pi$ is almost upper-bounded by the length of $\pi_f$ (this will be made precise below). So far this is nothing very unexpected, but what is more interesting is that if $f_d$ is non-authoritarian, then the clause space of $\pi_f$ is an upper bound on the number of variables mentioned simultaneously in $\pi$. Thus, the theorem says that we can convert (weak) lower bounds on variable space into (strong) lower bounds on clause space by making substitutions using non-authoritarian functions.

**Theorem 2.1 (Substitution space theorem for resolution).** *Let $F$ be any unsatisfiable CNF formula and $f_d$ be any non-constant Boolean function of arity $d$. Then $F[f_d]$ can be refuted in resolution in width*

$$W_{\mathfrak{R}}\big(F[f_d] \vdash 0\big) = \mathrm{O}\big(d \cdot W_{\mathfrak{R}}(F \vdash 0)\big) \ ,$$

*length*

$$L_{\mathfrak{R}}\big(F[f_d] \vdash 0\big) \leq \min_{\pi:F \vdash 0}\big\{L(\pi) \cdot \exp\big(\mathrm{O}(d \cdot W(\pi))\big)\big\} \ ,$$

*and total space*

$$TotSp_{\mathfrak{R}}\big(F[f_d] \vdash 0\big) \leq \min_{\pi:F \vdash 0}\big\{TotSp(\pi) \cdot \exp\big(\mathrm{O}(d \cdot W(\pi))\big)\big\} \ .$$

*In the other direction, any semantic resolution refutation $\pi_f : F[f_d] \vdash 0$ of the substitution formula can be transformed into a syntactic resolution refutation $\pi : F \vdash 0$ of the original formula such that the number of axiom downloads[4] in $\pi$ is at most the number of axiom downloads in $\pi_f$. If in addition $f_d$ is non-authoritarian, it holds that $Sp(\pi_f) > VarSp(\pi)$, i.e., the clause space of refuting the substitution formula $F[f_d]$ is lower-bounded by the variable space of refuting the original formula $F$.*

---

[4]It would have been nice if the bound in terms of number of axiom downloads could have be strengthened to a bound in terms of length, but this is *not* true. The reason for this is that the proof refuting $F[f_d]$ is allowed to use any arbitrarily strong *semantic* inference rules, and this can lead to exponential savings compared to syntactic resolution. To see this, just let $F$ be an encoding of, say, the pigeonhole principle and let $\pi_f$ be the refutation that downloads all axioms of $F[f_d]$ and then derives contradiction in one step. Luckily enough, though, the bound in terms of axiom downloads turns out to be exactly what we need for our applications.

Note that if $F$ is refutable simultaneously in linear length and constant width, then the bound in Theorem 2.1 on $L\bigl(F[f_d] \vdash 0\bigr)$ becomes linear in $L(F \vdash 0)$.

The substitution space theorem for $k$-DNF resolution extends Theorem 2.1 by telling us that for $k$-non-authoritarian functions $f$, we can translate back and forth between standard resolution refutations of $F$ and $\mathfrak{R}(k)$-refutations of the substitution formula $F[f]$ in a (reasonably) length- and space-preserving way. When the "proof blackboard" contains $k$-DNFs instead of disjunctive clauses, the analysis becomes much more challenging, however, and the bounds we are able to obtain become correspondingly worse. Below, we state the theorem with asymptotic factors hidden by the asymptotic notation to make it easier to parse. The complete version is given in Section 5.

**Theorem 2.2 (Substitution space theorem for $k$-DNF resolution).** *Let $F$ be any unsatisfiable $c$-CNF formula and $f_d$ be any non-constant Boolean function of arity $d$, and suppose furthermore that $c$, $d$, and $k$ are universal constants. Then the following two properties hold for the substitution formula $F[f_d]$:*

1. *If $F$ can be refuted in syntactic standard resolution in length $L$ and total space $s$ simultaneously, then $F[f_d]$ can be refuted in syntactic $\mathfrak{R}(d)$ in length $\mathrm{O}(L)$ and total space $\mathrm{O}(s)$ simultaneously.*

2. *If $f_d$ is $k$-non-authoritarian and $F[f_d]$ can be refuted by a semantic $\mathfrak{R}(k)$-refutation that requires formula space $s$ and makes $L$ axiom downloads, then $F$ can be refuted by a syntactic standard resolution refutation that requires variable space at most $\mathrm{O}\bigl(s^{k+1}\bigr)$ and makes at most $L$ axiom downloads.*

The proofs of Theorems 2.1 and 2.2 are inspired by our recent work [BN08] and indeed our main theorem there can be seen to follow from Theorem 2.1. Let us discuss the new aspects of the more general theorems presented in this paper. First and foremost, our results extend to $\mathfrak{R}(k)$ for $k > 1$ whereas the previous theorem applies only to resolution. Second, our previous statement only hold for a very special kind of formulas (namely the pebbling formulas discussed above) whereas Theorems 2.1 and 2.2 can be used to convert *any* CNF formula requiring large variable space into a new and closely related CNF formula requiring large formula space. Third, in this paper we get length-preserving as well as space-preserving reductions, whereas it was unclear how to derive similar reductions from our previous work. And length-preserving reductions are crucial for our length-space trade-offs described below.

We will return to these theorems and sketch the main ingredients in the proofs in Sections 2.4 and 2.5, but before that we want to describe why these tools will be so useful for us. We do so next.

## 2.2 Translating Resolution Refutations to Pebblings

The *pebble game* played on a DAG $G$ models the calculation described by $G$, where the source vertices contain the inputs and non-source vertices specify operations on the values of the predecessors. Placing a pebble on a vertex $v$ corresponds to storing in memory the partial result of the calculation described by the subgraph rooted at $v$. Removing a pebble from $v$ corresponds to deleting the partial result of $v$ from memory. Black pebbles correspond to deterministic computation and white pebbles to nondeterministic guesses. A *pebbling* $\mathcal{P}$ of $G$ is a sequence of moves starting with the graph being completely empty and ending with all vertices empty except for a black pebble on the (unique) sink vertex. The *time* of a pebbling is the number of pebbling moves and the *space* is the maximal number of pebbles needed at any point during the pebbling.

The pebble game on the graph $G$ can be encoded as an unsatisfiable CNF formula $Peb_G$ saying that the sources of $G$ are true and that truth propagates through the graph in accordance with the pebbling rules, but that the sink is false. Given any black-only pebbling $\mathcal{P}$ of $G$, we can mimic this pebbling in a resolution refutation of $Peb_G$ by deriving that a literal $v$ is true whenever the corresponding vertex in $G$ is pebbled (this was perhaps first observed in [BIW04]).

**Lemma 2.3 ([BIW04]).** *Let $G$ be a DAG with unique sink and bounded vertex indegree. Then given any complete black pebbling $\mathcal{P}$ of $G$, we can construct a standard resolution refutation $\pi : Peb_G \vdash 0$ such that $L(\pi) = \mathrm{O}(\text{time}(\mathcal{P}))$, $W(\pi) = \mathrm{O}(1)$, and $TotSp(\pi) = \mathrm{O}(\text{space}(\mathcal{P}))$.*

In the other direction, we start with the result of the first author [Ben09] that if we take any refutation of a pebbling contradiction and let positive and negative literals correspond to black and white pebbles respectively, then we get (essentially) a legal black-white pebbling of the underlying DAG. That is not quite what we need, however, since it only provides a weak bound in terms of variable space.

This is where Theorems 2.1 and 2.2 come into play. If we make substitutions in $Peb_G$ with suitably non-authoritarian functions, the upper bounds in Lemma 2.3 remain true (with adjustments in constant factors), while the lower bounds are lifted from variable space to formula space. For simplicity, we only state the lower bounds in the special case for standard resolution below.

**Theorem 2.4.** *Let $f$ be any non-authoritarian Boolean function and $G$ be any DAG with unique sink and bounded indegree. Then from any standard resolution refutation $\pi : Peb_G[f] \vdash 0$ we can extract a black-white pebbling strategy $\mathcal{P}_\pi$ for $G$ such that $\text{time}(\mathcal{P}_\pi) = \mathrm{O}(L(\pi))$ and $\text{space}(\mathcal{P}_\pi) = \mathrm{O}(Sp(\pi))$.*

## 2.3 Space Separations and Length-Space Trade-offs

Combining the theorems in Section 2.1 with the reductions between resolution and pebble games in Section 2.2, we can now establish our space separation and length-space trade-off results. Let us start by formally stating the space hierarchy theorem for $\mathfrak{R}(k)$.

**Theorem 2.5 ($k$-DNF resolution space hierarchy).** *For every $k \geq 1$ there exists an explicitly constructible family $\{F_n\}_{n=1}^\infty$ of CNF formulas of size $\Theta(n)$ and width $\mathrm{O}(1)$ such that*

- *there are $\mathfrak{R}(k+1)$-refutations $\pi_n : F_n \vdash 0$ in simultaneous length $L(\pi_n) = \mathrm{O}(n)$ and formula space $Sp(\pi_n) = \mathrm{O}(1)$, but*

- *any $\mathfrak{R}(k)$-refutation of $F_n$ requires formula space $\Omega\bigl(\sqrt[k+1]{n/\log n}\bigr)$.*

*The constants hidden by the asymptotic notation depend only on $k$.*

The families $\{F_n\}_{n=1}^\infty$ are obtained by considering pebbling formulas defined in terms of the graphs in [GT78] requiring pebbling space $\Theta(n/\log n)$, and substituting a $k$-non-authoritarian Boolean function $f$ of arity $k+1$, for instance XOR over $k+1$ variables, in these formulas.

Moving on to our length-space trade-offs, in the remainder of this section we try to highlight some of the results that we find to be the most interesting. A fuller and more detailed account of our collection of trade-off results is given in Section 7. We reiterate that all of our results are for explicitly constructible formulas, and that in addition most of the constructions are actually very clean and transparent in that they are obtainable from pebbling formulas over simple families of DAGs.

From the point of view of space complexity, the easiest formulas are those refutable in constant total space, i.e., formulas having so simple a structure that there are resolution refutations where we never need to keep more than a constant number of symbols on the proof blackboard. A priori, it is not even clear whether we should expect that any trade-off phenomena could occur for such formulas, but it turns out that there are quadratic length-space trade-offs.

**Theorem 2.6 (Quadratic trade-offs for constant space).** *For any fixed positive integer $K$ there are explicitly constructible CNF formulas $\{F_n\}_{n=1}^\infty$ of size $\Theta(n)$ and width $\mathrm{O}(1)$ such that the following holds (where all multiplicative constants hidden in the asymptotic notation depend only on $K$):*

- *The formulas $F_n$ are refutable in syntactic resolution in total space $TotSp_\mathfrak{R}(F_n \vdash 0) = \mathrm{O}(1)$.*

- *For any $s_{\text{hi}}(n) = \text{O}(\sqrt{n})$ there are syntactic resolution refutations $\pi_n$ of $F_n$ in simultaneous length $L(\pi_n) = \text{O}((n/s_{\text{hi}}(n))^2)$ and total space $TotSp(\pi_n) = \text{O}(s_{\text{hi}}(n))$.*

- *For any semantic resolution refutation $\pi_n : F_n \vdash 0$ in formula space (i.e., clause space) $Sp(\pi_n) \leq s_{\text{hi}}(n)$ it holds that $L(\pi_n) = \Omega((n/s_{\text{hi}}(n))^2)$.*

- *For any $k \leq K$, any semantic $k$-DNF resolution refutation $\pi_n$ of $F_n$ in formula space $Sp(\pi_n) \leq s_{\text{hi}}(n)$ must have length $L(\pi_n) = \Omega\left((n/(s_{\text{hi}}(n)^{1/(k+1)}))^2\right)$. In particular, any constant-space $\mathfrak{R}(k)$-refutation must also have quadratic length.*

Theorem 2.6 follows by combining our machinery with the seminal work on pebbling trade-offs by Lengauer and Tarjan [LT82] and the structural results on simulations of black-white pebblings by resolution in [Nor10b].

*Remark* 2.7. Notice that the trade-off applies to both formula space and total space. This is because the upper bound is stated in terms of the larger of these two measures (total space) while the lower bound is in terms of the smaller one (formula space). Note also that the upper bounds hold for the usual, syntactic versions of the proof systems, whereas the lower bounds hold for the much stronger semantic systems, and that for standard resolution the upper and lower bounds are tight up to constant factors. These properties of our results are inherited from the substitution space theorems, and they hold for all our trade-offs stated here and in Section 7. Finally, we remark that we have to pick some arbitrary but fixed limit $K$ for the size of the terms when stating the results for $k$-DNF resolution, since for any family of formulas we consider there will be very length- and space-efficient $\mathfrak{R}(k)$-refutation refutations if we allow terms of unbounded size.

Our next result relies on a new pebbling trade-off result in [Nor10b], building on earlier work by Carlson and Savage [CS80, CS82]. Using this new result, we can derive among other things the rather striking statement that for any *arbitrarily slowly growing* non-constant function, there are explicit formulas of such (arbitrarily small) space complexity that nevertheless exhibit *superpolynomial* length-space trade-offs.

**Theorem 2.8 (Superpolynomial trade-offs for arbitrarily slowly growing space).** *Let $s_{\text{lo}}(n) = \omega(1)$ be any arbitrarily slowly growing function[5] and fix any $\epsilon > 0$ and positive integer $K$. Then there are explicitly constructible CNF formulas $\{F_n\}_{n=1}^{\infty}$ of size $\Theta(n)$ and width $\text{O}(1)$ such that the following holds:*

- *The formulas $F_n$ are refutable in syntactic resolution in total space $TotSp_{\mathfrak{R}}(F_n \vdash 0) = \text{O}(s_{\text{lo}}(n))$.*

- *There are syntactic resolution refutations $\pi_n$ of $F_n$ in simultaneous length $L(\pi_n) = \text{O}(n)$ and total space $TotSp(\pi_n) = \text{O}\left((n/s_{\text{lo}}(n)^2)^{1/3}\right)$.*

- *Any semantic resolution refutation of $F_n$ in clause space $\text{O}\left((n/s_{\text{lo}}(n)^2)^{1/3-\epsilon}\right)$ must have superpolynomial length.*

- *For any $k \leq K$, any semantic $\mathfrak{R}(k)$-refutation of $F_n$ in formula space $\text{O}\left((n/s_{\text{lo}}(n)^2)^{1/(3(k+1))-\epsilon}\right)$ must have superpolynomial length.*

*All multiplicative constants hidden in the asymptotic notation depend only on $K$, $\epsilon$ and $s_{\text{lo}}$.*

---

[5] For technical reasons, let us also assume here that $s_{\text{lo}}(n) = \text{O}(n^{1/7})$, i.e., that $s_{\text{lo}}(n)$ does not grow too quickly. This restriction is inconsequential since for such fast-growing $s_{\text{lo}}(n)$ other trade-off results presented below will yield much stronger bounds.

Observe the robust nature of this trade-off, which is displayed by the long range of space complexity in standard resolution, from $\omega(1)$ up to $\approx n^{1/3}$, which requires superpolynomial length. Note also that the trade-off result for standard resolution is very nearly tight in the sense that the superpolynomial lower bound on length in terms of space reaches up to very close to where the linear upper bound kicks in.

The two theorems above focus on trade-offs for formulas of low space complexity, and the lower bounds on length obtained in the trade-offs are somewhat weak—the superpolynomial growth in Theorem 2.8 is something like $n^{s_{\text{lo}}(n)}$. We next present a theorem that has both a stronger superpolynomial length lower bounds than Theorem 2.8 and an even more robust trade-off covering a wider (although non-overlapping) space interval. This theorem again follows by applying our tools to the pebbling trade-offs in [LT82].

**Theorem 2.9 (Robust superpolynomial trade-off for medium-range space).** *For any positive integer $K$, there are explicitly constructible CNF formulas $\{F_n\}_{n=1}^{\infty}$ of size $\Theta(n)$ and width $\mathrm{O}(1)$ such that the following holds (where the hidden constants depend only on $K$):*

- *The formulas $F_n$ are refutable in syntactic resolution in total space $TotSp_{\mathfrak{R}}(F_n \vdash 0) = \mathrm{O}(\log^2 n)$.*

- *There are syntactic resolution refutations of $F_n$ in length $\mathrm{O}(n)$ and total space $\mathrm{O}(n/\log n)$.*

- *Any semantic resolution refutation of $F_n$ in clause space $Sp(\pi_n) = \mathrm{o}(n/\log n)$ must have length $L(\pi_n) = n^{\Omega(\log \log n)}$.*

- *For any $k \leq K$, any semantic $\mathfrak{R}(k)$-refutation in formula space $Sp(\pi_n) = \mathrm{o}\left((n/\log n)^{1/(k+1)}\right)$ must have length $L(\pi_n) = n^{\Omega(\log \log n)}$.*

Having presented trade-off results in the low-space and medium-space range, we conclude by presenting a result at the other end of the space spectrum. Namely, appealing one last time to yet another result in [LT82], we can show that there are formulas of nearly linear space complexity (recall that any formula is refutable in linear formula space) that exhibit not only superpolynomial but even exponential trade-offs.

We state this final theorem only for standard resolution since it is not clear whether it makes sense for $\mathfrak{R}(k)$. That is, we can certainly derive formal trade-off bounds in terms of the $(k+1)$st square root as in the theorems above, but we do not know whether there actually exist $\mathfrak{R}(k)$-refutation in sufficiently small space so that the trade-offs apply. Hence, such trade-off claims for $\mathfrak{R}(k)$, although impressive looking, might simply be vacuous. We can obtain other exponential trade-offs for $\mathfrak{R}(k)$ (see Section 7 for the details), but they are not quite as strong as the result below for resolution.

**Theorem 2.10 (Exponential trade-offs for nearly-linear space).** *Let $\kappa$ be any sufficiently large constant. Then there are CNF formulas $F_n$ of size $\Theta(n)$ and width $\mathrm{O}(1)$ and a constant $\kappa' \ll \kappa$ such that:*

- *The formulas $F_n$ have syntactic resolution refutations in total space $\kappa' \cdot n/\log n$.*

- *$F_n$ is also refutable in syntactic resolution in length $\mathrm{O}(n)$ and total space $\mathrm{O}(n)$ simultaneously.*

- *However, any semantic refutation of $F_n$ in clause space at most $\kappa \cdot n/\log n$ has length $\exp(n^{\Omega(1)})$.*

To get a feeling for this last trade-off result, note again that the lower bound holds for proof systems with arbitrarily strong derivation rules, as long as they operate with disjunctive clauses. In particular, it holds for proof systems that can in one step derive anything that is semantically implied by the current content of the blackboard. Recall that such a proof system can refute any unsatisfiable CNF formula $F$ with $n$ clauses in length $n+1$ simply by writing down all clauses of $F$ on the blackboard and then concluding, in one single derivation step, the contradictory empty clause implied by $F$. In Theorem 2.10 this proof system has space nearly sufficient for such an ultra-short refutation of the whole formula. But even so, when we feed this proof system the formulas $F_n$ and restrict it to having at most $\mathrm{O}(n/\log n)$ clauses on the blackboard at any one given time, it will have to keep going for an exponential number of steps before it is finished.

## 2.4 Proof Ingredients for Substitution Space Theorem for Resolution

Before embarking on the formal proofs of our theorems, we want to provide some intuition for the substitution space theorems that are the keys to our results. Let us first focus on the result for standard resolution and describe the proof structure in some detail. The analogous result for $k$-DNF resolution is proven in a similar way, but with the added technical complications that we need to prove size bounds on sets of $k$-DNF formulas. These issues, and in particular our result for minimally unsatisfiable sets of $k$-DNF formulas, are discussed in Section 2.5.

Thus, let $F$ be any unsatisfiable CNF formla and $f_d$ any non-authoritarian Boolean function (as described in Section 2.1), and let $F[f_d]$ denote the CNF formula obtained by substituting $f(x_1, \ldots, x_d)$ for every variable $x$ in $F$ and expanding the result to conjunctive normal form.

The first part of Theorem 2.1, that any resolution refutation of $F$ can be transformed into a refutation of $F[f_d]$ with similar parameters, is not hard to prove. Essentially, whenever the refutation of the original formula $F$ writes a clause $C$ on the blackboard, we write the corresponding set of clauses $C[f_d]$ on the blackboard where we are refuting the substitution formula. We make the additional observation that if we take any resolution refutation $\pi$ of $F$ and write down the new refutation $\pi_f$ of $C[f_d]$ resulting from this transformation—assuming for concreteness that the function $f_d$ is exclusive or, say—it is easy to verify that the number of variable occurrences in $\pi$, i.e., the *variable space*, translates into a lower bound on the number of clauses in $\pi_f$, i.e., the *formula space* (which as we recall is called *clause space* for standard resolution). Equation (2.1) provides an example of this variable-space-to-clause-space blow-up.

It is more challenging, however, to prove the reverse direction that we can get lower bounds on clause space for $F[f_d]$ from lower bounds on variable space for $F$. Ideally, we would like to claim that any prover refuting $F[f_d]$ had better write down to the blackboard clause sets on the form $C[f_d]$ corresponding to clauses $C$ in some refutation of the original CNF formula $F$, and that if he or she does not, then we can analyze the refutation as if that is what is happening anyway, just ignoring the clauses that do not fit into this framework.

To argue this more formally, we need to specify how sets of clauses in a refutation of $F[f_d]$ should be translated to clauses in a purported refutation of $F$. We do this by devising a way of "projecting" any refutation of $F[f_d]$ down on a refutation of $F$. These "projections" are defined in terms of a special kind of "precise implication" which we describe next. Recall that for Boolean functions $F$ and $G$, we say that $F$ *implies* $G$, denoted $F \vDash G$, if any truth value assignment satisfying $F$ must also satify $G$.

**Definition 2.11 (Precise implication and projected clauses (informal)).** Suppose that $\mathbb{D}$ is a set of clauses over variables in $Vars(F[f_d])$ and that $P$ and $N$ are (disjoint) subset of variables of $F$. If any truth value assignment satisfying $\mathbb{D}$ must also satisfy $\bigvee_{x \in P} f_d(\vec{x}) \vee \bigvee_{y \in N} \neg f_d(\vec{y})$ but this is not the case for strict subsets $P' \subsetneq P$ or $N' \subsetneq N$, we say that the clause set $\mathbb{D}$ implies $\bigvee_{x \in P} f_d(\vec{x}) \vee \bigvee_{y \in N} \neg f_d(\vec{y})$ *precisely*.

Let us write any clause $C$ as $C = C^+ \vee C^-$, where $C^+ = \bigvee_{x \in Lit(C)} x$ is the disjunction of the positive literals in $C$ and $C^- = \bigvee_{\overline{y} \in Lit(C)} \overline{y}$ is the disjunction of the negative literals. Then we say that $\mathbb{D}$ *projects* $C$ if $\mathbb{D}$ implies $\bigvee_{x \in C^+} f_d(\vec{x}) \vee \bigvee_{\overline{y} \in C^-} \neg f_d(\vec{y})$ precisely, and we write $proj_F(\mathbb{D})$ to denote the set of all clauses that $\mathbb{D}$ projects on $F$.

Given this definition, we want to take any refutation $\pi_f = \{\mathbb{D}_0, \mathbb{D}_1, \ldots, \mathbb{D}_\tau\}$ of $F[f_d]$ and argue that $\pi = \{proj_F(\mathbb{D}_0), proj_F(\mathbb{D}_1), \ldots, proj_F(\mathbb{D}_\tau)\}$ is (essentially) the refutation of $F$ that we are looking for.

It is not hard to see that for a "well-behaved" prover refuting $F[f_d]$ using a refutation $\pi$ of the original formula $F$ as a template but substituting the clause set $C[f_d]$ for every clause $C$ appearing on the blackboard, applying the projection in Definition 2.11 at every step in the derivation will give us back the refutation $\pi$ of $F$ that we started with (the reader can check that this is the case for instance for the clause set in (2.1)). What is more remarkable is that this projection of refutations of $F[f_d]$ always works no matter what the prover is doing, in the sense that the result is always a resolution refutation of the original formula $F$ and this projected refutation does not only (essentially) preserve length, which is not too complicated to show, but also space. We refer to Section 4 for the formal statements and proofs.

## 2.5 Substitution Space Theorem for $\Re(k)$ and Minimally Unsatisfiable $k$-DNF Sets

The proof of the first part of Theorem 2.2 is again reasonably straightforward and resembles our proof of the substitution theorem for the standard resolution proof system. For the second part, however, we require a result, described next, that bounds the number of variables appearing in a minimally unsatisfiable $k$-DNF set of a given size. Since this result addresses a combinatorial problem that appears to be interesting (and challenging) in its own right, we describe it in some detail below.

We start by recalling that a set of 1-DNF formulas, i.e., a CNF formula, is said to be *minimally unsatisfiable* if it is unsatisfiable but every proper subset of its clauses is satisfiable, and try to generalize this definition to the case of $k > 1$. Perhaps the first, naive, idea how to extend this notion is to define $\mathbb{D}$ to be minimally unsatisfiable if it is unsatisfiable but all proper subsets of it are satisfiable. This will not work, however, and the set of formulas

$$\{x, \big((\overline{x} \wedge y_1) \vee (\overline{x} \wedge y_2) \vee (\overline{x} \wedge y_3) \vee \cdots \vee (\overline{x} \wedge y_n)\big)\} \tag{2.2}$$

shows why this approach is problematic. The set (2.2), which consists of two 2-DNF formulas, is unsatisfiable but every proper subset of it is satisfiable. However, the number of variables appearing in the set can be arbitrarily large so there is no way of bounding $|Vars(\mathbb{D})|$ as a function of $|\mathbb{D}|$.

A more natural requirement is to demand minimality not only at the formula level but also at the term level, saying that not only do all DNF formulas in the set have to be there but also that no term in any formula can be shrunk to a smaller, weaker term without the set becoming satisfiable. Luckily enough, this also turns out to be the concept we need for our applications. The formal definition follows next.

**Definition 2.12 (Minimal implication and minimally unsatisfiable $k$-DNF sets).** Let $\mathbb{D}$ be a set of $k$-DNF formulas and let $G$ be a formula. We say that $\mathbb{D}$ *minimally implies* $G$ if $\mathbb{D} \vDash G$ and furthermore, replacing any single term $T$ appearing in a single DNF formula $D \in \mathbb{D}$ with a proper subterm of $T$, and calling the resulting DNF set $\mathbb{D}'$, results in $\mathbb{D}' \nvDash G$. If $G$ is unsatisfiable we say $\mathbb{D}$ is *minimally unsatisfiable*.

To see that this definition generalizes the notion of a minimally unsatisfiable CNF formula, notice that removing a clause $C'$ from a CNF formula $F$ is equivalent to replacing a term of $C'$, which is a single literal, with a proper subterm of it, which is the empty term. This is because the empty term evaluates to $1$ on all assignments, which means that the resulting clause also evaluates to $1$ on all assignments and hence can be removed from $F$. With this definition in hand, we are thus interested in understanding the following problem:

> *Given a minimally unsatisfiable set of $m$ $k$-DNF formulas, what is an upper bound on the number of variables that this set of formulas can contain?*

As was noted above, for $k = 1$ the set $\mathbb{D}$ is equivalent to a CNF formula, because it is a set of disjunctions of literals, and we have the following "folklore" result which seems to have been proved independently on several different occasions (see [AL86, BET01, CS88, Kul00]).

**Theorem 2.13.** *If $\mathbb{D}$ is a minimally unsatisfiable CNF formula, then $|Vars(\mathbb{D})| < |\mathbb{D}|$.*

Theorem 2.13 has a relatively elementary proof based on Hall's marriage theorem, but its importance to obtaining lower bounds on resolution length and space is hard to overemphasize. For instance, the seminal lower bound on refutation length of random CNFs given by Chvátal and Szemerédi in [CS88] makes crucial use of it, as does the proof of the "size-width trade-off" of [BW01]. Examples of applications of this theorem in resolution space lower bounds include [ABRW02, BG03, BN08, Nor09a, NH08].

For sets of $k$-DNF formulas with $k > 1$, we are not aware of any upper or lower bounds on minimally unsatisfiable sets prior to our work. The main technical result that we need in order to establish the $k$-DNF resolution space hierarchy is the following extension of Theorem 2.13 to the case of $k > 1$.

**Theorem 2.14.** *Suppose that $\mathbb{D}$ is a minimally unsatisfiable $k$-DNF set. Then the number of variables in $\mathbb{D}$ is at most $|Vars(\mathbb{D})| \leq (k \cdot |\mathbb{D}|)^{k+1}$.*

*Proof sketch.* Let us sketch the proof for $k = 2$. (The full proof is given in Section 5.) Suppose that we have a 2-DNF set $\mathbb{D}$ with $m$ formulas mentioning $\Omega(m^3)$ variables. Then there is at least one 2-DNF formula $D^*$ mentioning $\Omega(m^2)$ variables. By the definition of minimality, the set $\mathbb{D} \setminus \{D\}$ is satisfiable. Let $\alpha$ be some minimal partial assignment fixing $\mathbb{D} \setminus \{D\}$ to true, and note that $\alpha$ needs to set at most $2(m-1)$ variables (at most one 2-term per formula).

Consider the 2-terms in $D^*$. If there are $2m$ terms over completely disjoint pairs of variables, then there is some 2-term $a \wedge b$ untouched by $\alpha$. If so, $\alpha$ can be extended to a satisfying assignment for all of $\mathbb{D}$, which is a contradiction. Hence there are at most $O(m)$ terms over disjoint sets of variables.

But $D^*$ contains $\Omega(m^2)$ variables. By counting (and adjusting the implicit constant factors), there must exist some literal $a^*$ in $D^*$ occurring in a lot of terms $(a^* \wedge b_1) \vee (a^* \wedge b_2) \vee \cdots (a^* \wedge b_{2m})$. Again by minimality, there is a (partial) truth value assignment $\alpha'$ satisfying $\mathbb{D} \setminus \{D\}$ *and setting $a^*$ to true.* (To see this, note that shrinking, for instance, $a^* \wedge b_1$ to $a^*$ should make the whole set satisfiable). But if we pick such an $\alpha'$ of minimal size, there must exist some $b_i$ that is not falsified and we can extend $\alpha'$ to a satisfying assignment for $a^* \wedge b_i$ and hence for the whole set. Contradiction. $\square$

We want to point out that in contrast to Theorem 2.13, which is exactly tight (consider the set

$$\left\{\bigvee_{i=1}^{n} x_i,\ \neg x_1,\ \neg x_2,\ \ldots,\ \neg x_n\right\} \tag{2.3}$$

of $n+1$ clauses over $n$ variables), there is no matching lower bound on the number of variables in Theorem 2.14. The best explicit construction that we were able to obtain, stated next, has number of variables only *linear* in the number of $k$-DNF formulas (for $k$ constant), improving only by a factor $k^2$ over the bound for CNF formulas in Theorem 2.13.

**Lemma 2.15.** *There are minimally unsatisfiable $k$-DNF sets $\mathbb{D}$ with $|Vars(\mathbb{D})| \geq k^2(|\mathbb{D}| - 1)$.*

*Proof sketch.* Consider any minimally unsatisfiable CNF formula consisting of $n+1$ clauses over $n$ variables (for instance, the one in (2.3)). Substitute every variable $x_i$ with

$$\left(x_i^1 \wedge x_i^2 \wedge \cdots \wedge x_i^k\right) \vee \left(x_i^{k+1} \wedge x_i^{k+2} \wedge \cdots \wedge x_i^{2k}\right) \vee \cdots \vee \left(x_i^{k^2-k+1} \wedge x_i^{k^2-k+2} \wedge \cdots \wedge x_i^{k^2}\right) \tag{2.4}$$

and expand every clause to a $k$-DNF formula. Note that this is possible since the negation of (2.4) that we need to substitute for $\neg x_i$ can also be expressed as a $k$-DNF formula

$$\bigvee_{(j_1,\ldots,j_k) \in [1,k] \times \ldots \times [(k^2-k+1,k^2]} \left(\neg x_i^{j_1} \wedge \cdots \wedge \neg x_i^{j_k}\right)\ . \tag{2.5}$$

It is straightforward to verify that the result is a minimally unsatisfiable $k$-DNF set in the sense of Definition 2.12, and this set has $n+1$ formulas over $k^2n$ variables. $\square$

In our first preliminary report [BN09] on our results for $k$-DNF resolution, we stated that we saw no particular reason to believe that the upper bound in Theorem 2.14 should be tight, hinting that Lemma 2.15 might well be closer to the truth. Surprisingly to us, this turned out to be wrong. In a joint work [NR09] with Razborov, the second author recently showed that there are minimally unsatisfiable $k$-DNF sets with $m$ formulas and $\approx m^k$ variables, which means that Theorem 2.14 is tight up to an additive one in the exponent.

Concluding this section, we remark that the precise statement required to prove the second part of Theorem 2.2 is somewhat more involved than Theorem 2.14. However, the two proofs follow each other very closely. We refer to Section 5 for the details.

## 3 Preliminaries

In this section we give the basic formal definitions used in this paper and state a few well-known facts that we will need.

### 3.1 Formulas, Assignments and Restrictions

For the most part we use standard notation for formulas in conjunctive normal form (CNF) and disjunctive normal form (DNF). However, we will also use the very convenient, although perhaps slightly less standard, set notation to treat objects such as clauses, terms, restrictions and CNF and DNF formulas. We explain this notation and terminology next. Throughout this paper, we let $[n]$ denote the set $\{1, 2, \ldots, n\}$.

**DNF and CNF formulas as sets**  For $x$ a Boolean variable, a *literal over* $x$ is either a Boolean variable $x$, called a *positive literal over* $x$, or its negation, denoted $\neg x$ or $\overline{x}$ and called a *negative literal over* $x$. Sometimes it will also be convenient to write $x^1$ for unnegated variables and $x^0$ for negated ones. We define $\neg\neg x$ to be $x$. When $x$ is understood from context or unimportant we simply speak of a (positive, negative) *literal*. A *CNF formula* is a set of clauses, i.e., disjunctions of literals, and a *DNF formula* is a set of terms, i.e., conjunctions of literals. The *variable set* of a term $T$, denoted $Vars(T)$, is the set of Boolean variables over which there are literals in $T$ and $Lit(T)$ is the set of literals. The variable and literal sets of a clause are similarly defined and these definitions are extended to CNF and DNF formulas by taking unions. If $X$ is a set of Boolean variables and $Vars(T) \subseteq X$ we say $T$ is a term *over* $X$ and similarly define clauses, CNF formulas, and DNF formulas over $X$.

We will sometimes think of *clauses* and *terms* as sets of literals and borrow set-theoretic notation and terminology to discuss logical formulas. For instance, we say that the term $T'$ is a *subterm* of $T$, and write $T' \subseteq T$ to denote that the set of literals of $T'$ is contained in the set of literals of $T$. We similarly speak of, and denote, subclauses and subformulas. We say the clause $C$ (or term $T$) is a $k$-*clause* ($k$-*term*, respectively) if $|C| \leq k$ ($|T| \leq k$, respectively). A $k$-*DNF formula* $D$ is a set of $k$-terms and a $k$-*CNF formula* is a set of $k$-clauses. We define the *size* $S(F)$ of any formula $F$ to be the total number of literals in $F$ counted with repetitions. More often, we will be interested in the number of clauses $|F|$ of a CNF formula $F$ or the number of terms in a DNF formula. We write $|D|$ to denote is the number of terms in a $k$-DNF formula and correspondingly for the number of clauses in a CNF formula.

**Assignments and restrictions as sets**  As is the case with CNF and DNF formulas, we prefer to use in our proof a set-theoretic representation of restrictions and assignments, as defined next.

A *restriction* $\rho$ over a set of Boolean variables $X$ is a subset of literals over $X$ with the property that for each variable $x \in X$ there is at most one literal over $x$ in $\rho$. The *set of variables assigned* by $\rho$ is $Vars(\rho)$ and the *size* of $\rho$ is $|\rho| = |Vars(\rho)|$. We say the restriction $\rho'$ *extends* $\rho$ if $\rho' \supseteq \rho$, and in this case we also say that $\rho$ *agrees* with $\rho'$. An *assignment* $\alpha$ to $X$ is a restriction satisfying $|\alpha| = |X|$.

## 3  Preliminaries

For $a$ a literal over $X$ and $\rho$ a restriction over $X$, let the restriction of $a$ under $\rho$ be

$$a\restriction_\rho = \begin{cases} 1 & a \in \rho \\ 0 & \neg a \in \rho \\ a & \text{otherwise} \end{cases} \tag{3.1}$$

If $a\restriction_\rho = 1$ we say $\rho$ *satisfies* $a$, if $a\restriction_\rho = 0$ we say $\rho$ *falsifies* $a$ and otherwise we say $\rho$ leaves $a$ *unfixed*. We extend the definition of a restriction to a term $T = a_1 \wedge \cdots \wedge a_s$ and clause $C = a'_1 \vee \cdots \vee a'_s$ as follows. Let $\neg\rho = \{\neg a | a \in \rho\}$ denote the restriction obtained by replacing every literal in $\rho$ by its negation.

$$T\restriction_\rho = \begin{cases} 1 & T \subseteq \rho \\ 0 & T \cap \neg\rho \neq \emptyset \\ T \setminus \rho & \text{otherwise} \end{cases}, \quad C\restriction_\rho = \begin{cases} 0 & C \subseteq \neg\rho \\ 1 & C \cap \rho \neq \emptyset \\ C \setminus \neg\rho & \text{otherwise} \end{cases} \tag{3.2}$$

In words, we say $T$ is *satisfied* by $\rho$ if $\rho$ satisfies all literals in $T$, we say $T$ is *falsified* by $\rho$ if some literal of $\rho$ is falsified and otherwise $T$ is unfixed by $\rho$. Dually, $C$ is satisfied if some literal of it is satisfied by $\rho$, it is falsified if all its literals are falsified by $\rho$ and otherwise it remains unfixed. Notice that the *empty term*, i.e., the term of size $0$, is satisfied by every restriction and the *empty clause* is falsified by all of them. We extend the definition of a restriction to a DNF formula $D = D_1 \vee \cdots \vee D_m = \{D_1, \ldots, D_m\}$ by

$$D\restriction_\rho = \begin{cases} 1 & \exists i \in [m], D_i\restriction_\rho = 1 \\ 0 & D_i\restriction_\rho = 0, i \in [m] \\ \{D_i\restriction_\rho \,:\, D_i\restriction_\rho \neq 0\} & \text{otherwise,} \end{cases} \tag{3.3}$$

and to a CNF formula $F = C_1 \wedge \cdots \wedge C_m = \{C_1, \ldots, C_m\}$ by

$$F\restriction_\rho = \begin{cases} 0 & \exists i \in [m], C_i\restriction_\rho = 0 \\ 1 & C_i\restriction_\rho = 1, i \in [m] \\ \{C_i\restriction_\rho \,:\, C_i\restriction_\rho \neq 1\} & \text{otherwise.} \end{cases} \tag{3.4}$$

The notions of a restriction satisfying, falsifying and leaving unfixed a DNF or CNF formula are analogous to those defined for terms and clauses. If $\rho$ is a restriction satisfying a formula $F$, yet every proper subrestriction $\rho' \subsetneq \rho$ does not satisfy $F$, then we say $\rho$ is a *minimal* satisfying restriction. A minimal falsifying restriction is analogously defined. We write $\rho(\neg C)$ to denote the minimal restriction fixing a clause $C$ to false, i.e., $\rho(\neg C) = \{\overline{a} \mid a \in C\}$.

When $\alpha$ is a truth value assignment to the variables in a formula $F$, it will be notationally convenient to think of $\alpha$ as assigning a truth value to $F$. We write $\alpha(F)$ to denote this assigned value, or sometimes $F\restriction_\alpha$ when we think of $\alpha$ as a restriction. We will use these two notations interchangeably. We write

$$\alpha^{x=\nu}(y) = \begin{cases} \alpha(y) & \text{if } y \neq x, \\ \nu & \text{if } y = x, \end{cases} \tag{3.5}$$

to denote the truth value assignment that agrees with $\alpha$ everywhere except possibly at $x$, to which it assigns the value $\nu$.

A term (clause, respectively) is said to be *trivial* if it contains both a positive and a negative literal over the same variable. We may assume without loss of generality that all terms (clauses, respectively) appearing in our paper are nontrivial, because the value of a DNF (CNF, respectively) remains unchanged after addition or removal of trivial terms (clauses, respectively). We say that a DNF formula $D$ over $X$ *represents* a Boolean function $f : X \to \{0, 1\}$ if and only if for all assignments $\alpha \in \{0, 1\}^X$, we have $f(\alpha) = D(\alpha)$. The notion of a CNF formula representing $f$ is analogously defined. Clearly, every Boolean function $f$ can be represented by a DNF formula (for instance, the disjunction of all terms corresponding to satisfying truth value assignments) as well as by a CNF formula (for instance, the conjunction of all clauses ruling out falsifying assignments for $f$).

**Implication** If $\mathbb{C}$ is a set of formulas we say that a restriction (or assignment) *satisfies* $\mathbb{C}$ if and only if it satisfies every formula in $\mathbb{C}$. For $\mathbb{D}, \mathbb{C}$ two sets of formulas over a set of variables $X$, we say that $\mathbb{D}$ *implies* $\mathbb{C}$, denoted $\mathbb{D} \vDash \mathbb{C}$, if and only if every assignment $\alpha$ to $X$ that satisfies $\mathbb{D}$ also satisfies $\mathbb{C}$. In particular, $\mathbb{D} \vDash 0$ if and only if $\mathbb{D}$ is *unsatisfiable*, i.e., no assignment satisfies $\mathbb{D}$.

## 3.2  $k$-DNF Resolution

We now give a more precise description of the $k$-DNF resolution proof systems and the proof complexity measures for these systems that we are interested in studying.

**Definition 3.1 ($k$-DNF-resolution inference rules).** The $k$-*DNF-resolution* proof systems are a family of sequential proof systems parameterized by $k \in \mathbb{N}^+$. Lines in a $k$-DNF-resolution refutation are $k$-DNF formulas and the following inference rules are allowed (where $A, B, C$ denote $k$-DNF formulas, $T, T'$ denote $k$-terms, and $a_1, \ldots, a_k$ denote literals):

$k$-***cut***  $\dfrac{(a_1 \wedge \ldots \wedge a_{k'}) \vee B, \quad \neg a_1 \vee \ldots \vee \neg a_{k'} \vee C}{B \vee C}$, where $k' \leq k$.

$\wedge$-***introduction***  $\dfrac{A \vee T, \quad A \vee T'}{A \vee (T \wedge T')}$, as long as $|T \cup T'| \leq k$.

$\wedge$-***elimination***  $\dfrac{A \vee T}{A \vee T'}$ for any $T' \subseteq T$.

***Weakening***  $\dfrac{A}{A \vee B}$ for any $k$-DNF formula $B$.

The formulas above the line are called the *inference assumptions* and the formula below is called the *consequence*. For brevity we denote by $\mathfrak{R}(k)$ the proof system of $k$-DNF resolution.

When we want to study length and space simultaneously in resolution, we have to be slightly careful with the definitions in order to capture length-space trade-offs. Just listing the clauses used in a resolution refutation does not tell us *how* the refutation was performed, and essentially the same refutation can be carried out in vastly different time depending on the space constraints (as is shown in this paper). Following the exposition in [ET01], therefore, we view a resolution refutation as a Turing machine computation, with a special read-only input tape from which the axioms can be downloaded and a working memory where all derivation steps are made. Then the length of a proof is essentially the time of the computation and space measures memory consumption. The following definition is the straightforward generalization to $\mathfrak{R}(k)$ of the space-oriented definition of a refutation from [ABRW02].

**Definition 3.2 (Derivation).** A $k$-*DNF configuration* $\mathbb{D}$, or, simply, a *configuration*, is a set of $k$-DNF formulas. A sequence of configurations $\{\mathbb{D}_0, \ldots, \mathbb{D}_\tau\}$ is said to be a $\mathfrak{R}(k)$-*derivation* from a CNF formula $F$ if $\mathbb{D} = \emptyset$ and for all $t \in [\tau]$, the set $\mathbb{D}_t$ is obtained from $\mathbb{D}_{t-1}$ by one of the following *derivation steps*:

***Axiom Download***  $\mathbb{D}_t = \mathbb{D}_{t-1} \cup \{C\}$ for some $C \in F$.

***Inference***  $\mathbb{D}_t = \mathbb{D}_{t-1} \cup \{D\}$ for some $D$ inferred by one of the inference rules listed in Definition 3.1 from a set of assumptions that belongs to $\mathbb{D}_{t-1}$.

***Erasure***  $\mathbb{D}_t = \mathbb{D}_{t-1} \setminus \{D\}$ for some $D \in \mathbb{D}_{t-1}$.

A $\mathfrak{R}(k)$-derivation $\pi : F \vdash \mathbb{D}'$ of a $k$-DNF set $\mathbb{D}'$ from a formula $F$ is a derivation $\pi = \{\mathbb{D}_0, \ldots, \mathbb{D}_\tau\}$ such that $\mathbb{D}_\tau = \mathbb{D}'$. A $\mathfrak{R}(k)$-*refutation* of $F$ is a $\mathfrak{R}(k)$-derivation of the empty DNF (denoted by 0), i.e., the DNF formula with no terms, or, phrased differently, the unsatisfiable empty disjunction.

When the derived $k$-DNF set $\mathbb{D}'$ contains a single formula $D$, we will often abuse notation slightly by writing simply $\pi : F \vdash D$ instead of $\pi : F \vdash \{D\}$.

**Definition 3.3 (Refutation length and space).** The *formula space*, or simply *space*, of a configuration $\mathbb{D}$ is its size $|\mathbb{D}|$. The *variable space*[6] of $\mathbb{D}$, denoted $VarSp(\mathbb{D})$, is the number of literals appearing in $\mathbb{D}$, i.e., $VarSp(\mathbb{D}) = |Vars(\mathbb{D})|$ and the *total space* of $\mathbb{D}$, denoted $TotSp(\mathbb{D})$ is the number of literals appearing in $\mathbb{D}$ counted with repetitions. (Notice that $TotSp(\mathbb{D}) \geq VarSp(\mathbb{D})$.)

The *length* of a $\mathfrak{R}(k)$-derivation $\pi$ is the number of axiom downloads and inference steps in it.[7] The space (variable space, total space, respectively) of a derivation $\pi$ is defined as the maximal space (variable space, total space, respectively) of a configuration in $\pi$. If $\pi$ is a derivation of $\mathbb{D}$ from a formula $F$ of length $L$ and space $s$ then we say $\mathbb{D}$ can be derived from $F$ in length $L$ and space $s$ *simultaneously*.

We define the $\mathfrak{R}(k)$-*refutation length* of a formula $F$, denoted $L_{\mathfrak{R}(k)}(F \vdash 0)$, to be the minimum length of any $\mathfrak{R}(k)$-refutation of it. The $\mathfrak{R}(k)$-*refutation space* of $F$, denoted $Sp_{\mathfrak{R}(k)}(F \vdash 0)$, the $\mathfrak{R}(k)$-*refutation total space* $TotSp_{\mathfrak{R}(k)}(F \vdash 0)$, and the $\mathfrak{R}(k)$-*refutation variable space* of $F$, denoted $VarSp_{\mathfrak{R}(k)}(F \vdash 0)$, are analogously defined by taking minima over all $\mathfrak{R}(k)$-refutations of $F$. (When the proof system $\mathfrak{R}(k)$ in question is clear from context, we will drop the subindex in the proof complexity measures.)

Notice that the system $\mathfrak{R}(1) = \mathfrak{R}$ is the usual *resolution* proof system. We remark that in resolution, the $\wedge$-introduction and $\wedge$-elimination rules do not apply, and the cut rule reduces to the familiar *resolution rule* saying that the clauses $C_1 \vee x$ and $C_2 \vee \neg x$ can be combined to derive $C_1 \vee C_2$. The formula space measure in resolution is known as *clause space*. For resolution, it is also convenient to define the auxiliary measure of *width*.

**Definition 3.4 (Resolution width).** The *width* $W(C)$ of a clause $C$ is the number of literals in it and the width $W(F)$ of a formula $F$ is the size of a widest clause in $F$. The width $W(\pi)$ of a derivation $\pi$ is defined in the same way, and $W(F \vdash 0)$ denotes the minimum width of any resolution refutation of $F$.

Using the same terminology as in [ABRW02], we also define a stronger version of the $k$-DNF resolution proof systems.

**Definition 3.5 (Semantic $k$-DNF resolution).** The *semantic $k$-DNF resolution* proof systems are defined as in Definitions 3.1 and 3.2 but with the modification that the inference rule is that *any $k$-DNF formula* implied by $\mathbb{D}_t$ can be derived at time $t+1$. We refer to the systems in Definitions 3.1 and 3.2 as *syntactic* to distinguish them from the semantic ones.

Clearly, semantic systems are much stronger than syntactic ones. In particular, any unsatisfiable CNF formula with $m$ clauses has a semantic resolution refutation of length $m + 1$, since we can just list all clauses in the formula and then derive the empty clause in one step (which is implied since the formula is unsatisfiable). Note, however, that this refutation also has space $m + 1$. This is in stark contrast to the fact that there are exponential lower bounds known on length for $k$-DNF resolution for any $k$. The reason that we distinguish between semantic and syntactic proof systems is that all our upper bounds in the trade-offs are in terms of syntactic systems (i.e., the usual ones) whereas the lower bounds in the trade-offs hold even for semantic proof systems.

### 3.3 Some Structural Results for Resolution

Although the weakening rule is sometimes convenient for technical reasons, for the case of standard resolution it is easy to show that any weakening steps can always be eliminated from a refutation of an unsatifiable

---

[6]We remark that there is some terminological confusion in the literature here. The term "variable space" has also been used in previous papers (including by the current authors) to refer to what is here called "total space." The terminology adopted in this paper is due to Hertel and Urquhart (see [Her08]), and we feel that although this naming convention is as of yet less well-established, it feels much more natural than the alternative.

[7]As noted above, the reader who so prefers can instead define the length of a derivation $\pi = \{\mathbb{D}_0, \ldots, \mathbb{D}_\tau\}$ as the number of steps $\tau$ in it, since the difference is at most a factor of 2. We have chosen the definition above for consistency with previous papers defining length as the number of clauses in a listing of the derivation.

CNF formula without changing anything essential. Thus, while the results in this paper will be stated for resolution with the weakening rule, they also hold for resolution refutations using only axiom downloads, resolution rule applications and erasures. Let us highlight this fact in a (somewhat) formal proposition for the record.

**Proposition 3.6.** *Any resolution refutation $\pi : F \vdash 0$ using the weakening rule can be transformed into a refutation $\pi' : F \vdash 0$ without weakening in at most the same length, width, clause space, total space, and variable space, and performing at most the same number of axiom downloads, inferences and erasures as $\pi$.*

The proof of Proposition 3.6 is an easy forward induction over the resolution refutation (simply ignoring all weakening moves and keeping the subclauses instead, which can never increase any measure we are interested in). We omit the details.

Another convenient fact is that restrictions preserve resolution refutations in the following sense.

**Proposition 3.7.** *If $\pi$ is a resolution refutation of $F$ and $\rho$ is a restriction on $Vars(F)$, then $\pi\!\restriction_\rho$ can be transformed into a resolution refutation of $F\!\restriction_\rho$ in at most the same length, width, clause space, total space, and variable space as $\pi$.*

In fact, $\pi\!\restriction_\rho$ is a refutation of $F\!\restriction_\rho$ (removing all trivially true clauses), but possibly using weakening. The proof of this is again an easy induction over $\pi$.

We will also make use of the *implicational completeness* of resolution. Formally, this means that if $\mathbb{C}$ is a set of clauses and $C$ is a clause, then $\mathbb{C} \vDash C$ if and only if there exists a resolution derivation of $C$ from $\mathbb{C}$.

**Proposition 3.8.** *Suppose $\mathbb{C}$ is a set of clauses and $C$ is a clause, both over a set of variables of size $n$. Then $\mathbb{C} \vDash C$ if and only if there exists a resolution derivation of $C$ from $\mathbb{C}$. Furthermore, if $C$ can be derived from $\mathbb{C}$ then it can be derived in length at most $2^{n+1} - 1$ and total space at most $n(n + 2)$ simultaneously.*

*Proof sketch.* Suppose first that $C = 0$ is the contradictory empty clause. Build a search tree where all vertices on level $i$ query the $i$th variable and where we go to the left, say, if the variable is false under a given truth value assignment $\alpha$ and to the right if the variable is true. As soon as some clause in $\mathbb{C}$ is falsified by the partial assignment defined by the path to a vertex, we make that vertex into a leaf labelled by that clause. This tree has height $h \leq n$ and hence size at most $2^{h+1} - 1$, and if we turn it upside down we can obtain a legal tree-like refutation (without weakening) of $\mathbb{C}$ in this length. This refutation can be carried out in clause space $h + 2$ and total space upper-bounded by the clause space times the number of distinct variables, i.e., at most $n(n + 2)$. (We refer to, for instance, [Ben09, ET01] for more details.)

If $C \neq 0$, apply the unique minimal restriction $\rho$ falsifying $C$. Then $\mathbb{C}\!\restriction_\rho \vDash C\!\restriction_\rho = 0$, and we can construct a refutation of $\mathbb{C}\!\restriction_\rho$ from a search tree of height $h < n$, since $\mathbb{C}\!\restriction_\rho$ contains strictly fewer variables than $\mathbb{C}$. Removing the restriction $\rho$ from this refutation, and adding at most one extra weakening step for every other derivation step (this is an example of where the weakening rule comes in handy), we get a derivation of $C$ from $\mathbb{C}$. (See [BW01] for a formal proof of this fact.) This derivation has length at most $2 \cdot (2^{h+1} - 1) < 2^{n+1} - 1$ and total space at most $n(h + 2) < n(n + 2)$. □

In a resolution refutation of a formula $F$, there is nothing in Definition 3.2 that rules out that completely unnecessary derivation steps are made on the way, such as axioms being downloaded and then immediately erased again, or entire subderivations being made to no use. In our constructions it will be important that we can rule out some redundancies and enforce the following requirements for any resolution refutation:

- Every clause in memory is used in an inference step before being erased.

- Every clause is erased from memory immediately after having been used for the last time.

We say that a resolution refutation that meets these requirements is *frugal*. The formal definition, which is a mildly modified version of that in [Ben09], follows.

**Definition 3.9 (Frugal refutation).** Let $\pi = \{\mathbb{C}_0 = \emptyset, \mathbb{C}_1, \ldots, \mathbb{C}_\tau = \{0\}\}$ be a resolution refutation of some CNF formula $F$. The *essential clauses* in $\pi$ are defined by backward induction:

- If $\mathbb{C}_t$ is the first configuration containing $0$, then $0$ is essential at time $t$.

- If $D \in \mathbb{C}_t$ is essential and is inferred at time $t$ from $C_1, C_2 \in \mathbb{C}_{t-1}$ by resolution, then $C_1$ and $C_2$ are essential at time $t-1$.

- If $D$ is essential at time $t$ and $D \in \mathbb{C}_{t-1}$, then $D$ is essential at time $t-1$.

*Essential clause configurations* are defined by forward induction over $\pi$. The configuration $\mathbb{C}_t \in \pi$ is essential if all clauses $D \in \mathbb{C}_t$ are essential at time $t$, if $\mathbb{C}_t$ is obtained by inference of an essential clause from a configuration $\mathbb{C}_{t-1}$ containing only essential clauses at time $t-1$, or if $\mathbb{C}_t$ is obtained from an essential configuration $\mathbb{C}_{t-1}$ by an erasure step. Finally, $\pi = \{\mathbb{C}_0, \ldots, \mathbb{C}_r\}$ is a *frugal refutation* if all configurations $\mathbb{C}_t \in \pi$ are essential.

Without loss of generality, we can always assume that resolution refutations are frugal.

**Lemma 3.10.** *Any resolution refutation $\pi : F \vdash 0$ can be converted into a frugal refutation $\pi' : F \vdash 0$ without increasing the length, width, clause space, total space, or variable space. Furthermore, the axiom downloads, inferences and erasures performed in $\pi'$ are a subset of those in $\pi$.*

*Proof.* The construction of $\pi'$ is by backward induction over $\pi$. Set $s = \min\{t : 0 \in \mathbb{C}_t\}$ and $\mathbb{C}'_s = \{0\}$. Assume that $\mathbb{C}'_s, \mathbb{C}'_{s-1}, \ldots \mathbb{C}'_{t+2}, \mathbb{C}'_{t+1}$ have been constructed and consider $\mathbb{C}_t$ and the transition $\mathbb{C}_t \rightsquigarrow \mathbb{C}_{t+1}$.

*Axiom Download* $\mathbb{C}_{t+1} = \mathbb{C}_t \cup \{C\}$: Set $\mathbb{C}'_t = \mathbb{C}'_{t+1} \setminus \{C\}$. (If $C$ is not essential we get $\mathbb{C}'_t = \mathbb{C}'_{t+1}$.)

*Erasure* $\mathbb{C}_{t+1} = \mathbb{C}_t \setminus \{D\}$: Ignore, i.e., set $\mathbb{C}'_t = \mathbb{C}'_{t+1}$.

*Inference* $\mathbb{C}_{t+1} = \mathbb{C}_t \cup \{D\}$ inferred from $C_1, C_2 \in \mathbb{C}_t$: If $D \notin \mathbb{C}'_{t+1}$, ignore the step and set $\mathbb{C}'_t = \mathbb{C}'_{t+1}$. Otherwise (using fractional time steps for notational convenience) insert the configurations $\mathbb{C}'_t = \mathbb{C}'_{t+1} \cup \{C_1, C_2\} \setminus \{D\}$, $\mathbb{C}'_{t+\frac{1}{3}} = \mathbb{C}'_{t+1} \cup \{C_1, C_2\}$, $\mathbb{C}'_{t+\frac{2}{3}} = \mathbb{C}'_{t+1} \cup \{C_2\}$.

Finally go through $\pi'$ and eliminate any consecutive duplicate clause configurations.

It is straightforward to check that $\pi'$ is a legal resolution refutation. Let us verify that $\pi'$ is frugal. By backward induction, each $\mathbb{C}'_t$ for integral time steps $t$ contains only essential clauses. By forward induction, if $\mathbb{C}'_{t+1} = \mathbb{C}'_t \cup \{C\}$ is obtained by axiom download, all clauses in $\mathbb{C}'_{t+1}$ are essential. Erasures in $\pi$ are ignored. For inference steps, $\mathbb{C}'_t$ contains only essential clauses by induction, $\mathbb{C}'_{t+\frac{1}{3}}$ is essential by inference, and $\mathbb{C}'_{t+\frac{2}{3}}$ and $\mathbb{C}'_{t+1}$ are essential since they are derived by erasure from essential configurations. Finally, it is clear that $\pi'$ performs a subset of the derivation steps in $\pi$ and that the length, width, and space does not increase. □

## 3.4 Pebble Games and Pebbling Contradictions

Pebble games were devised for studying programming languages and compiler construction, but have found a variety of applications in computational complexity theory. In connection with resolution, pebble games have been employed both to analyze resolution derivations with respect to how much memory they consume (using the original definition of space in [ET01]) and to construct CNF formulas which are hard for different variants of resolution in various respects (see for example [AJPU02, BIW04, BEGJ00, BP03] and the sequence of papers [Nor09a, NH08, BN08] leading up to this work). An excellent survey of pebbling up to ca. 1980 is [Pip80]. We also refer the interested reader to the upcoming survey [Nor10a], which contains some later results and also describes connections between pebbling and proof complexity.

The black pebbling price of a DAG $G$ captures the memory space, i.e., the number of registers, required to perform the deterministic computation described by $G$. The space of a non-deterministic computation is measured by the black-white pebbling price of $G$. We say that vertices of $G$ with indegree 0 are *sources* and that vertices with outdegree 0 are *sinks* (or *targets*). In the following, unless otherwise stated we will assume that all DAGs under discussion have a unique sink and this sink will always be denoted $z$. The next definition is adapted from [CS76], though we use the established pebbling terminology introduced by [HPV77].

**Definition 3.11 (Black-white pebble game).** Suppose that $G$ is a DAG with sources $S$ and a unique sink $z$. The *black-white pebble game* on $G$ is the following one-player game. At any point in the game, there are black and white pebbles placed on some vertices of $G$, at most one pebble per vertex. A *pebble configuration* is a pair of subsets $\mathbb{P} = (B, W)$ of $V(G)$, comprising the black-pebbled vertices $B$ and white-pebbled vertices $W$. The rules of the game are as follows:

1. If all immediate predecessors of an empty vertex $v$ have pebbles on them, a black pebble may be placed on $v$. In particular, a black pebble can always be placed on any vertex in $S$.

2. A black pebble may be removed from any vertex at any time.

3. A white pebble may be placed on any empty vertex at any time.

4. If all immediate predecessors of a white-pebbled vertex $v$ have pebbles on them, the white pebble on $v$ may be removed. In particular, a white pebble can always be removed from a source vertex.

A *black-white pebbling* from $(B_0, W_0)$ to $(B_\tau, W_\tau)$ in $G$ is a sequence of pebble configurations $\mathcal{P} = \{\mathbb{P}_0, \ldots, \mathbb{P}_\tau\}$ such that $\mathbb{P}_0 = (B_0, W_0)$, $\mathbb{P}_\tau = (B_\tau, W_\tau)$, and for all $t \in [\tau]$, $\mathbb{P}_t$ follows from $\mathbb{P}_{t-1}$ by one of the rules above. A *(complete) pebbling of $G$*, also called a *pebbling strategy for $G$*, is a pebbling such that $(B_0, W_0) = (\emptyset, \emptyset)$ and $(B_\tau, W_\tau) = (\{z\}, \emptyset)$.

The *time* of a pebbling $\mathcal{P} = \{\mathbb{P}_0, \ldots, \mathbb{P}_\tau\}$ is simply $\textit{time}(\mathcal{P}) = \tau$ and the *space* is $\textit{space}(\mathcal{P}) = \max_{0 \leq t \leq \tau}\{|B_t \cup W_t|\}$. The *black-white pebbling price* (also known as the *pebbling measure* or *pebbling number*) of $G$, denoted $\textit{BW-Peb}(G)$, is the minimum space of any complete pebbling of $G$.

A *black pebbling* is a pebbling using black pebbles only, i.e., having $W_t = \emptyset$ for all $t$. The *(black) pebbling price* of $G$, denoted $\textit{Peb}(G)$, is the minimum space of any complete black pebbling of $G$.

For any DAG $G$ over $n$ vertices with bounded indegree, the black pebbling price (and thus also the black-white pebbling price) is at most $O(n/\log n)$ [HPV77], where the hidden constant depends on the indegree. A number of exact or asymptotically tight bounds on different graph families have been proven in the whole range from constant to $\Theta(n/\log n)$, for instance in [GT78, Kla85, LT80, Mey81, PTC77]. As to time, obviously any DAG $G$ over $n$ vertices can be pebbled in time $2n - 1$, and for all graphs we will study this is also a lower bound, so studying the time measure in isolation is not that exciting. A very interesting question, however, is how time and space are related in a single pebbling of $G$ if one wants to optimize both measures simultaneously, and this question turns out to be intimately connected with proof complexity trade-offs. Time-space trade-offs for pebble games have been studied in a long sequence of papers, but for our applications we will focus on [CS80, CS82, LT82] and the more recent results in [Nor10b]. Our way of relating pebbling trade-offs and proof complexity trade-offs goes via the CNF formulas defined next.

A *pebbling contradiction* over a DAG $G$ is a CNF formula that encodes the pebble game on $G$ by postulating the sources to be true and the target to be false, and specifying that truth propagates through the graph according to the pebbling rules. These formulas have previously been studied more or less implicitly in [RM99, BEGJ00] before being explicitly defined is [BW01].

**Definition 3.12 (Pebbling contradiction).** Suppose that $G$ is a DAG with sources $S$ and a unique sink $z$. Identify every vertex $v \in V(G)$ with a propositional logic variable $v$. The *pebbling contradiction* over $G$, denoted $Peb_G$, is the conjunction of the following clauses:

- for all $s \in S$, a unit clause $s$ (*source axioms*),

- For all non-source vertices $v$ with immediate predecessors $u_1, \ldots, u_\ell$, the clause $\overline{u}_1 \vee \cdots \vee \overline{u}_\ell \vee v$ (*pebbling axioms*),

- for the sink $z$, the unit clause $\overline{z}$ (*target* or *sink axiom*).

If $G$ has $n$ vertices and maximal indegree $\ell$, the formula $Peb_G$ is an unsatisfiable $(1+\ell)$-CNF formula with $n + 1$ clauses over $n$ variables.

## 3.5 Substitution Formulas

Throughout this paper, we will let $f_d$ denote any (non-constant) Boolean function $f_d : \{0,1\}^d \mapsto \{0,1\}$ of arity $d$. We use the shorthand $\vec{x} = (x_1, \ldots, x_d)$, so that $f_d(\vec{x})$ is just an equivalent way of writing $f_d(x_1, \ldots, x_d)$. Every function $f_d(x_1, \ldots, x_d)$ is equivalent to a CNF formula over $x_1, \ldots, x_d$ with at most $2^d$ clauses. Fix a canonical way to represent functions as CNF formulas and let $Cl[f_d(\vec{x})]$ denote the canonical set of clauses representing $f_d$. Similarly, let $Cl[\neg f_d(\vec{x})]$ denote the clauses in the canonical representation of the negation of $f$.

For instance, we choose to define

$$Cl[\vee_2(\vec{x})] = \{x_1 \vee x_2\} \quad \text{and} \quad Cl[\neg\vee_2(\vec{x})] = \{\overline{x}_1, \overline{x}_2\} \tag{3.6}$$

for logical or of two variables and

$$Cl[\oplus_2(\vec{x})] = \{x_1 \vee x_2, \overline{x}_1 \vee \overline{x}_2\} \quad \text{and} \quad Cl[\neg\oplus_2(\vec{x})] = \{x_1 \vee \overline{x}_2, \overline{x}_1 \vee x_2\} \tag{3.7}$$

for logical exclusive or of two variables. The general definitions for exclusive or are

$$Cl[\oplus_d(\vec{x})] = \{\bigvee_{i=1}^d x_i^{\nu_i} \mid \sum_{i=1}^d \nu_i \equiv d \pmod{2}\} \tag{3.8}$$

and

$$Cl[\neg\oplus_d(\vec{x})] = \{\bigvee_{i=1}^d x_i^{\nu_i} \mid \sum_{i=1}^d \nu_i \not\equiv d \pmod{2}\} \tag{3.9}$$

from which we can see that $Cl[\oplus_d(\vec{x})]$ and $Cl[\neg\oplus_d(\vec{x})]$ both are $d$-CNFs. We will also be interested in the threshold function saying that $k$ out of $d$ variables are true, which we will denote $thr_d^k$. To give an example, for $thr_4^2$ we have

$$Cl[thr_4^2(\vec{x})] = \begin{Bmatrix} x_1 \vee x_2 \vee x_3, \\ x_1 \vee x_2 \vee x_4, \\ x_1 \vee x_3 \vee x_4, \\ x_2 \vee x_3 \vee x_4 \end{Bmatrix} \tag{3.10}$$

and

$$Cl[\neg thr_4^2(\vec{x})] = \begin{Bmatrix} \overline{x}_1 \vee \overline{x}_2, \\ \overline{x}_1 \vee \overline{x}_3, \\ \overline{x}_1 \vee \overline{x}_4, \\ \overline{x}_2 \vee \overline{x}_3, \\ \overline{x}_2 \vee \overline{x}_4, \\ \overline{x}_3 \vee \overline{x}_4 \end{Bmatrix} \tag{3.11}$$

and in general we have

$$Cl[thr_d^k(\vec{x})] = \{\bigvee_{i \in S} x_i \,|\, S \subseteq [d], |S| = d - k + 1\} \tag{3.12}$$

and

$$Cl[\neg thr_d^k(\vec{x})] = \{\bigvee_{i \in S} \overline{x}_i \,|\, S \subseteq [d], |S| = k\} \ . \tag{3.13}$$

Clearly, $thr_d^1(x_1, \ldots, x_d)$ is just another way of writing the function $\bigvee_{i=1}^d x_i$, and $thr_d^d(x_1, \ldots, x_d) = \bigwedge_{i=1}^d x_i$.

In general, we could construct a canonical representation $Cl[f_d(\vec{x})]$ for $f_d$ as follows. For a truth value assignment $\alpha : \{x_1, \ldots, x_d\} \mapsto \{0, 1\}$ we define the clause $C_\alpha = x_1^{1-\alpha(x_1)} \vee \cdots \vee x_d^{1-\alpha(x_d)}$ that is true for all assignments to $x_1, \ldots, x_d$ except $\alpha$. Then we could define

$$Cl[f_d(\vec{x})] = \bigwedge_{\alpha : \alpha(f_d(\vec{x}))=0} C_\alpha \ . \tag{3.14}$$

But this way of representing the Boolean function can turn out to be unnecessarily involved. For instance, for binary logical and (3.14) yields $Cl[\wedge_2(\vec{x})] = \{x_1 \vee x_2, \, x_1 \vee \overline{x}_2, \, \overline{x}_1 \vee x_2\}$ instead of the arguably more natural representation $Cl[\wedge_2(\vec{x})] = \{x_1, x_2\}$. Therefore, we want the freedom to choose our own canonical representation when appropriate. Note, however, that (3.14) constitutes a proof of the fact that without loss of generality we can always assume that $|Cl[f_d(\vec{x})]| < 2^d$, since there are only $2^d$ truth value assignments and $f_d$ is assumed to be non-constant.

The following observation is rather immediate, but nevertheless it might be helpful to state it explicitly.

**Observation 3.13.** *Suppose for any non-constant Boolean function $f_d$ that $C \in Cl[f_d(\vec{x})]$ and that $\rho$ is any partial truth value assignment such that $\rho(C) = 0$. Then for all $D \in Cl[\neg f_d(\vec{x})]$ it holds that $\rho(D) = 1$.*

*Proof.* If $\rho(C) = 0$ this means that $\rho(f_d) = 0$. Then clearly $\rho(\neg f_d) = 1$, so, in particular, $\rho$ must fix all clauses $D \in Cl[\neg f_d(\vec{x})]$ to true. □

We want to define formally what it means to substitute $f_d$ for the variables $Vars(F)$ in a CNF formula $F$. For notational convenience, we assume that $F$ only has variables $x, y, z$, et cetera, without subscripts, so that $x_1, \ldots, x_d, y_1, \ldots, y_d, z_1, \ldots, z_d, \ldots$ are new variables not occurring in $F$. We will say that the variables $x_1, \ldots, x_d$, and any literals over these variables, all *belong* to the variable $x$.

**Definition 3.14 (Substitution formula).** For a positive literal $x$ and a non-constant Boolean function $f_d$, we define the $f_d$-*substitution* of $x$ to be $x[f_d] = Cl[f_d(\vec{x})]$, i.e., the canonical representation of $f_d(x_1, \ldots, x_d)$ as a CNF formula. For a negative literal $\neg y$, the $f_d$-substitution is $\neg y[f_d] = Cl[\neg f_d(\vec{y})]$. The $f_d$-substitution of a clause $C = a_1 \vee \cdots \vee a_k$ is the CNF formula

$$C[f_d] = \bigwedge_{C_1 \in a_1[f_d]} \cdots \bigwedge_{C_k \in a_k[f_d]} (C_1 \vee \ldots \vee C_k) \tag{3.15}$$

and the $f_d$-substitution of a CNF formula $F$ is $F[f_d] = \bigwedge_{C \in F} C[f_d]$.

The reader is reminded that an example of this definition was given in Equation (2.1) on page 6.

We note that $F[f_d]$ is a CNF formula over $d \cdot |Vars(F)|$ variables containing strictly less than $|F| \cdot 2^{d \cdot W(F)}$ clauses. (Recall that we defined a CNF formula as a set of clauses, which means that $|F|$ is the number of clauses in $F$.) We have the following easy observation, the proof of which is presented for completeness.

**Observation 3.15.** *For any non-constant Boolean function $f_d : \{0, 1\}^d \mapsto \{0, 1\}$, it holds that $F[f_d]$ is unsatisfiable if and only if $F$ is unsatisfiable.*

*Proof.* Suppose that $F$ is satisfiable and let $\alpha$ be a truth value assignment such that $\alpha(F) = 1$. Then we can satisfy $F[f_d]$ by choosing an assignment $\alpha'$ for $\mathit{Vars}(F[f_d])$ in such a way that $f_d(\alpha'(x_1), \ldots, \alpha'(x_d)) = \alpha(x)$. For if $C \in F$ is satisfied by some literal $a_i$ set to true by $\alpha$, then $\alpha'$ will satisfy all clauses $C_i \in a_i[f_d]$ and thus also the whole CNF formula $C[f_d]$ in (3.15).

Conversely, suppose $F$ is unsatisfiable and consider any truth value assignment $\alpha'$ for $F[f_d]$. Then $\alpha'$ defines a truth value assignment $\alpha$ for $F$ in the natural way by setting $\alpha(x) = f_d(\alpha'(x_1), \ldots, \alpha'(x_d))$, and we know that there is some clause $C \in F$ that is not satisfied by $\alpha$. That is, for every literal $a_i \in C = a_1 \vee \cdots \vee a_k$ it holds that $\alpha(a_i) = 0$. But then $\alpha'$ does not satisfy $a_i[f_d]$, so there is some clause $C'_i \in a_i[f_d]$ such that $\alpha'(C'_i) = 0$. This shows that $\alpha'$ falsifies the disjunction $C'_1 \vee \cdots \vee C'_k \in C[f_d]$, and consequently $F[f_d]$ must also be unsatisfiable. □

For our present purposes, a particularly interesting kind of Boolean functions $f(x_1, \ldots, x_d)$ are those having the property that no subset of $k$ variables can determine the value of $f(x_1, \ldots, x_d)$.

**Definition 3.16 (Non-authoritarian function).** We say that a Boolean function $f$ over variables $X = \{x_1, \ldots, x_d\}$ is *k-non-authoritarian*[8] if no restriction to $X$ of size $k$ can fix the value of $f$. In other words, for every restriction $\rho$ to $X$ with $|\rho| \leq k$ there exist two assignments $\alpha_0, \alpha_1 \supset \rho$ such that $f(\alpha_0) = 0$ and $f(\alpha_1) = 1$.

Observe that a function on $d$ variables can be $k$-non-authoritarian only if $k < d$. As noted above, two natural examples of $d$-non-authoritarian functions are exclusive or $\oplus$ of $d + 1$ variables and majority of $2d + 1$ variables, i.e., $thr_{2d+1}^{d+1}$.

# 4  Substitution Space Theorem for Resolution

We now present the full proof of our first main technical contribution, which describes how the space complexity of a formula in resolution changes under substitution. Let us start by restating the theorem for reference.

**Theorem 2.1 (restated).** *Let $F$ be any unsatisfiable CNF formula and $f_d : \{0,1\}^d \mapsto \{0,1\}$ be any non-constant Boolean function. Then it holds that the substitution formula $F[f_d]$ can be refuted in standard syntactic resolution in width*
$$W(F[f_d] \vdash 0) = \mathrm{O}(d \cdot W(F \vdash 0)) \ ,$$

*length*
$$L_{\mathfrak{R}}(F[f_d] \vdash 0) \leq \min_{\pi : F \vdash 0} \{ L(\pi) \cdot \exp(\mathrm{O}(d \cdot W(\pi))) \} \ ,$$

*and total space*
$$\mathit{TotSp}_{\mathfrak{R}}(F[f_d] \vdash 0) \leq \min_{\pi : F \vdash 0} \{ \mathit{TotSp}(\pi) \cdot \exp(\mathrm{O}(d \cdot W(\pi))) \} \ .$$

*In the other direction, any semantic resolution refutation $\pi_f : F[f_d] \vdash 0$ of the substitution formula can be transformed into a syntactic resolution refutation $\pi : F \vdash 0$ of the original formula such that the number of axiom downloads in $\pi$ is at most the number of axiom downloads in $\pi_f$. If in addition $f_d$ is non-authoritarian, it holds that $Sp(\pi_f) > \mathit{VarSp}(\pi)$, i.e., the clause space of refuting the substitution formula $F[f_d]$ is lower-bounded by the variable space of refuting the original formula $F$.*

---

[8]Although related notions occur in the literature, we have not been able to locate any previous definition that captures what we require in Definition 3.16. We therefore introduce a (hopefully evocative) new name that decribes the property that we need.

## 4.1 Two Corollaries of the Substitution Space Theorem

Before embarking on the proof of the theorem, we want to make a couple of quick remarks. Although this might not be immediately obvious, Theorem 2.1 is remarkably powerful as a tool for understanding space in resolution. It will take some more work before we can present our main applications of this theorem, which are the strong time-space trade-off results discussed in Section 7. Let us note for starters, however, that without any extra work we immediately get lower bounds on space.

Esteban and Torán [ET01] proved that the clause space of refuting $F$ is upper-bounded by the formula size. In the papers [ABRW02, BG03, ET01] it was shown, using quite elaborate arguments, that there are polynomial-size $k$-CNF formulas with lower bounds on clause space matching this upper bound up to constant factors. Using Theorem 2.1 we can get a different proof of this fact.

**Corollary 4.1 ([ABRW02, BG03, ET01]).** *There are families of $k$-CNF formulas $\{F\}_{n=1}^{\infty}$ with $\Theta(n)$ clauses over $\Theta(n)$ variables such that $Sp(F_n \vdash 0) = \Theta(n)$.*

*Proof.* Just pick any formula family for which it is shown that any refutation of $F_n$ must at some point in the refutation mention $\Omega(n)$ variables at the same time (e.g., from [BW01]), and then apply Theorem 2.1. $\square$

It should be noted, though, that when we apply Theorem 2.1 the formulas in [ABRW02, BG03, ET01] are changed. We want to point out that there is another, and even more elegant way to derive Corollary 4.1 from [BW01] without changing the formulas, namely by using the lower bound on clause space in terms of width in [AD08].

For our next corollary, however, there is no other, simpler way known to prove the same result. Instead, our proof in this paper actually improves the constants in the result.

**Corollary 4.2 ([BN08]).** *There are families $\{F_n\}_{n=1}^{\infty}$ of $k$-CNF formulas of size $\mathrm{O}(n)$ refutable in linear length $L(F_n \vdash 0) = \mathrm{O}(n)$ and constant width $W(F_n \vdash 0) = \mathrm{O}(1)$ such that the minimum clause space required is $Sp(F_n \vdash 0) = \Omega(n/\log n)$.*

This follows, just as in [BN08], by considering graphs of size $\Theta(n)$ with maximal pebbling price $\Theta(n/\log n)$ and studying refutations of pebbling contradictions over such graphs. Since the variable space of such refutations is at least the pebbling price, Corollary 4.2 follows immediately from Theorem 2.1 (and, as noted, with better constants than in [BN08]), by using, say, exclusive or over two variables in the substitution formula.

## 4.2 Proof of the Substitution Space Theorem—Main Components

We divide the proof of Theorem 2.1 into three parts in Lemmas 4.3, 4.6, and 4.7 below. In this subsection, we state these three theorems and show how they combine to yield Theorem 2.1. The rest of Section 4 is then spent proving these three auxiliary lemmas.

**Lemma 4.3.** *For any CNF formula $F$ and any non-constant Boolean function $f_d$, it holds that*

$$W\big(F[f_d] \vdash 0\big) = \mathrm{O}\big(d \cdot W(F \vdash 0)\big) \ ,$$
$$L_{\mathfrak{R}}\big(F[f_d] \vdash 0\big) \leq \min_{\pi:F\vdash 0} \big\{L(\pi) \cdot \exp\big(\mathrm{O}(d \cdot W(\pi))\big)\big\} \ ,$$

*and*

$$TotSp_{\mathfrak{R}}\big(F[f_d] \vdash 0\big) \leq \min_{\pi:F\vdash 0} \big\{TotSp(\pi) \cdot \exp\big(\mathrm{O}(d \cdot W(\pi))\big)\big\} \ .$$

These upper bounds for resolution refutations of $F[f_d]$ are not hard to show. The proof proceeds along the following lines. Given a resolution refutation $\pi$ of $F$, we construct a refutation $\pi_f : F[f_d] \vdash 0$ mimicking the derivation steps in $\pi$. When $\pi$ downloads an axiom $C$, we download the $\exp\bigl(O(d \cdot W(C))\bigr)$ axiom clauses in $C[f_d]$. When $\pi$ resolves $C_1 \vee x$ and $C_2 \vee \overline{x}$ to derive $C_1 \vee C_2$, we use the fact that resolution is implicationally complete to derive $(C_1 \vee C_2)[f_d]$ from $(C_1 \vee x)[f_d]$ and $(C_2 \vee \overline{x})[f_d]$ in at most $\exp\bigl(O(d \cdot W(C_1 \vee C_2))\bigr)$ steps. We return to the details of the proof in Section 4.3.

Before stating the next lemma, we need to give a slightly more formal description than in Definition 2.11 of how clauses derived from $F[f_d]$ are translated into clauses over $Vars(F)$ using *projections* defined in terms of *precise implication*.

**Definition 4.4 (Precise implication).** Let $F$ be a CNF formula and $f_d$ a non-constant Boolean function, and suppose that $\mathbb{D}$ is a set of clauses over $Vars(F[f_d])$ and that $P$ and $N$ are (disjoint) subset of variables of $F$. If

$$\mathbb{D} \vDash \bigvee_{x \in P} f_d(\vec{x}) \vee \bigvee_{y \in N} \neg f_d(\vec{y}) \tag{4.1a}$$

but for all strict subsets $P' \subsetneq P$ and $N' \subsetneq N$ it holds that

$$\mathbb{D} \nvDash \bigvee_{x \in P'} f_d(\vec{x}) \vee \bigvee_{y \in N} \neg f_d(\vec{y}) \ , \tag{4.1b}$$

and

$$\mathbb{D} \nvDash \bigvee_{x \in P} f_d(\vec{x}) \vee \bigvee_{y \in N'} \neg f_d(\vec{y}) \ , \tag{4.1c}$$

we say that the clause set $\mathbb{D}$ implies $\bigvee_{x \in P} f_d(\vec{x}) \vee \bigvee_{y \in N} \neg f_d(\vec{y})$ *precisely* and write

$$\mathbb{D} \triangleright \bigvee_{x \in P} f_d(\vec{x}) \vee \bigvee_{y \in N} \neg f_d(\vec{y}) \ . \tag{4.2}$$

Note that $P = N = \emptyset$ in Definition 4.4 corresponds to $\mathbb{D}$ being unsatisfiable.

Recalling the convention that any clause $C$ can be written $C = C^+ \vee C^-$, where $C^+ = \bigvee_{x \in Lit(C)} x$ is the disjunction of the positive literals in $C$ and $C^- = \bigvee_{\overline{y} \in Lit(C)} \overline{y}$ is the disjunction of the negative literals, and choosing not to write $x \in Lit(C^+)$ or $\overline{y} \in Lit(C^-)$ below, but instead $x \in C^+$ and $\overline{y} \in C^-$ for brevity (which is still formally correct since a clause is a set of literals), we define projection of clauses as follows.

**Definition 4.5 (Projected clauses).** Let $F$ be a CNF formula and $f_d$ a non-constant Boolean function, and suppose that $\mathbb{D}$ is a set of clauses derived from $F[f_d]$. Then we say that $\mathbb{D}$ *projects* the clause $C = C^+ \vee C^-$ if there is a subset $\mathbb{D}_C \subseteq \mathbb{D}$ such that

$$\mathbb{D}_C \triangleright \bigvee_{x \in C^+} f_d(\vec{x}) \vee \bigvee_{\overline{y} \in C^-} \neg f_d(\vec{y}) \tag{4.3}$$

and we write

$$proj_F(\mathbb{D}) = \bigl\{C \bigm| \exists \mathbb{D}_C \subseteq \mathbb{D} \text{ s.t. } \mathbb{D}_C \triangleright \bigvee_{x \in C^+} f_d(\vec{x}) \vee \bigvee_{\overline{y} \in C^-} \neg f_d(\vec{y})\bigr\} \tag{4.4}$$

to denote the set of all clauses that $\mathbb{D}$ projects on $F$.

Given Definitions 4.4 and 4.5, which tell us how to translate clauses derived from $F[f_d]$ into clauses over $Vars(F)$, the next step is to show that this translation preserves resolution refutations.

**Lemma 4.6.** *Suppose that $\pi_f = \{\mathbb{D}_0, \ldots, \mathbb{D}_\tau\}$ is a semantic resolution refutation of $F[f_d]$ for some arbitrary unsatisfiable CNF formula $F$ and some arbitrary non-constant function $f_d$. Then the sets of projected clauses $\{proj_F(\mathbb{D}_0), \ldots, proj_F(\mathbb{D}_\tau)\}$ form the "backbone" of a resolution refutation $\pi$ of $F$ in the sense that:*

- *$proj_F(\mathbb{D}_0) = \emptyset$.*

- *$proj_F(\mathbb{D}_\tau) = \{0\}$.*

- *All transitions from $proj_F(\mathbb{D}_{t-1})$ to $proj_F(\mathbb{D}_t)$ for $t \in [\tau]$ can be accomplished by axiom downloads from $F$, inferences, erasures, and possibly weakening steps in such a way that the variable space in $\pi$ during these intermediate derivation steps never exceeds $\max_{\mathbb{D} \in \pi_f} \{VarSp(proj_F(\mathbb{D}))\}$.*

- *The only time $\pi$ performs a download of some axiom $C$ in $F$ is when $\pi_f$ downloads some axiom $D \in C[f_d]$ in $F[f_d]$.*

Note that by Proposition 3.6, we can get rid of the weakening moves in a postprocessing step, but allowing them in the statement of Lemma 4.6 makes the proof much cleaner. Accepting Lemma 4.6 on faith for the moment (deferring the proof to Section 4.4), the final missing link in the proof of the Substitution Space Theorem is the following lower bound.

**Lemma 4.7.** *Suppose that $\mathbb{D} \neq \emptyset$ is a set of clauses over $Vars(F[f_d])$ for some arbitrary CNF formula $F$ and some non-authoritarian function $f_d$. Then $Sp(\mathbb{D}) = |\mathbb{D}| > VarSp(proj_F(\mathbb{D}))$.*

Combining Lemmas 4.3, 4.6, and 4.7, the substitution space theorem for resolution follows. This is immediate, but for the convenience of the reader we write out the details.

*Proof of Theorem 2.1.* The first part of Theorem 2.1, i.e., that any refutation $\pi$ of $F$ can be converted to a refutation $\pi_f$ of the substitution formula $F[f_d]$, is Lemma 4.3 verbatim. For the second part of Theorem 2.1, Lemma 4.6 describes how any refutation $\pi_f$ of the substitution formula $F[f_d]$ can be translated back into a refutation $\pi$ of the original formula $F$. This is true regardless of what kind of function $f_d$ is used for the substitution. If in addition $f_d$ is non-authoritarian, Lemma 4.7 says that the clause space of $\pi_f$ provides an upper bound for the variable space of $\pi$. The theorem follows. □

It remains to prove Lemmas 4.3, 4.6, and 4.7. For convenience of notation in the proofs, let us define the disjunction $\mathbb{C} \vee \mathbb{D}$ of two clause sets $\mathbb{C}$ and $\mathbb{D}$ to be the clause set

$$\mathbb{C} \vee \mathbb{D} = \{C \vee D \mid C \in \mathbb{C}, D \in \mathbb{D}\} \ . \tag{4.5}$$

This notation extends to more than two clause sets in the natural way. Rewriting (3.15) in Definition 3.14 using this notation, we have that

$$(D \vee a)[f_d] = D[f_d] \vee a[f_d] = \bigwedge_{C_1 \in D[f_d]} \bigwedge_{C_2 \in a[f_d]} (C_1 \vee C_2) \ . \tag{4.6}$$

### 4.3 Refuting Substitution Formulas $F[f_d]$ by Simulating Refutations of $F$

To prove Lemma 4.3, we show how to construct, given $\pi : F \vdash 0$, a resolution refutation $\pi_f : D[f_d] \vdash 0$ by maintaining the invariant that if we have $\mathbb{C}$ in memory for $\pi$, then we have $\mathbb{C}[f_d]$ in memory for $\pi_f$ and this configuration has total space at most $\sum_{C \in \mathbb{C}} 2^{d \cdot W(C)} \leq |\mathbb{C}|2^{d \cdot W(\pi)}$. We get the following case analysis.

**Axiom download** If $\pi$ downloads $C$, we download all of $C[f_d]$, i.e., less than $2^{d \cdot W(C)}$ clauses which all have width at most $d \cdot W(C)$.

**Erasure** If $\pi$ erases $C$, we erase all of $C[f_d]$ in less than $2^{d \cdot W(C)}$ erasure steps.

**Inference** This is the only interesting case. Suppose that $\pi$ infers $C_1 \vee C_2$ from $C_1 \vee x$ and $C_2 \vee \overline{x}$. Then by induction we have $(C_1 \vee x)[f_d]$ and $(C_2 \vee \overline{x})[f_d]$ in memory in $\pi_f$. It is a straightforward extension of Observation 3.15 that if $\mathbb{C} \vDash D$, then $\mathbb{C}[f_d] \vDash D[f_d]$, so in particular it holds that $(C_1 \vee x)[f_d]$ and $(C_2 \vee \overline{x})[f_d]$ imply $(C_1 \vee C_2)[f_d]$. By the implicational completeness of resolution, these clauses can all be derived.

An upper bound (not necessarily tight) for the width of this derivation in $\pi_f$ is $d \cdot (W(C_1 \vee x) + W(C_2 \vee \overline{x}) + W(C_1 \vee C_2)) = O(d \cdot W(\pi))$, as claimed.

To bound the length, note that $(C_1 \vee C_2)[f_d]$ contains less than $2^{d \cdot W(C_1 \vee C_2)}$ clauses. For every clause $D \in (C_1 \vee C_2)[f_d]$, consider the minimal restriction $\rho(\neg D)$ falsifying $D$. Since

$$(C_1 \vee x)[f_d] \wedge (C_2 \vee \overline{x})[f_d] \vDash D \tag{4.7}$$

we have that

$$(C_1 \vee x)[f_d]\!\restriction_{\rho(\neg D)} \wedge\, (C_2 \vee \overline{x})[f_d]\!\restriction_{\rho(\neg D)} \vDash 0 \ . \tag{4.8}$$

The number of variables is at most $d \cdot (W(C_1 \vee C_2) + 1) = N$, and by Proposition 3.8 there is a refutation of $(C_1 \vee x)[f_d]\!\restriction_{\rho(\neg D)} \wedge (C_2 \vee \overline{x})[f_d]\!\restriction_{\rho(\neg D)}$ in length at most $2^{N+1} - 1$ and total space at most $2^{N+1}(2^{N+1} + 2) = 2^{O(N)}$. Looking at this refutation and removing the restriction $\rho(\neg D)$, it is straightforward to verify that we get a derivation of $D$ from $(C_1 \vee x)[f_d] \wedge (C_2 \vee \overline{x})[f_d]$ in the same length and same asymptotic total space (see, for instance, the inductive proof in [BW01]). We can repeat this for every clause $D \in (C_1 \vee C_2)[f_d]$ to derive all of the less than $2^{d \cdot (W(C_1 \vee C_2))}$ clauses in this set in total length at most

$$2^{d \cdot (W(C_1 \vee C_2))} \cdot 2^{d \cdot (W(C_1 \vee C_2) + 2)} \leq 2^{3d \cdot W(\pi)} = 2^{O(d \cdot W(\pi))} \ . \tag{4.9}$$

Taken together, we see that we get a refutation $\pi_f$ in length at most $L(\pi) \cdot 2^{O(d \cdot W(\pi))}$ and width at most $O(d \cdot W(\pi))$ and total space at most $TotSp(\pi) \cdot \exp(O(d \cdot W(\pi)))$. Lemma 4.3 follows.

## 4.4 Translating Refutations of Substitution Formulas $F[f_d]$ to Refutations of $F$

We next prove Lemma 4.6. Let us use the convention that $\mathbb{D}$ and $D$ denote clause sets and clauses derived from $F[f_d]$ while $\mathbb{C}$ and $C$ denote clause sets and clauses derived from $F$.

Let us also overload the notation and write $\mathbb{D} \vDash C$, $\mathbb{D} \nvDash C$, and $\mathbb{D} \rhd C$ for $C = C^+ \vee C^-$ when the corresponding implications hold or do not hold for $\mathbb{D}$ with respect to $\bigvee_{x \in C^+} f_d(\vec{x}) \vee \bigvee_{\overline{y} \in C^-} \neg f_d(\vec{y})$ (with precise implication $\rhd$ defined as in Definition 4.4). Note that it will always be clear when we use the notation in this overloaded sense since $\mathbb{D}$ and $C$ are defined over different sets of variables.

Recall from Definition 4.5 that $proj_F(\mathbb{D}) = \{C \,|\, \exists\, \mathbb{D}_C \subseteq \mathbb{D} \text{ s.t. } \mathbb{D}_C \rhd \bigvee_{x \in C^+} f_d(\vec{x}) \vee \bigvee_{\overline{y} \in C^-} \neg f_d(\vec{y})\}$ is the set of clauses projected by $\mathbb{D}$. In the spirit of the notational convention just introduced, we will let $\mathbb{C}_t$ be a shorthand for $proj_F(\mathbb{D}_t)$.

Suppose now that $\pi_f = \{\mathbb{D}_0, \ldots, \mathbb{D}_\tau\}$ is a resolution refutation of $F[f_d]$ for some arbitrary unsatisfiable CNF formula $F$ and some arbitrary non-constant function $f_d$.

The first two bullets in Lemma 4.6 are immediate. For $\mathbb{D}_0 = \emptyset$ we have $\mathbb{C}_0 = proj_F(\mathbb{D}_0) = \emptyset$, and it is easy to verify that $\mathbb{D}_\tau = \{0\}$ yields $\mathbb{C}_\tau = proj_F(\mathbb{D}_\tau) = \{0\}$. As an aside, we note that the empty clause will have appeared in $\mathbb{C}_t = proj_F(\mathbb{D}_t)$ earlier, namely for the first $t$ such that $\mathbb{D}_t$ is contradictory.

The tricky part is to show that all transitions from $\mathbb{C}_{t-1} = proj_F(\mathbb{D}_{t-1})$ to $\mathbb{C}_t = proj_F(\mathbb{D}_t)$ can be performed in such a way that the variable space in our refutation under construction $\pi : F \vdash 0$ never exceeds $\max\{VarSp(\mathbb{C}_{t-1}), VarSp(\mathbb{C}_t)\}$ during the intermediate derivation steps needed in $\pi$. The proof is by a case analysis of the derivation steps. Before plunging into the proof, let us make a simple but useful observation.

**Observation 4.8.** *Using the overloaded notation just introduced, if $\mathbb{D}_t \vDash C$ then $C = C^+ \vee C^-$ is derivable from $\mathbb{C}_t = \mathit{proj}_F(\mathbb{D}_t)$ by weakening.*

*Proof.* Pick $C_1^+ \subseteq C^+$ and $C_2^- \subseteq C^-$ minimal so that $\mathbb{D} \vDash C_1^+ \vee C_2^-$ still holds. Then by definition $\mathbb{D} \triangleright C_1^+ \vee C_2^-$ so $C_1^+ \vee C_2^- \in \mathbb{C}_t$ and $C \supseteq C_1^+ \vee C_2^-$ can be derived from $\mathbb{C}_t$ by weakening as claimed. □

Consider now the rule applied in $\pi_f$ at time $t$ to get from $\mathbb{D}_{t-1}$ to $\mathbb{D}_t$. We analyze the three possible cases—inference, erasure and axiom download—in this order.

**Inference** Suppose $\mathbb{D}_t = \mathbb{D}_{t-1} \cup \{D\}$. Since $\mathbb{D}_{t-1} \subseteq \mathbb{D}_t$, it is immediate from Definition 4.5 that no clauses in $\mathbb{C}_{t-1}$ can disappear at time $t$, i.e., $\mathbb{C}_{t-1} \setminus \mathbb{C}_t = \emptyset$. There can appear new clauses in $\mathbb{C}_t$, but by Observation 4.8 all such clauses are derivable by weakening from $\mathbb{C}_{t-1}$ since $\mathbb{D}_{t-1}$ implies $D$ and hence all of $\mathbb{D}_t$. During such weakening moves the variable space increases monotonically and is bounded from above by $\mathit{VarSp}(\mathbb{C}_t)$.

**Erasure** Since $\mathbb{D}_t \subseteq \mathbb{D}_{t-1}$, it is immediate from Definition 4.5 that no new clauses can appear at time $t$. Any clauses in $\mathbb{C}_{t-1} \setminus \mathbb{C}_t$ can simply be erased, which decreases the variable space monotonically.

**Axiom download** This is the only place in the case analysis where we need to do some work. Suppose that $\mathbb{D}_t = \mathbb{D}_{t-1} \cup \{D\}$ for some axiom clause $D \in A[f_d]$, where $A$ in turn is an axiom of $F$. If $C \in \mathbb{C}_t \setminus \mathbb{C}_{t-1}$ is a new projected clause, $D$ must be involved in projecting it so there is some subset $\mathbb{D} \subseteq \mathbb{D}_{t-1}$ such that

$$\mathbb{D} \cup \{D\} \triangleright C \; . \tag{4.10}$$

Also note that if $\mathbb{D}_{t-1} \vDash C$ we are done since $C$ can be derived from $\mathbb{C}_{t-1}$ by weakening, so we can assume that

$$\mathbb{D}_{t-1} \nvDash C \; . \tag{4.11}$$

We want to show that all clauses $C$ satisfying (4.10) and (4.11) can be derived from $\mathbb{C}_{t-1} = \mathit{proj}_F(\mathbb{D}_{t-1})$ by downloading $A \in F$, making inferences, and then possibly erasing $A$, and that this can be done without the variable space exceeding $\max\{\mathit{VarSp}(\mathbb{C}_{t-1}), \mathit{VarSp}(\mathbb{C}_t)\}$. The key to our proof is the next lemma.

**Lemma 4.9.** *Suppose that the clause set $\mathbb{D}$ over $\mathit{Vars}(F[f_d])$, the clause $D \in A[f_d]$, and the clause $C$ over $\mathit{Vars}(F)$ are such that $\mathbb{D} \cup \{D\} \triangleright C$ but $\mathbb{D} \nvDash C$. Then if we write the clause $A$ as $A = a_1 \vee \cdots \vee a_k$, it holds for every $a_i \in A \setminus C$ that there is a clause subset $\mathbb{D}^i \subseteq \mathbb{D}$ and a subclause $C^i \subseteq C$ such that $\mathbb{D}^i \triangleright C^i \vee \overline{a}_i$. That is, all clauses $C \vee \overline{a}_i$ for $a_i \in A \setminus C$ can be derived from $\mathbb{C} = \mathit{proj}_F(\mathbb{D})$ by weakening.*

*Proof.* Consider any truth value assignment $\alpha$ such that $\alpha(\mathbb{D}) = 1$ but $\alpha(\bigvee_{x \in C^+} f_d(\vec{x}) \vee \bigvee_{\overline{y} \in C^-} \neg f_d(\vec{y})) = 0$. Such an assignment exists since $\mathbb{D} \nvDash C$ by assumption. Also, since by assumption $\mathbb{D} \cup \{D\} \triangleright C$ we must have $\alpha(D) = 0$. If $A = a_1 \vee \cdots \vee a_k$, we can write $D \in A[f_d]$ on the form $D = D_1 \vee \cdots \vee D_k$ for $D_i \in a_i[f_d]$ (compare with Equation (4.6)). Fix any $a \in A$ and suppose for the moment that $a = x$ is a positive literal. Then $\alpha(D_i) = 0$ implies that $\alpha(f_d(\vec{x})) = 0$. By Observation 3.13, this means that $\alpha(\neg f_d(\vec{x})) = 1$. Since exactly the same argument holds if $a = \overline{y}$ is a negative literal, we conclude that

$$\mathbb{D} \vDash \bigvee_{x \in (C \vee \overline{a}_i)^+} f_d(\vec{x}) \vee \bigvee_{\overline{y} \in (C \vee \overline{a}_i)^-} \neg f_d(\vec{y}) \tag{4.12}$$

or, rewriting (4.12) using our overloaded notation, that

$$\mathbb{D} \vDash C \vee \overline{a}_i \; . \tag{4.13}$$

If $a_i \in C$, the clause $C \vee \overline{a}_i$ is trivially true and thus uninteresting, but otherwise we pick $\mathbb{D}^i \subseteq \mathbb{D}$ and $C^i \subseteq C$ minimal such that (4.13) still holds (and notice that since $\mathbb{D} \nvDash C$, the literal $\overline{a}_i$ cannot be dropped from the implication). Then by Definition 4.5 we have $\mathbb{D}^i \triangleright C^i \vee \overline{a}_i$ as claimed. □

We remark that Lemma 4.9 can be seen to imply that $Vars(A) \subseteq Vars(\mathbb{C}_t) = Vars(proj_F(\mathbb{D}_t))$. For $x \in Vars(A) \cap Vars(C)$ this is of course trivially true, but for $x \in Vars(A) \setminus Vars(C)$ Lemma 4.9 tells us that already at time $t-1$, there is a clause in $\mathbb{C}_{t-1} = proj_F(\mathbb{D}_{t-1})$ containing $x$, namely the clause $C^i \vee \overline{a}_i$ found in the proof above. Since $\mathbb{D}_t \supseteq \mathbb{D}_{t-1}$, this clause does not disappear at time $t$. This means that if we download $A \in F$ in our refutation $\pi : F \vdash 0$ under construction, we have $VarSp(\mathbb{C}_{t-1} \cup \{A\}) \leq VarSp(\mathbb{C}_t)$. Thus, we can download $A \in F$, and then possibly erase this clause again at the end of our intermediate resolution derivation to get from $\mathbb{C}_{t-1}$ to $\mathbb{C}_t$, without the variable space ever exceeding $\max\{VarSp(\mathbb{C}_{t-1}), VarSp(\mathbb{C}_t)\}$.

Let us now argue that all new clauses $C \in \mathbb{C}_t \setminus \mathbb{C}_{t-1}$ can be derived from $\mathbb{C}_{t-1} \cup \{A\}$. If $A \setminus C = \emptyset$, then the weakening rule applied on $A$ is enough. Suppose therefore that this is not the case and let $A' = A \setminus C = \bigvee_{a \in Lit(A) \setminus Lit(C)} a$. Appealing to Lemma 4.9 we know that for every $a \in A$ there is a $C_a \subseteq C$ such that $C_a \vee \overline{a} \in \mathbb{C}_{t-1}$. Note that by assumption (4.11) this means that if $x \in Vars(A) \cap Vars(C)$, then $x$ occurs with the same sign in $A$ and $C$, since otherwise we would get the contradiction $\mathbb{D} \models C \vee \overline{a} = C$. Summing up, $\mathbb{C}_{t-1}$ contains $C_a \vee \overline{a}$ for some $C_a \subseteq C$ for all $a \in Lit(A) \setminus Lit(C)$ and in addition we know that $Lit(A) \cap \{\overline{a} \mid a \in Lit(C)\} = \emptyset$. Let us write $A' = a_1 \vee \cdots \vee a_m$ and do the following weakening derivation steps from $\mathbb{C}_{t-1} \cup \{A\}$:

$$\begin{aligned} A &\rightsquigarrow C \vee A' \\ C_{a_1} \vee \overline{a}_1 &\rightsquigarrow C \vee \overline{a}_1 \\ C_{a_2} \vee \overline{a}_2 &\rightsquigarrow C \vee \overline{a}_2 \\ &\vdots \\ C_{a_m} \vee \overline{a}_m &\rightsquigarrow C \vee \overline{a}_m \end{aligned} \qquad (4.14)$$

Then resolve $C \vee A'$ in turn with all clauses $C \vee \overline{a}_1, C \vee \overline{a}_2, \ldots, C_{a_m} \vee \overline{a}_m$, finally yielding the clause $C$.

In this way all clauses $C \in \mathbb{C}_t \setminus \mathbb{C}_{t-1}$ can be derived one by one, and we note that we never mention any variables outside of $Vars(\mathbb{C}_{t-1} \cup \{A\}) \subseteq Vars(\mathbb{C}_t)$ in these derivations.

**Wrapping up the Proof of Lemma 4.6** We have proven that no matter what derivation step is made in the transition $\mathbb{D}_{t-1} \rightsquigarrow \mathbb{D}_t$, we can perform the corresponding transition $\mathbb{C}_{t-1} \rightsquigarrow \mathbb{C}_t$ for our projected clause sets without the variable space going above $\max\{VarSp(\mathbb{C}_{t-1}), VarSp(\mathbb{C}_t)\}$. Also, the only time we need to download an axiom $A \in F$ in our projected refutation $\pi$ of $F$ is when $\pi_f$ downloads some axiom $D \in A[f_d]$. This completes the proof of Lemma 4.6.

## 4.5 Lifting Variable Space Lower Bounds to Clause Space Lower Bounds

Finally, the turn has come to the proof of Lemma 4.7. Recall the convention that $x, y, z$ refer to variables in $F$ while $x_1, \ldots, x_d, y_1, \ldots, y_d, z_1, \ldots, z_d$ refer to variables in $F[f_d]$. Also recall that we use overloaded notation $\mathbb{D} \models C$, $\mathbb{D} \not\models C$, and $\mathbb{D} \triangleright C$ for $C = C^+ \vee C^-$ (where $C^+ = \bigvee_{x \in C} x$ and $C^- = \bigvee_{\overline{y} \in C} \overline{y}$) when the corresponding implications hold or do not hold for $\mathbb{D}$ with respect to $\bigvee_{x \in C^+} f_d(\vec{x}) \vee \bigvee_{\overline{y} \in C^-} \neg f_d(\vec{y})$.

We start with an intuitively plausible lemma saying that for all variables $x$ appearing in some clause projected by $\mathbb{D}$, the clause set $\mathbb{D}$ itself must contain at least one of the variables $x_1, \ldots, x_d$.

**Lemma 4.10.** *Suppose that $\mathbb{D}$ is a set of clauses derived from $F[f_d]$ and that $C \in proj_F(\mathbb{D})$. Then for all variables $x \in Vars(C)$ it holds that $\{x_1, \ldots, x_d\} \cap Vars(\mathbb{D}) \neq \emptyset$.*

*Proof.* Fix any $\mathbb{D}' \subseteq \mathbb{D}$ such that $\mathbb{D}$ precisely implies $C$ in the sense of Definition 4.4. By this definition, for all $z \in Vars(C)$ we have $\mathbb{D}' \not\models C \setminus \{z, \overline{z}\}$. Suppose that $z$ appears as a positive literal in $C$ (the case of a negative literal is completely analogous). This means that there is an assignment $\alpha$ such that $\alpha(\mathbb{D}') = 1$ but $\alpha\bigl(\bigvee_{x \in C^+ \setminus \{z\}} f_d(\vec{x}) \vee \bigvee_{y \in C^-} \neg f_d(\vec{y})\bigr) = 0$. Since $\mathbb{D}' \rhd C$, it must hold that $\alpha(f_d(\vec{z})) = 1$. Modify $\alpha$ into $\alpha'$ by changing the assignments to $z_1, \ldots, z_d$ in such a way that $\alpha'(f_d(\vec{z})) = 0$. Then $\alpha'\bigl(\bigvee_{x \in C^+} f_d(\vec{x}) \vee \bigvee_{y \in C^-} \neg f_d(\vec{y})\bigr) = 0$, so we must have $\alpha'(\mathbb{D}') = 0$. Since we only changed the assignments to (a subset of) the variables $z_1, \ldots, z_d$, the clause set $\mathbb{D}' \subseteq \mathbb{D}$ must mention at least one of these variables. □

With Lemma 4.10 in hand, we are ready to prove Lemma 4.7. Note that everything said so far in Section 4 (in particular, all of the proofs) applies to any non-constant Boolean function. In the proof of Lemma 4.7, however, it will be essential that we are dealing with non-authoritarian functions, i.e., functions $f_d$ having the property that no single variable $x_i$ can fix the the value of $f_d(x_1, \ldots, x_d)$.

Suppose that $\mathbb{D}$ is a set of clauses derived from $F[f_d]$ and write $V^* = Vars(proj_F(\mathbb{D}))$ to denote the set of all variables in $Vars(F)$ appearing in any clause projected by $\mathbb{D}$. We want to prove that $Sp(\mathbb{D}) = |\mathbb{D}| > |V^*|$ provided that $f_d$ is non-authoritarian.

To this end, consider the bipartite graph with the clauses in $\mathbb{D}$ labelling the vertices on the left-hand side and variables in $V^*$ labelling the vertices on the right-hand side. We draw an edge between $D \in \mathbb{D}$ and $x \in V^*$ if $Vars(D) \cap \{x_1, \ldots, x_d\} \neq \emptyset$. By Lemma 4.10 it holds that $Vars(\mathbb{D}) \cap \{x_1, \ldots, x_d\} \neq \emptyset$ for all variables $x \in V^*$, so in particular every variable $x \in V^*$ is the neighbour of at least one clause $D \in \mathbb{D}$. Let us write $N(D)$ to denote the neighbours of a left-hand vertex $D$ and extend this notation to sets of vertices by taking unions.

We claim that if $V^* = Vars(proj_F(\mathbb{D})) \neq \emptyset$, then there must exist some clause set $\mathbb{D}' \subseteq \mathbb{D}$ satisfying $|\mathbb{D}'| > N(\mathbb{D}')$. Suppose on the contrary that $|\mathbb{D}'| \leq N(\mathbb{D}')$ for all $\mathbb{D}' \subseteq \mathbb{D}$. Then by Hall's marriage theorem there is a matching of the clauses in $\mathbb{D}$ into the variable set $V^*$. Assume that $C = C^+ \vee C^-$ is any clause projected by $\mathbb{D}$ (such a clause exists since $V^* \neq \emptyset$). Then surely

$$\mathbb{D} \vDash \bigvee_{x \in C^+} f_d(\vec{x}) \vee \bigvee_{\overline{y} \in C^-} \neg f_d(\vec{y}) \tag{4.15}$$

(there is even a subset of $\mathbb{D}$ such that this implication is precise). But using the matching between $\mathbb{D}$ and $V^*$, we can satisfy $\mathbb{D}$ without assigning values to more than one variable $x_i \in Vars(\mathbb{D})$ corresponding to any $x \in Vars(F)$. Since $f_d$ is non-authoritarian, we can then extend this assignment to another assignment falsifying $f_d(\vec{x})$ for all $x \in C^+$ and satisfying $f_d(\vec{y})$ for all $\overline{y} \in C^-$. This means that our assignment satisfies the left-hand side of the implication (4.15) but falsifies the right-hand side, which is a contradiction. The claim follows.

Hence, fix any largest subset $\mathbb{D}_1 \subseteq \mathbb{D}$ such that $|\mathbb{D}_1| > N(\mathbb{D}_1)$. Clearly, if $\mathbb{D}_1 = \mathbb{D}$ we are done (remember that $N(\mathbb{D}) = V^*$), so suppose $\mathbb{D}_1 \neq \mathbb{D}$. In much the same way as above, we show that this assumption leads to a contradiction.

Let $\mathbb{D}_2 = \mathbb{D} \setminus \mathbb{D}_1 \neq \emptyset$ and define the vertex sets $V_1^* = N(\mathbb{D}_1)$ and $V_2^* = V^* \setminus V_1^*$. Note that we must have $V_2^* \subseteq N(\mathbb{D}_2)$ since $N(\mathbb{D}) = N(\mathbb{D}_1) \cup N(\mathbb{D}_2) = V^*$. By the maximality of $\mathbb{D}_1$ it must hold for all $\mathbb{D}' \subseteq \mathbb{D}_2$ that $|\mathbb{D}'| \leq |N(\mathbb{D}') \setminus V_1^*|$, because otherwise $\mathbb{D}'' = \mathbb{D}_1 \cup \mathbb{D}'$ would be a larger set with $|\mathbb{D}''| > |N(\mathbb{D}'')|$. But this implies that, again by Hall's marriage theorem, there is a matching $M$ of $\mathbb{D}_2$ into $N(\mathbb{D}_2) \setminus V_1^* = V_2^*$. Consider any clause $C \in proj_F(\mathbb{D})$ such that $Vars(C) \cap V_2^* \neq \emptyset$ and let $\mathbb{D}' \subseteq \mathbb{D}$ be any clause set such that

$$\mathbb{D}' \rhd \bigvee_{x \in C^+} f_d(\vec{x}) \vee \bigvee_{\overline{y} \in C^-} \neg f_d(\vec{y}) \tag{4.16}$$

(the existence of which is guaranteed by Definition 4.5). We claim that we can construct an assignment $\alpha$ that makes $\mathbb{D}'$ true but $\bigvee_{x \in C^+} f_d(\vec{x}) \vee \bigvee_{\overline{y} \in C^-} \neg f_d(\vec{y})$ false. This is clearly a contradiction, so if we can prove this claim it follows that our assumption $\mathbb{D}_1 \neq \mathbb{D}$ is false and that it instead must hold that $\mathbb{D}_1 = \mathbb{D}$ and thus $|N(\mathbb{D})| = |V^*| < |\mathbb{D}|$, which proves the theorem.

To establish the claim, let $\mathbb{D}'_i = \mathbb{D}' \cap \mathbb{D}_i$ for $i = 1, 2$ and let $C_i = C_i^+ \vee C_i^-$ for

$$C_i^+ = \bigvee_{\substack{x \in C \\ x \in V_i^*}} x \quad \text{and} \quad C_i^- = \bigvee_{\substack{\overline{y} \in C \\ y \in V_i^*}} \overline{y} \tag{4.17}$$

and $i = 1, 2$. We construct the assignment $\alpha$ satisfying $\mathbb{D}'$ but falsifying $\bigvee_{x \in C^+} f_d(\vec{x}) \vee \bigvee_{\overline{y} \in C^-} \neg f_d(\vec{y})$ in three steps:

1. Since $C_1^+ \vee C_i^- = C_1 \subsetneq C$ by construction (recall that we chose our clause $C$ in such a way that $Vars(C) \cap V_2^* \neq \emptyset$), the minimality condition in Definition 4.5 yields that

$$\mathbb{D}'_1 \nvDash \bigvee_{x \in C_1^+} f_d(\vec{x}) \vee \bigvee_{\overline{y} \in C_1^-} \neg f_d(\vec{y}) \tag{4.18}$$

   and hence we can find a truth value assignment $\alpha_1$ that sets $\mathbb{D}'_1$ to true, all $f_d(x_1, \ldots, x_d)$, $x \in C_1^+$, to false, and all $f_d(y_1, \ldots, y_d)$, $\overline{y} \in C_1^-$, to true. Note that $\alpha_1$ need only assign values to $\{z_1, \ldots, z_d \mid z \in Vars(C_1)\}$.

2. For $\mathbb{D}'_2$, we use the matching $M$ into $V_2^*$ found above to pick a distinct variable $x(D) \in Vars(F)$ for every $D \in \mathbb{D}'_2$ and then a variable $x(D)_i \in Vars(F[f_d])$ appearing in $D$, the existence of which is guaranteed by the edge between $D$ and $x(D)$. Let $\alpha_2$ be the assignment that sets all these variables $x(D)_i$ to the values that fix all $D \in \mathbb{D}'_2$ to true. We stress that $\alpha_2$ assigns a value to at most one variable $x(D)_i$ for every $x(D) \in Vars(F)$.

3. But since $f_d$ is non-authoritarian, this means that we can extend $\alpha_2$ to an assignment to all variables $x(D)_1, \ldots, x(D)_d$ that still satisfies $\mathbb{D}'_2$ but sets all $f_d(x_1, \ldots, x_d)$, $x \in C_2^+$, to false and all $f_d(y_1, \ldots, y_d)$, $\overline{y} \in C_2^-$, to true.

Hence, $\alpha = \alpha_1 \cup \alpha_2$ is an assignment such that $\alpha(\mathbb{D}') = 1$ but $\alpha(\bigvee_{x \in C^+} f_d(\vec{x}) \vee \bigvee_{\overline{y} \in C^-} \neg f_d(\vec{y})) = 0$, which proves the claim. This concludes the proof of Lemma 4.7.

Since Lemmas 4.3, 4.6, and 4.7 have now all been established, the proof of Theorem 2.1 is complete.

## 5 Substitution Space Theorem for $k$-DNF Resolution

To extend the substitution space theorem to $k$-DNF resolution, we construct a proof along very much the same lines as the proof for standard resolution in Section 4. The details of the proof gets much trickier, however, since the ideas used to obtain the (essentially tight) analysis for Theorem 2.1 no longer work. Indeed, they provably break down in view of the recent results in [NR09]. Instead, new, somewhat cruder, tools have to be developed.

Before proving the theorem, let us state it in its full generality with explicit constants. To parse the theorem more easily, the reader might be helped by thinking of $c$, $d$, and $k$ as constants. (This is the way the result was stated in Theorem 2.2, and this is also the case that will be of interest in our applications of the theorem).

**Theorem 5.1.** *Let $F$ be any unsatisfiable $c$-CNF formula and $f_d$ be any non-constant Boolean function of arity $d$. Then the following two properties hold for the substitution formula $F[f_d]$:*

1. *If $F$ can be refuted in syntactic standard resolution in length $L$ and total space $s$ simultaneously, then $F[f_d]$ can be refuted in syntactic $\mathfrak{R}(d)$ in length $L \cdot d^{4cd} \cdot 4^{cd}$ and total space $s \cdot d2^d + (cd+2)^3 \cdot 4^{cd} + O(1)$ simultaneously.*

2. *If $f_d$ is $k$-non-authoritarian and $F[f_d]$ can be refuted by a semantic $\mathfrak{R}(k)$-refutation that requires formula space $s$ and makes $L$ axiom downloads, then $F$ can be refuted by a syntactic standard resolution refutation that requires variable space at most $(2sk)^{k+1} \cdot 4^{k^2 d}$ and makes at most $L$ axiom downloads.*

The proof of the first part of Theorem 5.1 is straightforward, if a bit technically involved, and closely follows the corresponding proof for standard resolution. The second part of the proof is also very similar up to the point when we need to relate the formula space in $\mathfrak{R}(k)$-refutations of the substituted formula to the variable space of resolution refutations of the original formula. This boils down to bounding how many different Boolean functions of a particular form can be implied by a small set of $k$-DNF formulas.

A related, more narrow, problem, although strictly speaking *not* a special case of this general problem, is how many variables can appear in a minimally unsatisfiable set of $k$-DNF formulas of a given size (Definition 2.12 on page 12). Although there is no formal connection between the two problems as far as we are aware, the proof for standard resolution in Section 4 relies heavily on the techniques to prove tight bounds for minimally unsatisfiable CNF formulas. In this section, we find that this is the case for $k$-DNF resolution as well, as our techniques for proving bounds on minimally unsatisfiable $k$-DNF sets carry over to the proof of the $\mathfrak{R}(k)$ substitution space theorem. And interestingly enough, the proof in [NR09] that the analysis of our techniques for $\mathfrak{R}(k)$ is almost tight, and hence that any further substantial improvements seem to require fundamentally different ideas, are also derived from bounds on minimally unsatisfiable $k$-DNF sets.

Hence, minimal unsatisfiability for $k$-DNF sets appears to be an interesting and useful concept. Since in addition the underlying combinatorial problem is natural and appealing, we do not focus exclusively on $k$-DNF resolution in this section but also discuss our results for this purely combinatorial problem in some detail below.

## 5.1 Proof of $\mathfrak{R}(k)$ Substitution Space Theorem—Main Components

Following the pattern set in Section 4, let us start by presenting the main components of the proof and how they fit together to yield the theorem. As the reader will see, on a high level the proof is very similar to that of Theorem 2.1, but there are also some crucial differences in the details that we comment on below.

The first part of Theorem 5.1, which is Lemma 5.2 below, is established in Section 5.2.

**Lemma 5.2.** *Let $F$ be any $c$-CNF formula and $f_d : \{0,1\}^d \mapsto \{0,1\}$ be any non-constant Boolean function. Then it holds that if $F$ can be refuted in standard syntactic resolution in length $L$ and total space $s$ simultaneously, the substitution formula $F[f_d]$ can be refuted in syntactic $\mathfrak{R}(d)$ in length $L \cdot d^{4cd} \cdot 4^{cd}$ and total space $s \cdot d2^d + (cd+2)^3 \cdot 4^{cd} + O(1)$ simultaneously.*

To prove the second part of the theorem, we need to show how to convert a $\mathfrak{R}(k)$-refutation $\pi_f$ of $F[f_d]$ into a resolution refutation $\pi$ of $F$ such that the variable space of $\pi$ is bounded by the space of $\pi_f$, raised to the power of $k+1$. This part of the proof splits into two components. In Lemma 5.4 we claim that each $k$-DNF set $\mathbb{D} \in \pi_f$ can be "projected" onto a set of clauses over $Vars(F)$, such that the sequence of projected clause sets forms the "backbone" of a resolution refutation of $F$. This is just as in Theorem 2.1, except that here we lose a constant factor in the space bound when completing the backbone to a full proof. Then, in Lemma 5.5, we claim that if $\mathbb{C}$ is a set of clauses projected by a $k$-DNF set $\mathbb{D}$, the variable space of $\mathbb{C}$ is at most $\approx |\mathbb{D}|^{k+1}$.

The reason that we lose a constant factor when filling in the backbone is that the definition of projected clauses for $\mathfrak{R}(k)$, though still phrased in terms of precise implications (Definition 4.4 on page 25), is slightly different from (the stronger) Definition 4.5. This in turn depends on the fact that we cannot get this latter definition to work in the proof of Lemma 5.5 below.

**Definition 5.3 (Projections by $\mathfrak{R}(k)$-derivations).** Let $F$ be a CNF formula and $f_d$ a non-constant Boolean function, and suppose that $\mathbb{D}$ is a $k$-DNF set over $Vars(F[f_d])$. Writing $C = C^+ \vee C^-$ as a disjunction of the positive and negative literals, respectively, we say that $\mathbb{D}$ *projects* the clause $C$ if $\mathbb{D} \rhd \bigvee_{x \in C^+} f_d(\vec{x}) \vee \bigvee_{y \in C^-} \neg f_d(\vec{y})$ holds. We let

$$proj_F^*(\mathbb{D}) = \left\{ C \,\middle|\, \mathbb{D} \rhd \bigvee_{x \in C^+} f_d(\vec{x}) \vee \bigvee_{\overline{y} \in C^-} \neg f_d(\vec{y}) \right\} \tag{5.1}$$

denote the set of all clauses that $\mathbb{D}$ projects on $F$.

The difference from Definition 4.5 is that here, projected clauses are defined in terms of implications by the whole formula set $\mathbb{D}$ and not by arbitrary subsets of $\mathbb{D}$. Definition 4.5 is stronger than Definition 5.3 in that it gives us more projected clauses, which can be used to prove sharper lower bounds, but as mentioned above it turns out to be a bit too strong for our techniques to work in the $k$-DNF resolution case.

Informally, the next lemma states that the projection of a $k$-DNF resolution refutation of $F[f_d]$ is essentially a refutation of the original formula $F$ in standard resolution. The proof of this lemma appears in Section 5.3.

**Lemma 5.4.** *Let $k \geq 1$. Suppose that $\pi_f = \{\mathbb{D}_0, \ldots, \mathbb{D}_\tau\}$ is a $\mathfrak{R}(k)$-refutation of $F[f_d]$ for some arbitrary unsatisfiable CNF formula $F$ and some arbitrary non-constant function $f_d$. Then the sets of projected clauses $\{proj_F^*(\mathbb{D}_0), \ldots, proj_F^*(\mathbb{D}_\tau)\}$ form the "backbone" of a resolution refutation $\pi$ of $F$ in the sense that:*

- $proj_F^*(\mathbb{D}_0) = \emptyset$.

- $proj_F^*(\mathbb{D}_\tau) = \{0\}$.

- *All transitions from $proj_F^*(\mathbb{D}_{t-1})$ to $proj_F^*(\mathbb{D}_t)$ for $t \in [\tau]$ can be accomplished by axiom downloads from $F$, resolution inferences, erasures, and possibly resolution weakening steps in such a way that the variable space in $\pi$ during these intermediate steps never exceeds $2 \cdot \max_{\mathbb{D} \in \pi_f} \{VarSp(proj_F^*(\mathbb{D}))\}$.*

- *The only time $\pi$ performs a download of some axiom $C$ in $F$ is when $\pi_f$ downloads some axiom $D \in C[f_d]$ in $F[f_d]$.*

The following statement is the main tecnical novelty in the extension of the substitution space theorem to $k$-DNF resolution. It is proven in Section 5.4.

**Lemma 5.5.** *Suppose that $f_d$ is a $k$-non-authoritarian function of arity $d > k$ and that $\mathbb{D}$ is a $k$-DNF set over $Vars(F[f_d])$ for some CNF formula $F$. Then it holds that $VarSp(proj_F^*(\mathbb{D})) \leq 4^{k^2 d} \cdot (k \cdot Sp(\mathbb{D}))^{k+1}$.*

Putting together Lemmas 5.2, 5.4, and 5.5, the $\mathfrak{R}(k)$ substitution space theorem follows exactly as in the proof of the corresponding theorem for resolution in Section 4.2. We leave it to the reader to fill in the details and instead turn to establishing the lemmas.

## 5.2 Converting Resolution Refutations of $F$ to $\mathfrak{R}(k)$-refutations of $F[f]$

To prove Lemma 5.2, we convert a resolution refutation $\pi$ of $F$ into a $\mathfrak{R}(d)$-refutation of the substituted formula $F[f_d]$ while (roughly) preserving the length and total space simultaneously. This is done in two steps. First, we substitute each positive literal $x$ appearing in a clause $C$ in $\pi$ with some $d$-DNF representing $f_d(\vec{x})$ and similarly substitute $\neg x$ with a $d$-DNF representing $\neg f_d(\vec{x})$. The sequence of sets of clauses that was $\pi$ is transformed under this substitution into a sequence of $d$-DNF sets that forms the "backbone" of a $\mathfrak{R}(d)$-refutation. Then, we convert the backbone into a proper $\mathfrak{R}(d)$-refutation by simulating resolution inferences and axiom downloads. Consider a resolution inference step in $\pi$ which involved inferring $C \vee C'$ from $C \vee x, C \vee \neg x$. After substitution what we need to show is that $C[f_d] \vee C'[f_d]$ can be inferred from $C[f_d] \vee x[f_d], C'[f_d] \vee \neg x[f_d]$ in $\mathfrak{R}(d)$. This is shown in Lemma 5.6 below. The simulation of an axiom download is similarly addressed in Lemma 5.7, where we show that we can derive any $d$-DNF representation of $A[f_d]$ for an axiom $A \in F$ via a $\mathfrak{R}(k)$-derivation of bounded length and space. Given these two lemmas, the proofs of which follow below, we can complete the proof of Lemma 5.2.

**Lemma 5.6.** *Suppose $D_1, D_2$ are two k-DNF formulas over r variables. If $D_1 \wedge D_2 \vDash 0$, then the k-DNF set $\{D_1, D_2\}$ has a $\mathfrak{R}(k)$-refutation of length $|D_1| \cdot |D_2|$ and total space at most $2(TotSp(D_1) + TotSp(D_2))$ simultaneously.*

**Lemma 5.7.** *Suppose F is a CNF formula and D is a k-DNF formula and $|Vars(F) \cup Vars(D)| = r$. If $F \vDash D$ then D can be derived from F via a $\mathfrak{R}(k)$-derivation of length less than $k^{|D|} \cdot 2^{r+1}$ and total space at most $((r+2)\,TotSp(D))^2$ simultaneously.*

Postponing the proofs for a moment, let us see how these two lemmas yield Lemma 5.2.

*Proof of Lemma 5.2.* Let $\pi = \{\mathbb{C}_0, \ldots, \mathbb{C}_\tau\}$ be a resolution refutation of the CNF formula $F$. Let $\pi_f = \{\mathbb{D}_0, \ldots, \mathbb{D}_\tau\}$ denote the sequence of $d$-DNF sets obtained by substituting $\pi$ with $f_d$ in the following way. We start by fixing for each literal $a$ a $d$-DNF formula representing $a[f_d]$. For a clause $C = \bigvee_i a_i$ appearing in $\mathbb{C}_t$, construct a $d$-DNF formula $D_C$ which represents $C[f_d]$ by taking the disjunction of the $d$-DNF formulas representing $a_i$. Finally, set $\mathbb{D}_t = \{D_C \mid C \in \mathbb{C}_t\}$. In this way, every clause in $\mathbb{C}_t$ turns into a $d$-DNF formula in $\mathbb{D}_t$. Notice that the total space of $\mathbb{D}_t$ is less than $d \cdot 2^d$ times the total space of $\mathbb{C}_t$ because every literal appearing in $\mathbb{C}_t$ turns under substitution into a $d$-DNF with less than $2^d$ terms. To complete the proof of Lemma 5.2 it suffices to show for $0 \le t < \tau$ that $\mathbb{D}_{t+1}$ can be derived from $\mathbb{D}_t$ via a $\mathfrak{R}(d)$-derivation of length $\le d^{4cd} \cdot 4^{cd}$ and extra total space $(cd+2)^3 \cdot 4^{cd} + O(1)$. We divide into cases according to the type of the derivation step at time $t$.

**Erasure** If $\mathbb{C}_{t+1} = \mathbb{C}_t \setminus \{C\}$ then by construction we have $\mathbb{D}_{t+1} \subset \mathbb{D}_t$, so $\mathbb{D}_{t+1}$ can be derived in $\mathfrak{R}(d)$ from $\mathbb{D}_t$ by erasures.

**Axiom download** Let $A \in F$ be the axiom downloaded at time $t+1$, i.e., $\mathbb{C}_{t+1} = \mathbb{C}_t \cup \{A\}$. Let $A'$ be an arbitrary $d$-DNF representation of $A[f_d]$, recalling that $A[f_d]$ is a set of axioms of $F[f_d]$. This set involves at most $c \cdot d$ many variables and $A[f_d] \vDash A'$. Furthermore, $A'$ is a DNF formula over $2cd$ many literals so it has at most $4^{c \cdot d}$ many terms and has total space at most $cd4^{cd}$. Applying Lemma 5.7 we conclude $A'$ can be derived from $A[f_d]$ in length $d^{4cd} \cdot 2^{cd+1}$ and variable space $(cd+2)^3 \cdot 4^{cd}$.

**Inference** Suppose $\mathbb{C}_{t+1} = \mathbb{C}_t \cup \{C \vee C'\}$ where $C \vee C'$ is derived from $C \vee x, C' \vee \neg x \in \mathbb{C}_t$. Notice that $(C \vee x)[f_d] = (C[f_d]) \vee x[f_d]$ and $(C' \vee \neg x)[f_d] = C'[f_d] \vee \neg x[f_d]$. Since we can bound the number of terms in a $d$-DNF formula representing $x[f_d]$ by $2^d$, by Lemma 5.6 we can derive the empty DNF formula 0 from $d$-DNF formulas representing $x[f_d]$ and $\neg x[f_d]$ via a derivation of length at most $2^{2d}$ and total space at most $2^{2d+1}$. Applying weakening steps, when necessary, to the formulas involved in this refutation, we conclude that the $d$-DNF formula representing $(C \vee C')[f_d]$ can be derived from the $d$-DNF formulas representing $(C \vee x)[f_d]$ and $(C' \vee \neg x)[f_d]$ via a derivation of length at most $2^{2d}$ and $2^{2d+1}$ extra total space.

**Weakening** Suppose $\mathbb{C}_{t+1} = \mathbb{C}_t \cup \{C \vee C'\}$ for $C \in \mathbb{C}_t$. Then the $d$-DNF formula representing $(C \vee C')[f_d]$ can be derived in a single step from the $d$-DNF formula representing $C[f_d]$ using weakening.

We have shown how to complete the conversion of $\pi_f$ into a $\mathfrak{R}(d)$-refutation of $F[f_d]$ that is longer by at most a factor of $d^{4cd} \cdot 2^{cd}$ and uses at most $(cd+2)^3 \cdot 4^{cd} + O(1)$ extra total space. Taking into account the upper bound of $S \cdot d \cdot 2^d$ on the total space of $\mathbb{D}_t$, the lemma now follows. □

It remains to prove Lemmas 5.6 and 5.7. We attend to them in order.

*Proof of Lemma 5.6.* First we claim that for every term $T \in D_1$ and for every term $T' \in D_2$ we have

$$T' \cap \{\neg a \mid a \in T\} \neq \emptyset \ . \tag{5.2}$$

To see this, assume by way of contradiction that (5.2) fails to hold for $T \in D_1$ and $T' \in D_2$. Consider the minimal restriction $\rho$ that satisfies $T$. We see that $\rho$ satisfies $D_1$ and can be extended to an assignment that satisfies $T'$ as well, contradicting the assumption $D_1 \wedge D_2 \vDash 0$.

The refutation of $\{D_1, D_2\}$ proceeds by sequentially removing from $D_1$ all its terms. Let $T$ be a term of $D_1$ that we wish to remove. By (5.2) each term $T' \in D_2$ contains a literal $\neg a$ such that $a \in T$. Apply $\wedge$-elimination to replace $T'$ by $\neg a$. Repeating this process for each term $T' \in D_2$ we derive from $D_2$ in extra total space at most $TotSp(D_2)$ the clause $\bigvee_{a \in T} \neg a$. Resolve this clause with $D_1$ to remove $T$. This step requires extra total space at most $TotSp(D_1) + TotSp(D_2)$. Repeat the process for all $T \in D_1$ to obtain the empty DNF. This process required total space at most $2(TotSp(D_1) + TotSp(D_2))$ and the refutation length is $|D_1| \cdot |D_2|$ so the lemma follows. □

Before giving the formal proof of Lemma 5.7, we explain the main ideas by working out a concrete example. Suppose $F$ is the CNF that expresses the exclusive-or function over $k$ variables and let us derive, in $\mathfrak{R}(k)$, the DNF formula $D$ that expresses the same exclusive-or over $k$ variables. Recall that $F$ has $2^{k-1}$ clauses, each of size $k$, and similarly $D$ has $2^{k-1}$ terms of size $k$ each. Our starting point is the observation that whenever a CNF $F$ implies a DNF $D$, it holds that if we pick any one literal from each term of $D$ and call this set of literals $C$, viewing it as a clause, then $F \vDash C$. Indeed, if this were not the case we could find an assignment that satisfied $F$ but falsified $C$. But if $C$ is falsified this means we have set one literal in each term of $D$ to false, hence $D$ is set to false, contradicting $F \vDash D$.

Now, to derive $D$ from $F$ we proceed as follows. Let us write $D$ for the sake of notational convenience as

$$D = (a_1 \wedge a_2 \wedge \ldots \wedge a_k) \vee (b_1 \wedge b_2 \wedge \ldots \wedge b_k) \vee \ldots \vee (z_1 \wedge z_2 \wedge \ldots, \wedge z_k) \ ,$$

where $a_1, \ldots, a_k, b_1, \ldots, b_k, \ldots, z_k$ are names of literals over the $k$ variables of our exclusive or. (Notice that many of these names refer to the same literal.) By the discussion in the previous paragraph we know that

$$F \vDash (a_1 \vee b_1 \vee \ldots \vee z_1) \ .$$

By the implicational completeness of resolution we can derive this clause from $F$ in resolution. Similarly, we can derive each of the clauses

$$(a_2 \vee b_1 \vee \ldots \vee z_1), \ldots, (a_k \vee b_1 \vee \ldots \vee z_1)$$

with the only difference being in the value of the very first literal in the clause. So using $k-1$ applications of the $\wedge$-introduction rule we can obtain

$$((a_1 \wedge a_2 \wedge \ldots \wedge a_k) \vee b_1 \vee \ldots \vee z_1) \ .$$

Using the same strategy we can obtain the $k-1$ formulas

$$((a_1 \wedge a_2 \wedge \ldots \wedge a_k) \vee b_2 \vee \ldots \vee z_1), \ldots, ((a_1 \wedge a_2 \wedge \ldots \wedge a_k) \vee b_k \vee \ldots \vee z_1)$$

and once we have them, another sequence of $k-1$ applications of $\wedge$-introduction leads to

$$((a_1 \wedge a_2 \wedge \ldots \wedge a_k) \vee (b_1 \wedge b_2 \wedge \ldots \wedge b_k) \vee c_1 \vee \ldots \vee z_1) \ .$$

Continuing in this fashion we keep increasing the number of size-$k$ terms in our formula until we obtain all of $D$.

Note that using this construction, we get a length bound that is doubly exponential in the number of variables in downloaded axiom clauses. This happens when the number of terms in the DNF formula $D$ is exponential in the number of variables $r$, in which case the resulting length bound is of the form $k^{2^{\Omega(r)}}$. This is so since our proof strategy outlined above derives each clause that is obtained by selecting one literal per term of $D$. If, as in this example, $D$ is the $k$-DNF that computes the exclusive or, the number of terms is $2^{k-1}$ and each term has size $k$, thus, we derive $k2^{k-1}$ clauses in total. We note that this seems very inefficient given the fact that there are only $k$ underlying variables. Since we are not primarily interesting in optimizing what in the end will be constant factors, we leave as an open problem the task of reducing the length bound stated in Lemma 5.7 to a singly-exponential one and instead proceed with the formal proof of the lemma.

*Proof of Lemma 5.7.* As explained above, we derive from $F$ in resolution a set of clauses that is equivalent to the $k$-DNF formula $D$. From this set of clauses we derive $D$ using a sequence of $\wedge$-introduction inference rule applications. The key idea is to do all of this in a space-efficient manner by deriving the clauses one by one in a particular order and "merging" each derived clause into a DNF formula that, at the end of this process, turns out to be $D$. Details follow.

Denote $|D|$ by $s$. Suppose $D = \bigvee_{i=1}^{s} \bigwedge_{j=1}^{k_i} a_{i,j}$ where $k_i \leq k$ and $a_{i,j}$ denotes a literal (belonging to a set of $r$ variables). By the distributivity of disjunction over conjunction, $D$ is equivalent to the CNF formula

$$G_D := \bigwedge_{j_1,\ldots,j_s \in [k_1] \times \ldots \times [k_s]} \bigvee_{i=1}^{s} a_{i,j_i} \ . \tag{5.3}$$

Each clause of $G_D$ is implied by $F$ because otherwise there would be an assignment satisfying $F$ but falsifying $G_D$, thereby falsifying $D$ as well, in contradiction to the assumption $F \vDash D$. By the implicational completeness of resolution (Proposition 3.8) there is a resolution derivation of each clause of $G_D$ from $F$. This derivation has length less than $2^{r+1}$ and space at most $(r+2)^2$ because it involves at most $r$ variables. We now show how to construct $D$ from the clauses of $G_D$.

For $s' \in [s]$ and $\vec{j} = (j_{s'+1}, \ldots, j_s) \in [k_{s'+1}] \times \ldots \times [k_s]$, let

$$D_{s',\vec{j}} = \left(\bigvee_{i=1}^{s'} \bigwedge_{j=1}^{k_i} a_{i,j}\right) \vee \bigvee_{i=s'+1}^{s} a_{i,j_i} \ . \tag{5.4}$$

We prove by induction on $s' \geq 0$ that $D_{s',\vec{j}}$ can be derived in total space

$$\bigl((r+2)(\mathit{TotSp}(D_{s',\vec{j}}))\bigr)^2 = \left((r+2)\left(\sum_{i=1}^{s'} k_i + (s - s')\right)\right)^2 \tag{5.5}$$

and length less than $k^{s'} 2^{r+1}$. The base case ($s' = 0$) follows from the discussion in the previous paragraph because $D_{0,\vec{j}}$ is a single clause that is implied by $F$. For the inductive step assume the claim holds for $s' - 1$. We show how to derive, for $k' = 1, \ldots, k_{s'}$, the formula

$$D'_{k'} := \left(\bigvee_{i=1}^{s'-1} \bigwedge_{j=1}^{k_i} a_{i,j}\right) \vee \left(\bigwedge_{j=1}^{k'} a_{s',j}\right) \vee \bigvee_{i=s'+1}^{s} a_{i,j_i} \tag{5.6}$$

in length less than $k' k^{s'-1} 2^{r+1}$ and total space

$$\bigl((r+2)(\mathit{TotSp}(D'_{k'}))\bigr)^2 = \left((r+2)\left(\sum_{i=1}^{s'-1} k_i + k' + (s - s')\right)\right)^2 \ . \tag{5.7}$$

This is shown by induction on $k' \geq 1$. For $k' = 1$ notice (5.6) is nothing but $D_{s'-1,(1,j_{s'+1},\ldots,j_s)}$ so by the inductive hypothesis with respect to $s'-1$ it can be derived in length less than $k^{s'-1}2^{r+1}$ and total space

$$\left((r+2)\left(\sum_{i=1}^{s'-1} k_i + (s - (s'-1))\right)\right)^2 = \left((r+2)\left(\sum_{i=1}^{s'-1} k_i + k' + (s - s')\right)\right)^2 \tag{5.8}$$

For the inductive step assume we have derived $D'_{k'}$ using at most the total space stated in (5.7). Erase all formulas in the memory but for $D'_{k'}$ and notice this remaining formula has total space

$$\sum_{i=1}^{s'-1} k_i + k' + (s - s') \ . \tag{5.9}$$

Using the inductive hypothesis on $s'-1$ again, derive the DNF formula

$$\left(\bigvee_{i=1}^{s'-1} \bigwedge_{j=1}^{k_i} a_{i,j}\right) \vee a_{s',k'+1} \vee \bigvee_{i=s'+1}^{s} a_{i,j_i} \tag{5.10}$$

in total space as in (5.8) and length less than $k^{s'-1}2^{r+1}$. Notice that the total total space used is bounded by the sum given in (5.8) plus the sum in (5.9) (this latter space is required to save the formula $D'_{k'}$) so the combined total space is at most

$$\left((r+2)\left(\sum_{i=1}^{s'-1} k_i + (s - (s'-1))\right)\right)^2 + \sum_{i=1}^{s'-1} k_i + k' + (s - s')$$

$$\leq \left((r+2)\left(\sum_{i=1}^{s'-1} k_i + (k'+1) + (s - s')\right)\right)^2 . \tag{5.11}$$

Now combine $D'_{k'}$ and (5.10) using a single $\bigwedge$-introduction step to obtain $D_{k'+1}$. We see that $D_{k'+1}$ can be derived in total space bounded by (5.8) and length less than $k'k^{s'-1}2^{r+1}$. Summing over $k' = 1, \ldots, k$ we conclude that the derivation of $D_{s'+1,\vec{j}}$ is of length less than $k \cdot k^{s'-1}2^{r+1}$ and total space

$$\left((r+2)\left(\sum_{i=1}^{s'+1} k_i + (s - (s'+1))\right)\right)^2 \tag{5.12}$$

as claimed. Setting $s' = s$ and noticing $TotSp(D) = \sum_{i=1}^{s} k_i$ finishes the proof of the lemma. □

The following easy corollary of the proofs above will also be useful to us.

**Corollary 5.8.** *Let $F$ be any c-CNF formula and $f_d : \{0,1\}^d \mapsto \{0,1\}$ be any non-constant Boolean function. Then it holds that if $F$ can be refuted in standard syntactic resolution in length $L$ and formula space $s$ simultaneously, the substitution formula $F[f_d]$ can be refuted in syntactic $\mathfrak{R}(d)$ in length $L \cdot d^{4^{cd}} \cdot 4^{cd}$ and formula space $s + 2^d + O(1)$ simultaneously.*

*Proof.* This follows from inspection of the proofs of Lemmas 5.6 and 5.7, noting that every clause in the standard resolution proof can be represented by a single corresponding $d$-DNF formula in the $\mathfrak{R}(d)$-proof. We omit the details. □

### 5.3 Projected $\Re(k)$-refutations Are (Essentially) Resolution Refutations

We next prove the converse of Lemma 5.2, namely that $\Re(k)$-refutations of substituted formulas can be translated back to standard resolution refutations of the original formulas as claimed in Lemma 5.4.

This part of the substitution space theorem for $k$-DNF resolution is the one that is the most similar to the theorem for standard resolution, and the proof of Lemma 5.4 closely follows the structure of that of the corresponding Lemma 4.6. We note, though, that there are a few subtle differences between the two proofs due to the fact that our definition of precise implication (Definition 5.3) is somewhat different than the one used in the substitution space theorem for resolution (Definition 4.5). Definition 4.5 appears to be "the right definition" and yields tighter results for standard resolution, but for technical reasons we are forced to relax it a bit in order to obtain the results for $k$-DNF resolution.

We first fix some notation. In analogy with Section 4.4, we use the convention that $\mathbb{D}$ and $D$ denote $k$-DNF sets and $k$-DNF formulas over $Vars(F[f_d])$ while $\mathbb{C}$ and $C$ denote clause sets and clauses over $Vars(F)$. Let us also overload the notation and write $\mathbb{D} \vDash C$, $\mathbb{D} \nvDash C$, and $\mathbb{D} \rhd C$ for $C = C^+ \vee C^-$ when the corresponding implications hold or do not hold for $\mathbb{D}$ with respect to $\bigvee_{x \in C^+} f_d(\vec{x}) \vee \bigvee_{\overline{y} \in C^-} \neg f_d(\vec{y})$. Finally, let $\mathbb{C}_t$ be a shorthand for $proj_F^*(\mathbb{D}_t)$.

Suppose now that $\pi_f = \{\mathbb{D}_0, \ldots, \mathbb{D}_\tau\}$ is a $k$-DNF resolution refutation of $F[f_d]$ for some arbitrary unsatisfiable CNF formula $F$ and some arbitrary non-constant function $f_d$. The first two bullets in Lemma 5.4 are immediate. The hard part is to show that all transitions from $\mathbb{C}_{t-1} = proj_F^*(\mathbb{D}_{t-1})$ to $\mathbb{C}_t = proj_F^*(\mathbb{D}_t)$ can be carried out in such a way that the variable space never exceeds $VarSp(\mathbb{C}_{t-1}) + VarSp(\mathbb{C}_t) \leq 2 \cdot \max_{s \in [\tau]} \{VarSp(\mathbb{C}_s)\}$ during the intermediate derivation steps needed. The proof is by a case analysis of the derivation steps. In the case analysis we will again need Observation 4.8, which is easily seen to hold in this context as well with exactly the same proof.

Consider now the rule applied in $\pi_f$ at time $t$ to get from $\mathbb{D}_{t-1}$ to $\mathbb{D}_t$. We analyze the three possible cases—inference, erasure and axiom download—in this order.

**Inference** Note that $\mathbb{D}_{t-1} \vDash \mathbb{D}_t$ since all inference rules are sound. Moreover, since $\mathbb{D}_t \supseteq \mathbb{D}_{t-1}$ we have $\mathbb{D}_t \vDash \mathbb{D}_{t-1}$. It follows from Definition 5.3 that the set of projected clauses does not change, i.e., $\mathbb{C}_{t-1} = \mathbb{C}_t$, and nothing needs to be done.

**Erasure** If $C \in \mathbb{C}_t \setminus \mathbb{C}_{t-1}$ is a new projected clause appearing at time $t$ as a result of an erasure $\mathbb{D}_t = \mathbb{D}_{t-1} \setminus \{D\}$, it clearly holds that $\mathbb{D}_{t-1} \vDash C$. Hence, all such clauses $C \in \mathbb{C}_t \setminus \mathbb{C}_{t-1}$ can be derived by weakening from $\mathbb{C}_{t-1}$ by Observation 4.8, after which all clauses in $\mathbb{C}_{t-1} \setminus \mathbb{C}_t$ can be erased. During these intermediate steps the variable space is upper-bounded by $VarSp(\mathbb{C}_{t-1} \cup \mathbb{C}_t) \leq VarSp(\mathbb{C}_{t-1}) + VarSp(\mathbb{C}_t)$.

**Axiom download** This is the place in the case analysis where we need to do some serious work. Suppose that $\mathbb{D}_t = \mathbb{D}_{t-1} \cup \{D\}$ for some axiom clause $D \in A[f_d]$, where $A$ in turn is an axiom of $F$. If $C \in \mathbb{C}_t \setminus \mathbb{C}_{t-1}$ is a new projected clause then we must have $\mathbb{D}_{t-1} \nvDash C$ and $\mathbb{D}_{t-1} \cup \{D\} \rhd C$.

We want to show that all such clauses $C$ can be derived from $\mathbb{C}_{t-1} = proj_F^*(\mathbb{D}_{t-1})$ by downloading $A \in F$, making inferences, and then possibly erasing $A$, and that this can be done without the variable space exceeding $VarSp(\mathbb{C}_{t-1}) + VarSp(\mathbb{C}_t)$. Our tool for proving this is the next technical lemma, which is the analogue of Lemma 4.9.

**Lemma 5.9.** *Let $\mathbb{D}$ be a $k$-DNF set derived from $D \in F[f_d]$, $D \in A[f_d]$ be an axiom clause of $F[f_d]$, and $C$ be a clause over $Vars(F)$. If $\mathbb{D}$, $D$, and $C$ are such that $\mathbb{D} \cup \{D\} \rhd C$ but $\mathbb{D} \nvDash C$. Then if $A = a_1 \vee \cdots \vee a_k$, for every $a_i \in A \setminus C$ there is a subclause $C^i \subseteq C$ such that $\mathbb{D} \rhd C^i \vee \overline{a}_i$. That is, all clauses $C \vee \overline{a}_i$ for $a_i \in A \setminus C$ can be derived from $\mathbb{C} = proj_F^*(\mathbb{D})$ by weakening.*

*Proof.* Consider any assignment $\alpha$ such that $\alpha(\mathbb{D}) = 1$ but $\alpha(\bigvee_{x \in C^+} f_d(\vec{x}) \vee \bigvee_{\overline{y} \in C^-} \neg f_d(\vec{y})) = 0$. Such an assignment exists since $\mathbb{D} \not\models C$ by assumption. Also, since by assumption $\mathbb{D} \cup \{D\} \rhd C$ we must have $\alpha(D) = 0$. If $A = a_1 \vee \cdots \vee a_s$, we can write $D \in A[f_d]$ on the form $D = D_1 \vee \cdots \vee D_s$ for $D_i \in a_i[f_d]$. Fix any $a \in A$ and suppose $a = x$ is a positive literal. Then $\alpha(D_i) = 0$ implies that $\alpha(f_d(\vec{x})) = 0$ which means that $\alpha(\neg f_d(\vec{x})) = 1$. Since exactly the same argument holds if $a = \overline{y}$ is a negative literal, we conclude (using our overloaded notation) that

$$\mathbb{D} \models C \vee \overline{a}_i \ . \tag{5.13}$$

If $a_i \in C$, the clause $C \vee \overline{a}_i$ is trivially true, but otherwise we pick $C^i \subseteq C$ minimal such that (5.13) still holds (where since $\mathbb{D} \not\models C$, the literal $\overline{a}_i$ cannot be dropped from the implication). Then by Definition 5.3 we have $\mathbb{D} \rhd C^i \vee \overline{a}_i$ as claimed. □

Lemma 5.9 tells us that every $x \in \mathit{Vars}(A) \setminus \mathit{Vars}(C)$ appears in some clause at time $t-1$, namely, in the clause $C^i \vee \overline{a}_i$ found in the proof above. Since in addition obviously $\mathit{Vars}(A) \cap \mathit{Vars}(C) \subseteq \mathit{Vars}(\mathbb{C}_t)$ this means that if we download $A \in F$ in our refutation $\pi : F \vdash 0$ under construction, we have $\mathit{Vars}(A) \subseteq \mathit{Vars}(\mathbb{C}_{t-1}) \cup \mathit{Vars}(\mathbb{C}_t)$ and hence $\mathit{VarSp}(\mathbb{C}_{t-1} \cup \{A\}) \leq \mathit{VarSp}(\mathbb{C}_{t-1}) + \mathit{VarSp}(\mathbb{C}_t)$.

Thus, we can download $A \in F$, and then possibly erase this clause again at the end of our intermediate resolution derivation to get from $\mathbb{C}_{t-1}$ to $\mathbb{C}_t$, without the variable space ever exceeding $\mathit{VarSp}(\mathbb{C}_{t-1}) + \mathit{VarSp}(\mathbb{C}_t)$. The concluding argument that all new clauses $C \in \mathbb{C}_t \setminus \mathbb{C}_{t-1}$ can be derived from $\mathbb{C}_{t-1} \cup \{A\}$ is exactly as in Section 4.4, but using Lemma 5.9 instead of Lemma 4.9, and it is easily verified that we never mention any variables outside of $\mathit{Vars}(\mathbb{C}_{t-1}) \cup \mathit{Vars}(A) \cup \mathit{Vars}(C)$ in these derivations.

Concluding the proof of Lemma 5.4, we have established that no matter what derivation step is made in the transition $\mathbb{D}_{t-1} \rightsquigarrow \mathbb{D}_t$, we can perform the corresponding transition $\mathbb{C}_{t-1} \rightsquigarrow \mathbb{C}_t$ for our projected clause sets without the variable space going above $\mathit{VarSp}(\mathbb{C}_{t-1}) + \mathit{VarSp}(\mathbb{C}_t) \leq 2 \cdot \max_{\mathbb{D} \in \pi_f} \{\mathit{VarSp}(\mathit{proj}_F^*(\mathbb{D}))\}$. Also, the only time we need to download an axiom $A \in F$ in our projected refutation $\pi$ of $F$ is when $\pi_f$ downloads some axiom $D \in A[f_d]$. This completes the proof of Lemma 5.4.

## 5.4 On the Size of Minimally Unsatisfiable and Minimally Implicating $k$-DNF Sets

We now prove the third and final lemma in Section 5.1, namely Lemma 5.5 bounding the number of variables appearing in a $k$-DNF set that minimally implies a formula. We first deal with the "special case" of minimally unsatisfiable sets (Theorem 2.14 on page 13). The actual result in Lemma 5.5 needed to prove the substitution space theorem follows the outline of this simpler case.

The following simple but important lemma will be used both in the proof of Theorem 2.14 and of Lemma 5.5.

**Lemma 5.10.** *Suppose that $\mathbb{D}$ is a $k$-DNF set that minimally implies a formula $G$. Then for every literal $a$ appearing in any term $T$ in a $k$-DNF formula $D \in \mathbb{D}$ there exists a restriction $\rho$ to $\mathit{Vars}(\mathbb{D})$ satisfying*

- $|\rho| \leq k|\mathbb{D}|$.

- $D'\!\restriction_\rho = 1$ *for all* $D' \in \mathbb{D} \setminus \{D\}$.

- $(T \setminus \{a\})\!\restriction_\rho = 1$.

- $G\!\restriction_\rho \neq 1$.

The point here is that, intuitively speaking, the restriction $\rho$ is very nearly satisfying the $k$-DNF set $\mathbb{D}$ (except for a single literal in a single term) but still has not fixed the formula $G$ implied by $\mathbb{D}$ to true. Also, $\rho$ assigns values to comparatively few variables.

*Proof of Lemma 5.10.* By Definition 2.12, there exists an assignment $\alpha$ to $Vars(\mathbb{D})$ such that $\alpha(D') = 1$ for all $D' \in \mathbb{D}' \setminus \{D\}$ and $\alpha(T \setminus \{a\}) = 1$ but $\alpha(G) = 0$. Let $\rho$ be a restriction of minimal size that agrees with $\alpha$ and satisfies the second and third bullet in the statement of the lemma. Such a restriction can be found by selecting one term $T'$ satisfied by $\alpha$ in each $D' \in \mathbb{D}' \setminus D$ and setting $\rho$ to agree with $\alpha$ on $\bigcup_j Vars(T_j) \cup Vars(T \setminus \{a\})$ and be unfixed elsewhere. Since $|T'| \leq k$ we see $\rho$ has size $\leq k|\mathbb{D}'|$. The last bullet stated above holds because $G(\alpha) = 0$ and $\rho$ agrees with $\alpha$ on all variables fixed by $\rho$. □

Let us now restate and then prove Theorem 2.14.

**Theorem 2.14 (restated).** *Suppose that $\mathbb{D}$ is a minimally unsatisfiable $k$-DNF set. Then the number of variables in $\mathbb{D}$ is at most $|Vars(\mathbb{D})| \leq (k \cdot |\mathbb{D}|)^{k+1}$.*

*Proof.* Let $\mathbb{D} = \{D_1, \ldots, D_m\}$ be a $k$-DNF formula set with $m = |\mathbb{D}|$. For $S$ a set of literals, let $D_i(S)$ be the set of terms in $D_i$ that contain $S$ (recall that we identify a term with the set of literals appearing in it). Formally,

$$D_i(S) = \{T \in D_i : T \supseteq S\} \ . \tag{5.14}$$

Let $Vars(D_i(S))$ denote the set of variables appearing in the set of terms $D_i(S)$. Our theorem follows from the next claim.

**Claim 5.11.** *If $S$ is a set of literals and $|S| = k - r$ then $|Vars(D_i(S))| \leq k \cdot (km)^r$.*

Before proving the claim, let us complete the proof of the theorem. Take $S = \emptyset$ for which we get $r = k$ and notice that $D_i(\emptyset) = D_i$. Claim 5.11 gives

$$|Vars(D_i)| = |Vars(D_i(\emptyset))| \leq k(km)^k \tag{5.15}$$

and summing over all all $m$ formulas in the set we get

$$|Vars(\mathbb{D})| \leq \sum_{i=1}^{m} |Vars(D_i)| \leq m \cdot k(km)^k = (km)^{k+1} = (k|\mathbb{D}|)^{k+1} \tag{5.16}$$

which concludes the proof. □

*Proof of Claim 5.11.* By induction on $r \geq 0$. For the base case of $r = 0$ notice $|S| = k$ so there can be at most one term in $D_i$ that contains all literals in $S$ implying $|Vars(D_i(S))|$ is either 0 or $k$.

For the inductive step we may assume the existence of some term $T \in D_i$ that strictly contains $S$, because otherwise $S$ appears at most once as a term in $D_i$ and the claim holds as in the base case. Assuming $T \supsetneq S$, let $a$ be a literal in $T \setminus S$. Lemma 5.10 guarantees the existence of a restriction $\rho$ of size at most $km$ such that $D_j\!\restriction_\rho = 1$ for all $j \in [m], j \neq i$ and $(T \setminus \{a\})\!\restriction_\rho \neq 0$. By the unsatisfiability of $\mathbb{D}$ we conclude $\rho$ falsifies every term $T' \in D_i$ for which $T' \supsetneq S$. Since $(T \setminus \{a\})\!\restriction_\rho = 1$ and $(T \setminus \{a\}) \supseteq S$ we conclude that every term in $D_i$ that contains $S$ must also contain a literal set to false by $\rho$, because otherwise $\rho$ could be extended to an assignment satisfying $\mathbb{D}$. Recall that $\neg\rho$ is the set of literals set to false by $\rho$. We have just shown that

$$D_i(S) = \bigcup_{a' \in \neg\rho} D_i(S \cup \{a'\}) \ . \tag{5.17}$$

So to bound $Vars(D_i(S))$ we need only bound $Vars(D_i(S \cup \{a'\}))$ for all $a' \in \neg\rho$. We use the inductive hypothesis. Notice $(\neg\rho) \cap S = \emptyset$ because $\rho$ satisfies $S$. Thus, for $a' \in \neg\rho$ we have $|S \cup \{a'\}| = k - (r-1)$. Apply the inductive hypothesis to $S \cup \{a'\}$ to conclude

$$|Vars(D_i(S))| \leq \sum_{a' \in \neg\rho} |Vars(D_i(S \cup \{a'\}))| \leq k(km)^{r-1} \ . \tag{5.18}$$

Summing over all $a \in \neg\rho$ and recalling that $|\neg\rho| = |\rho| \leq km$ establishes the claim. □

To prove Lemma 5.5, we need to address two issues that did not appear in the previous proof. First, our starting point is a $k$-DNF set $\mathbb{D}$ that is satisfiable and implies a set of projected clauses. We deal with this by constructing a formula (to be denoted $G'$) that is the conjunction of all clauses projected by $\mathbb{D}$. The second issue, which is more subtle, is that $\mathbb{D}$ is a set of formulas defined over $Vars(F[f_d])$ whereas the clauses projected by $\mathbb{D}$ are over the different variable set $Vars(F)$. The following definition provides some convenient notation for connecting the two sets of variables.

**Definition 5.12 (Shadow).** For $a$ a literal over a variable $y \in Vars(F[f_d])$ let the *shadow* of $a$, denoted $\mathbf{V}(a)$, be the variable $x \in Vars(F)$ to which $a$ belongs, i.e., the shadow of $y$ is the variable $x$ such that $y \in Vars(x[f_d])$. For $T$ a set of literals (which we will later identified with a term or a restriction) let its shadow be $\mathbf{V}(T) = \bigcup_{a \in T} \mathbf{V}(a)$ and for $D$ a set of terms we define its shadow as $\mathbf{V}(D) = \bigcup_{T \in D} \mathbf{V}(T)$.

The following technical lemma, which will be proven later on, is the analog of Claim 5.11, accounting for the needed modifications which were discussed in the beginning of this subsection. The claim in this lemma is also the central point in our proof of Lemma 5.5. We now state the lemma and promptly use it to prove Lemma 5.5.

**Lemma 5.13.** *Suppose $\mathbb{D} = \{D_1, \ldots, D_m\}$ is a $k$-DNF set over $Vars(F[f_d])$ and $G$ is a CNF formula over $Vars(F)$ such that $\mathbb{D}$ minimally implies the substituted formula $G' = G[f_d]$. Suppose furthermore that $S \subseteq Vars(F)$ and $|S| = k - r$ for $r \geq 0$. Then, letting $D_i(S) = \{T \in D_i \mid \mathbf{V}(T) \supseteq S\}$ denote the set of terms in $D_i$ whose shadow contains $S$, we have $|\mathbf{V}(D_i(S))| \leq k \cdot \left(4^{kd} \cdot k|\mathbb{D}|\right)^r$..*

*Proof of Lemma 5.5.* Let $\mathbb{D} = \{D_1, \ldots, D_m\}$ and $G' = \bigwedge_{C \in proj_F^*(\mathbb{D})} C[f_d]$. Notice that by Definition 5.3, $G'$ is of the form $G' = G[f_d]$ for some CNF formula $G$ over $Vars(F)$ so $G'$ conforms to the assumptions of Lemma 5.13.

First we argue that we may assume without loss of generality that $\mathbb{D}$ minimally implies $G'$. If this is not the case, there must exist a term $T$ appearing in $D_i \in \mathbb{D}$ and a proper subterm $T' \subsetneq T$ such that replacing $T$ by $T'$ and calling the replaced $k$-DNF set by $\mathbb{D}'$, we still have $\mathbb{D}' \vDash G'$. In this case set $\mathbb{D}' = \mathbb{D}$ and repeat the process. Notice that repeating the process does not increase the size of $\mathbb{D}$ (in fact, the size can shrink if some $k$-DNF formula includes an empty term). Since each repetition of this process strictly shrinks the number of literals in $\mathbb{D}$ (counted with repetitions), we see it must terminate. Upon termination the remaining $k$-DNF set, denoted $\hat{\mathbb{D}}$, which is of size at most $m$, minimally implies $G'$.

Our next observation is that for every variable $x$ appearing in $G$ there must exist a literal $a$ belonging to it that appears in $\hat{\mathbb{D}}$. To see this, argue by way of contradiction. Let $C = C' \vee x$ be a clause appearing in $G$ and assume for simplicity that $x$ is a positive literal (the case of a negative literal is identical). Conditions (4.1a) and (4.1b) of Definition 5.3 imply that there exists an assignment $\alpha$ to $Vars(F[f_d])$ such that $\alpha(\mathbb{D}) = \alpha(x[f_d]) = 1$ but $\alpha(C'[f_d]) = 0$. By construction, $\mathbb{D} \vDash \hat{\mathbb{D}}$ so $\alpha(\hat{\mathbb{D}}) = 1$ as well. By assumption, no variable belonging to $x$ appears in $\hat{\mathbb{D}}$, so by changing the value of $\alpha$ on $Vars(x[f_d])$ as to falsify $x[f_d]$ we reach an assignment that satisfies $\hat{\mathbb{D}}$ but falsifies $G[f_d]$, contradiction. To simplify notation from here on we assume without loss of generality that $\mathbb{D}$ minimally implies $G'$ and note for the record that

$$VarSp(G) = |Vars(G)| \leq |\mathbf{V}(\mathbb{D})| \; . \tag{5.19}$$

Next, for $D \in \mathbb{D}$ we bound $|\mathbf{V}(D)|$ using Lemma 5.13 and get

$$|\mathbf{V}(D)| = |\mathbf{V}(D(\emptyset))| \leq k \cdot \left(4^{kd} \cdot k|\mathbb{D}'|\right)^k \leq k \cdot \left(4^{kd} \cdot k|\mathbb{D}|\right)^k \; . \tag{5.20}$$

Summing over all $D \in \mathbb{D}$ gives

$$|\mathbf{V}(\mathbb{D})| \leq \sum_{D \in \mathbb{D}} |\mathbf{V}(D)| \leq |\mathbb{D}| \cdot k \cdot \left(4^{kd} \cdot k|\mathbb{D}|\right)^k = 4^{k^2 d} \cdot \left(k|\mathbb{D}|\right)^{k+1} \tag{5.21}$$

and this, together with (5.19), establishes Lemma 5.5. $\square$

Hence, all that remains now to complete the proof of Theorem 5.1 is to show Lemma 5.13. We end this section by presenting a proof of this final technical lemma.

*Proof of Lemma 5.13.* By induction on $r \geq 0$. For the base case of $r = 0$ we have $|S| = k$. Since $D_i$ is a $k$-DNF formula then any term $T$ for which $\mathbf{V}(T) \supseteq S$ must have $\mathbf{V}(T) = S$. Thus, $|\mathbf{V}(D_i(S))| = k$ and the inequality claimed in the lemma holds.

For the inductive case of $r > 0$, let $\bar{S}$ denote the set of literals that belong to $S$, and let $\text{terms}(S)$ denote the set of terms over $\bar{S}$. We bound the number of terms by $|\text{terms}(S)| = 2^{2|Vars(\bar{S})|} \leq 4^{dk}$ because each term is a set of literals coming from a set of literals of size $2|Vars(\bar{S})|$. Partition the terms in $D_i(S)$ according to their intersection with $\bar{S}$. Formally, for every $s \in \text{terms}(S)$ let

$$D_i(s) = \{T \in D_i(S) \mid T \cap \bar{S} = s\} \ . \tag{5.22}$$

We have partitioned $D_i(S)$ into $4^{kd}$ partitions so to prove the claim in the lemma it is sufficient to show for each partition that

$$|\mathbf{V}(D_i(s))| \leq km \left(k \cdot \left(4^{kd} km\right)^{r-1}\right) \ . \tag{5.23}$$

Consider one term $s \in \text{terms}(S)$. If $\mathbf{V}(D_i(s)) = S$ then clearly (5.23) holds so we assume $\mathbf{V}(D_i(s)) \supsetneq S$. In this case there exists $T \in D_i$ such that $\mathbf{V}(T) \supsetneq S$ which implies the existence of a literal $a \in T \setminus \bar{S}$. Let $\rho$ be a restriction satisfying the properties of Lemma 5.10 with respect to $a, T, \mathbb{D}'$ and $G'$.

**Proposition 5.14.** *Every term $T'$ appearing in $D_i(s)$ must include a literal $a \notin \bar{S}$ whose shadow belongs to the shadow of $\rho$ as well. Formally, $(\mathbf{V}(T') \setminus S) \cap (\mathbf{V}(\rho) \setminus S) \neq \emptyset$.*

*Proof of Proposition 5.14.* By way of contradiction. Assume $T'$ falsifies the proposition. By assumption $T'$ has the same set of literals as $T$ within $\bar{S}$ and the third property of $\rho$ listed in Lemma 5.10 implies $\rho$ satisfies all literals of $T'$ inside $\bar{S}$. Assuming that the intersection in the statement of the proposition is empty, we can extend $\rho$ to a restriction $\rho'$ that satisfies $T'$ by setting at most $k$ "new" variables on top of those set by $\rho$. The crucial observation is that none of the "new" variables set by $\rho'$ have their shadow in $\mathbf{V}(\rho)$. More to the point, suppose $x_i$ is a "new" variable whose value is set by $\rho'$ but is not set by $\rho$. Let $x$ denote the shadow of $x_i$ and let $\vec{x} = \{x_1, \ldots, x_d\}$ be the set of variables whose shadow is $x$. Our crucial observation, restated in different words, is that $\rho$ does not set the value of *any* variable $x_1, \ldots, x_k$. This is where the $k$-non-authoritarianism of $f_d$ comes into play, because it implies that $\rho'$ cannot fix the value of $f_d(\vec{x})$ by just setting a single variable $x_i$. But this means that we can extend $\rho'$ so that $f_d(\vec{x})$ will obtain any truth value we find fit. We conclude that the fourth property listed in Lemma 5.10 holds for $\rho'$ as well as for $\rho$. This property implies that $\rho'$ can be extended to an assignment $\alpha'$ such that $G'(\alpha') = 0$. So $\alpha'$ is an assignment that satisfies $\mathbb{D}'$ but falsifies $G'$. We have reached a contradiction, and the proposition follows. □

We continue with the proof of the inequality (5.23). The second property of Lemma 5.10 implies that $|\mathbf{V}(\rho)| \leq km$. Thus, Proposition 5.14 shows that there exists a set $V_s \subseteq Vars(F) \setminus S$ of size at most $km$ such that

$$\mathbf{V}(D_i(s)) \subseteq \bigcup_{v \in V_s} \mathbf{V}(D_i(S \cup \{v\})) \ . \tag{5.24}$$

Since $v \notin S$ we have $|S \cup \{v\}| = k - (r-1)$ so we may apply the inductive hypothesis of the inequality in Lemma 5.13 to $S \cup \{v\}$ which gives

$$|\mathbf{V}(D_i(s))| \leq \sum_{v \in V_s} |\mathbf{V}(D_i(S \cup \{v\}))| \leq km \left(k \cdot \left(4^{kd} \cdot km\right)^{r-1}\right) \ . \tag{5.25}$$

We have shown that the inequality (5.23) holds for all $s \in \text{terms}(S)$. Summing over all terms, there are at most $4^{kd}$ of them. The lemma follows. □

# 6 Reductions Between $k$-DNF Resolution and Pebbling

It is not hard to see how a black pebbling $\mathcal{P}$ of a DAG $G$ can be used to construct a resolution refutation of the pebbling contradiction $Peb_G$ in Definition 3.12 in length and space upper-bounded by *time*($\mathcal{P}$) and *space*($\mathcal{P}$), respectively. It is straightforward to show that this translation from pebblings to refutations works even if we do an $f_d$-substitution in the pebbling contradiction. We present a proof of this fact in Section 6.1.

Using our new results in Section 4, we can prove the more surprising fact that there is also a fairly tight reduction in the other direction: provided that the function $f_d$ is non-authoritarian, any resolution refutation of $Peb_G[f_d]$ translates into a black-white pebbling of $G$ with the same time-space properties (adjusting for constant factors depending on the function $f_d$ and the maximal indegree of $G$). This new reduction is given in Section 6.2.

In Section 6.3, corresponding versions of both theorems are given for $k$-DNF resolution. Finally, in Section 6.4 we put everything together to prove a meta-theorem saying that for DAGs $G$ having the right time-space trade-off properties, we can prove that pebbling contradictions defined over such DAGs inherit (roughly) the same trade-off properties. This will allow us to use pebbling time-space trade-offs to obtain a wealth of strong trade-offs for both formula space and total space for standard resolution and $\mathfrak{R}(k)$ in Section 7.

## 6.1 From Black Pebblings to Resolution Refutations

In Lemma 2.3 we recalled that given any black-only pebbling $\mathcal{P}$ of a DAG $G$, we can mimic this pebbling in a resolution refutation of $Peb_G$ by deriving that a literal $v$ is true whenever the corresponding vertex in $G$ is pebbled. This construction carries over also to substitution formulas $Peb_G[f_d]$ and we have the following theorem.

**Theorem 6.1.** *Let $f_d$ be a non-constant Boolean function of arity $d$ and let $G$ be a DAG with indegree at most $\ell$ and unique sink. Then given any complete black pebbling $\mathcal{P}$ of $G$, we can construct a resolution refutation $\pi : Peb_G[f_d] \vdash 0$ such that*

$$L(\pi) \leq \textit{time}(\mathcal{P}) \cdot \exp\bigl(\mathrm{O}(d(\ell+1))\bigr) \ ,$$
$$W(\pi) \leq d(\ell+1) \ , \textit{ and}$$
$$TotSp(\pi) \leq \textit{space}(\mathcal{P}) \cdot \exp\bigl(\mathrm{O}(d(\ell+1))\bigr) \ .$$

*Proof.* Follows directly from Lemma 2.3 and Theorem 2.1. □

We note that in our applications we will have the function arity $d$ and the DAG indegree $\ell$ fixed (for standard resolution, we can pick $d = \ell = 2$), which means that the bounds on length and space above turns into $L(\pi) = \mathrm{O}\bigl(\textit{time}(\mathcal{P})\bigr)$ and $TotSp(\pi) = \mathrm{O}\bigl(\textit{space}(\mathcal{P})\bigr)$. We also remark that for concrete functions $f_d$, such as for instance XOR over two variables, we can easily compute explicit upper bounds on the constants hidden in the asymptotic notation if we so wish, and these constants are small.

## 6.2 From Resolution Refutations to Black-White Pebblings

In the other direction, i.e., from resolution refutations to pebbling strategies, the current paper establishes the following correspondence for resolution (which is a slightly more general version of the statement made in Theorem 2.4).

**Theorem 6.2.** *Let $f$ be any non-authoritarian Boolean function and $G$ be any DAG with unique sink and bounded indegree $\ell$. Then from any resolution refutation $\pi : Peb_G[f] \vdash 0$ we can extract a black-white pebbling strategy $\mathcal{P}_\pi$ for $G$ such that $\textit{time}(\mathcal{P}_\pi) \leq (\ell+1) \cdot L(\pi)$ and $\textit{space}(\mathcal{P}_\pi) \leq Sp(\pi)$.*

Before proving this theorem, we want to highlight the fact that Theorems 6.1 and 6.2 are far from being perfect converses of one another. This is so since the reduction in one direction uses black pebbling (Theorem 6.1) while the reduction in the other direction is in terms of black-white pebbling (Theorem 6.2). It was shown in [KS91] that there can be a quadratic gap in pebbling price depending on whether white pebbles may be used or not. At first sight this might not look to bad, since [Mey81] proved that the gap can never be worse than quadratic. However, the translation given in [Mey81] of a black-white pebbling in space $s$ to a black-only pebbling in space $\mathrm{O}(s^2)$ incurs an exponential blow-up in pebbling time, destroying all hope of obtaining nontrivial tradeoff results in this way. Hence, to get meaningful tradeoffs from Theorems 6.1 and 6.2 we need a graph family where either the trade-off properties with respect to black-white and black-only pebbling are closely related, or where we can translate not only black but also black-white pebbling strategies into resolution refutations in a way that preserves both time and space.

In this paper, we will use both of these approaches to get tight trade-off results. Some pebbling results in the literature, notably several results in the monumental paper [LT82], have been shown to hold asymptotically for both black-white and black-only pebbling, and we also appeal to [Nor10b] to get some new such graphs. Also, in certain limited settings it turns out to be possible to simulate black-white pebblings in resolution, and for such graphs we can again get tight proof complexity trade-offs drawing on structural results in [Nor10b]. These results hold only for a particular type of black-white pebblings, however, and it remains an interesting open problems whether or not black-white pebblings can be simulated efficiently in resolution in general.

*Proof of Theorem 6.2.* The proof is in three steps:

1. First, we convert $\pi : Peb_G[f] \vdash 0$ to a refutation $\pi'$ of $Peb_G$ such that $VarSp(\pi') \leq Sp(\pi)$ and the number of axiom downloads in $\pi'$ is upper-bounded by the number of axiom downloads in $\pi$. This is Theorem 2.1.

2. The refutation $\pi' : Peb_G \vdash 0$ can contain weakening moves, which we do not want, so we appeal to Proposition 3.6 to get a refutation $\pi'' : Peb_G \vdash 0$ without any weakening steps. By Lemma 3.10, without loss of generality we can assume that $\pi''$ is frugal (Definition 3.9). This part of the proof just uses standard techniques, and the number of axiom downloads and the variable space can only decrease when going from $\pi'$ to $\pi''$.

3. Finally, we show that $\pi''$ corresponds to a black-white pebbling strategy $\mathcal{P}$ for $G$ such that *time*($\mathcal{P}$) is upper-bounded by the number of axiom downloads and *space*($\mathcal{P}$) is upper-bounded by the maximal number of variables occurring simultaneously in $\pi''$. This final part was essentially proven by the first author in [Ben09], but since we need a more detailed result than can be read off from that paper, we present the full construction below for completeness.

Putting together these three steps, Theorem 6.2 clearly follows. □

What remains is thus to show the following, slightly more detailed version of a lemma in [Ben09], which serves to establish part 3 in the proof above.

**Lemma 6.3 ([Ben09]).** *Let $G$ be any DAG with unique sink and bounded indegree $\ell$, and suppose that $\pi$ is any resolution refutation of $Peb_G$ without weakening that is also frugal. Then there is a black-white pebbling strategy $\mathcal{P}_\pi$ for $G$ such that* **space**$(\mathcal{P}_\pi) \leq VarSp(\pi)$ *and* **time**$(\mathcal{P}_\pi)$ *is at most $(\ell + 1)$ times the number of axiom downloads in $\pi$.*

*Proof.* Given a refutation $\pi = \{\mathbb{C}_0 = \emptyset, \mathbb{C}_1, \ldots, \mathbb{C}_\tau = \{0\}\}$ of $Peb_G$, we translate every clause set $\mathbb{C}_t$ into a black-white pebble configuration $\mathbb{P}_t = (B_t, W_t)$ using a slightly modified version of the ideas in [Ben09], and then show that $\mathcal{P} = \{\mathbb{P}_0, \ldots, \mathbb{P}_\tau\}$ is essentially a legal black-white pebbling of $G$ as in the statement of the lemma. The translation will satisfy the invariant that $B_t \cup W_t = Vars(\mathbb{C}_t)$ which yields the upper bound on pebbling space in terms of variable space. The first configuration $\mathbb{C}_0 = 0$ is thus translated into $\mathbb{P}_0 = (\emptyset, \emptyset)$.

Suppose inductively that $(B_{t-1}, W_{t-1})$ has been constructed from $\mathbb{C}_{t-1}$ and consider all the variables $x \in Vars(\mathbb{C}_t)$ one by one. If $x \in Vars(\mathbb{C}_t) \cap B_{t-1}$, keep $x$ in $B_t$. Otherwise, if $x \in Lit(\mathbb{C}_t)$ appears as a positive literal, add $x$ to $B_t$. Otherwise, if $\overline{x} \in Lit(\mathbb{C}_t)$, add $x$ to $W_t$. This is our translation of $\mathbb{C}_t$ into black pebbles $B_t$ and white pebbles $W_t$. To see that this translation yields a legal pebbling, consider the derivation rule applied to get from $\mathbb{C}_{t-1}$ to $\mathbb{C}_t$.

***Axiom download***   Suppose that we download the pebbling axiom or source axiom for a vertex $v$ with immediate predecessors $u_1, \ldots, u_{\ell'}$ (where we have $\ell' = 0$ for a source $v$). All predecessors $u_i$ not having pebbles on them at time $t-1$ get white pebbles. Then $v$ gets a black pebble, if it did not already have one. Note that this is a legal pebble placement since all immediate predecessors of $v$ (if any) have pebbles at this point. We remark that to black-pebble $v$, we might have to remove a white pebble from $v$ first, but since all immediate predecessors have pebbles on them this poses no problems. Also, downloading the sink axiom places a white pebble on the sink $z$ if this vertex is empty, which is a legal pebbling move. By the bound on the indegree, this step involves placing at most $\ell + 1$ pebbles.

***Inference***   In this case $Vars(\mathbb{C}_{t-1}) = Vars(\mathbb{C}_t)$, so nothing happens.

***Erasure***   Suppose that the clause erased in $C$. Just apply the translation function. Suppose that this results in a pebble on $x$ disappearing. Then we have $x \in Vars(C)$ but $x \notin Vars(\mathbb{C}_t)$. Before being erased, $C$ has been resolved with some other clause (recall that $\pi$ is frugal). But as long as we did not resolve over the variable $x$, we will still have $x \in Vars(\mathbb{C}_t)$, and hence $\mathbb{C}$ must have been resolved over $x$ at some time $t' < t$. At this time $x$ appeared both positively and negatively in $\mathbb{C}_{t'}$, and in view of how we defined the translation from clauses to pebbles, this means that the vertex $x$ has contained a black pebble in the interval $[t', t-1]$. Thus the pebble disappearing at time $t$ is black, and black pebbles can always be removed freely.

To conclude the proof, note that during the course of the refutation all axioms must have been downloaded at least once, since $Peb_G$ is easily seen to be minimally unsatisfiable. In particular, this means that the sink $z$ is black-pebbled at some time during the proof, and we can decide to keep the black pebble on $z$ from that moment onwards. (This potentially adds one pebble extra to the pebbling space, but this is fine since the inequality in Theorem 2.1 is strict so there is margin for this.)

Since every time an axiom is downloaded it must also be erased at some later time, we get the time bound of $(\ell + 1)$ times the number of axiom downloads (and in fact it is easy to see that this bound can be improved by taking into account the inference steps, when nothing happens in the pebbling). The lemma follows. □

## 6.3   Reductions Between Pebbling and $k$-DNF Resolution

For $k$-DNF resolution, we get the following, slightly different, reductions. The next theorem is particularly interesting when we have $d = K + 1$, which we can get, for instance, by picking $f_d(x_1, \ldots, x_d) = \bigoplus_{i=1}^{d} x_i$.

**Theorem 6.4.** *Let $K$ be any fixed positive integer. Suppose that $f_d : \{0,1\}^d \mapsto \{0,1\}$ is any $K$-non-authoritarian Boolean function and that $G$ is any DAG with unique sink of size $\Theta(n)$ and bounded vertex indegree $\ell$. Then the following holds.*

1. *The formula $Peb_G[f_d]$ is refutable in syntactic $\mathfrak{R}(d)$ in length $O(n)$ and formula space $O(1)$.*

2. For any $k \leq K$, from any semantic $\Re(k)$-refutation $\pi$ of $Peb_G[f_d]$ we can extract a black-white pebbling $\mathcal{P}_\pi$ of $G$ such that $\mathsf{time}(\mathcal{P}_\pi) = \mathrm{O}(L(\pi))$ and $\mathsf{space}(\mathcal{P}_\pi) = \mathrm{O}\big(Sp(\pi)^{k+1}\big)$.

*Any constants hidden in the asymptotic notation depends only on $K$, $d$, and $\ell$.*

*Proof.* Part 2 of the theorem follows just as in the proof of Theorem 6.2, but appealing to Theorem 5.1 for $\Re(k)$ instead of Theorem 2.1 for standard resolution. For part 1, we use the result in [Ben09] that any formula $Peb_G$ can be refuted in linear length and constant clause space simultaneously, and then appeal to Corollary 5.8. □

### 6.4 Obtaining Resolution Trade-offs from Pebbling

Combining the theorems above, we can now prove that if we can find DAGs $G$ with appropriate pebbling trade-off properties, such DAGs immediately yield trade-off results in resolution. And as we will see in Section 7, there are (explicitly constructible) DAGs with the needed properties.

In order not to clutter the statement of the next theorem, we assume that the indegree $\ell$ of the DAGs, the arity $d$ of the Boolean functions $f$, and the maximal size of terms $K$ in the $k$-DNF resolution proof systems that we consider are all fixed, so that any dependence on $K$, $d$ and $\ell$ can be hidden in the asymptotical notation. (This is not much of a restriction since we will always be able to choose $\ell = 2$, and also we will have $d = K + 1$ in the applications that we care about.)

**Theorem 6.5 ($\Re(k)$-trade-offs from pebbling).** *Let $K$, $d$, and $\ell$, be arbitrary but fixed positive integers such that $K < d$, and let $f : \{0,1\}^d \mapsto \{0,1\}$ be some $K$-non-authoritarian function. Suppose that $G$ is a DAG with $n$ vertices, unique sink, and bounded indegree $\ell$, and that $g, h : \mathbb{N}^+ \mapsto \mathbb{N}^+$ are functions satisfying the following properties:*

- *For every $s \geq \mathsf{Peb}(G)$ there is a black pebbling $\mathcal{P}$ of $G$ with $\mathsf{space}(\mathcal{P}) \leq s$ and $\mathsf{time}(\mathcal{P}) \leq g(s)$.*

- *For every $s \geq \mathsf{BW\text{-}Peb}(G)$ and every black-white pebbling $\mathcal{P}$ of $G$ with $\mathsf{space}(\mathcal{P}) \leq s$ it holds that $\mathsf{time}(\mathcal{P}) \geq h(s)$.*

*Then the following holds for the substitution pebbling formula $Peb_G[f]$:*

1. *$Peb_G[f]$ is a CNF formula of size $\Theta(n)$ and width $\mathrm{O}(1)$.*

2. *$Peb_G[f]$ is refutable in standard syntactic resolution in length $L_\Re(Peb_G[f] \vdash 0) = \mathrm{O}(n)$ and width $W(Peb_G[f] \vdash 0) = \mathrm{O}(1)$ simultaneously, and is also refutable in syntactic resolution in total space $TotSp_\Re(Peb_G[f] \vdash 0) = \mathrm{O}\big(\mathsf{Peb}(G)\big)$.*

3. *$Peb_G[f]$ is refutable in syntactic $\Re(d)$ in length $L_{\Re(d)}(Peb_G[f] \vdash 0) = \mathrm{O}(n)$ and formula space $Sp_{\Re(d)}(Peb_G[f] \vdash 0) = \mathrm{O}(1)$ simultaneously.*

4. *For every $s \geq \mathsf{Peb}(G)$ there is a standard syntactic resolution refutation $\pi_s : Peb_G[f] \vdash 0$ in length $L(\pi_s) = \mathrm{O}(g(s))$ and total space $TotSp(\pi_s) = \mathrm{O}(s)$.*

5. *The clause space of any semantic resolution refutation is lower-bounded by $Sp_\Re(Peb_G[f] \vdash 0) \geq \mathsf{BW\text{-}Peb}(G)$, and for every $s \geq \mathsf{BW\text{-}Peb}(G)$ and every semantic refutation $\pi_s : Peb_G[f] \vdash 0$ in clause space $Sp(\pi_s) \leq s$, it holds that $L(\pi_s) = \Omega(h(s))$.*

6. *For every $s \geq \mathsf{BW\text{-}Peb}(G)$ and every $k \leq K$, the formula space of any semantic $\Re(k)$-refutation is lower-bounded by $Sp_{\Re(k)}(Peb_G[f] \vdash 0) = \Omega\big(\sqrt[k+1]{\mathsf{BW\text{-}Peb}(G)}\big)$, and any semantic $\Re(k)$-refutation $\pi_s : Peb_G[f] \vdash 0$ in formula space $Sp(\pi_s) = \mathrm{o}\big(\sqrt[k+1]{s}\big)$, it holds that $L(\pi_s) = \Omega(h(s))$.*

*All hidden constants in the asymptotical notation depend only on $K$, $d$, and $\ell$, and are independent of $G$.*

*Proof.* Item 1 on the list is an easy consequence of Definition 3.14. The rest is just a matter of putting together all the theorems proven in this section. Hence, items 2 and 4 both follow from Theorem 6.1 (to get item 2, consider the trivial pebbling that black-pebbles all vertices of $G$ in topological order). Theorem 6.2 yields item 5. Finally, items 3 and 6 follow from Theorem 6.4. □

This theorem will be of particular interest when we can find graph families $\{G_n\}_{n=1}^{\infty}$ with $\mathsf{Peb}(G_n) = \Theta\big(\mathsf{BW\text{-}Peb}(G_n)\big)$ having trade-off functions $g_n(s) = \Theta(h_n(s))$. For such families of DAGs, Theorem 6.5 yields asymptotically tight trade-offs in standard resolution. As we can see, these trade-offs hold for both clause space and total space simultaneously with respect to length, since the upper bounds are in terms of total space and the lower bounds in terms of clause space.

For $k$-DNF resolution, $k \geq 2$, the trade-offs obtained by using Theorem 6.5 are *not* tight, since there is a gap of a $(k+1)$st root between the upper and lower bounds on space. We do not know what the "correct" result for $\mathfrak{R}(k)$ is, and a priori it could well be the case that qualitatively the same trade-offs should hold for $\mathfrak{R}(k)$ as for standard resolution. Such a strengthening of Theorem 6.5 provably *cannot* be obtained by our methods, however. In view of the recent results in [NR09], our techniques will always leave at least a $k$th root gap, so in order to make any more substantial improvements to Theorem 6.5 it seems that fundamentally new ideas would be needed.

# 7 Separation and Trade-off Results for $k$-DNF Resolution

We have finally reached the point where we can state and prove our time-space trade-off results for standard resolution and $\mathfrak{R}(k)$. Given all the work done so far, the proofs are all variations of the following pattern: pick some suitable graph family, make the appropriate choices of parameters, consider the corresponding pebbling contradiction CNF formulas, do $f$-substitution for some non-authoritarian function $f$, and apply Theorem 6.5 (which we obtained with the help of the "substitution space theorems" in Theorems 2.1 and 5.1).

## 7.1 More Detailed Discussion of Previous Trade-off Results for Resolution

For completeness, we start this section by a slightly more detailed discussion than in Section 1 of previous work in this area. The question of length-space trade-offs in resolution was first studied by the first author in [Ben09] and more recently by the second author in [Nor09b]. (A proceedings version work of Hertel and Pitassi [HP07] claimed a trade-off result that was simplified and improved by [Nor09b]. The journal version [HP10] retracted this claim due to an error in the proof.) Let us describe these works and how they compare to our results.

The paper [Ben09] contains a number of results for general resolution. For instance, it shows a strong trade-off between clause space and width, establishing that there are formulas refutable in constant clause space as well as constant width, but where for any particular refutation $\pi$ it holds that $Sp(\pi) \cdot W(\pi) = \Omega(n/\log n)$. However, the lengths-space trade-offs in [Ben09] are limited to the very restricted case of tree-like resolution, and do not extend to general resolution.

In contrast, [Nor09b], studies general, unrestricted resolution. The results therein apply to resolution refutation of bounded-width formulas, but again every theorem is restricted to dealing with a particular space measure. An unsatisfying aspect of all results in [Nor09b] is that they use quite artificial constructions of formulas "glued together" from two different unsatisfiable subformulas over disjoint variable sets. In particular, these constructions are non-explicit. Moreover, in [Nor09b] the length-space trade-off results apply only for a very carefully selected ratio of space to formula size, and display an abrupt decay of proof length when space is increased even by small amounts.

In contrast, all trade-off results presented in the current paper have the following properties:

- They apply to general, unrestricted resolution, and even extend to $k$-DNF resolution.

- They are stated for explicitly constructible formulas that have bounded width and are (minimally) unsatisfiable.

- They apply simultaneously to both total space and clause space (formula space for $k$-DNF resolution), since the upper bounds are in terms of total space and the lower bounds in terms of clause/formula space. Moreover, recalling Definition 3.5 on page 17, we stress that all our upper bounds are in terms of *syntactic* proof systems (i.e., the usual ones) whereas the lower bounds in the trade-offs hold even for the much stronger *semantic* versions of the proof systems.

- Finally, the trade-offs are very robust in the sense that they are not sensitive to small perturbations of either length or space.

In the rest of this section, we state and prove our collection of separation and trade-off results.

## 7.2 Space Hiearchy for $k$-DNF Resolution

We gave a sketch of the proof of the space hierarchy for $k$-DNF resolution already in Section 2.3, and the missing details in this sketch are easily filled in from the material in Section 6. Let us nevertheless write out the details here for the convenience of the reader.

**Theorem 2.5 (restated).** *For every $k \geq 1$ there exists an explicitly constructible family of $(3(k+1))$-CNF formulas $\{F_n\}_{n=1}^{\infty}$ of size $\Theta(n)$ such that*

1. *there are $\Re(k+1)$-refutations $\pi_n : F_n \vdash 0$ in simultaneous length $L(\pi_n) = \mathrm{O}(n)$ and formula space $Sp(\pi_n) = \mathrm{O}(1)$, but*

2. *any $\Re(k)$-refutation of $F_n$ requires formula space $\Omega\bigl(\sqrt[k+1]{n/\log n}\bigr)$.*

*The constants hidden by the asymptotic notation depend only on $k$.*

The theorem is established with the help of the following graph family.

**Lemma 7.1 ([GT78]).** *There is an explicitly constructible family of DAGs $\{G_n\}_{n=1}^{\infty}$ of size $\Theta(n)$ having a unique sink and vertex indegree 2 such that $\mathsf{BW\text{-}Peb}(G_n) = \Theta(n/\log n)$.*

*Proof of Theorem 2.5.* Take the pebbling formulas defined in terms of the graphs in Lemma 7.1, substitute a $k$-non-authoritarian Boolean function $f$ of arity $k+1$, say XOR over $k+1$ variables for concreteness, and appeal to Theorem 6.5. The size and width of the formula follows from part 1 (and Definition 3.14), the upper bound on formula space is part 3, and the lower bound follows from part 6 of Theorem 6.5. □

We note that for standard resolution, it is possible to get rid of the square root in the separation and prove that any resolution refutation of $F_n$ requires formula space $\Theta(n/\log n)$ (as was done in [BN08]). This follows by appealing to part 5 of Theorem 6.5 instead.

## 7.3 Trade-offs for Constant Space

Shifting focus to our length-space trade-offs, which occupy the bulk of this section, our first result is that trade-offs can occur even for formulas refutable in constant space. What is more, there are such formulas for which we can prove not only a trade-off threshold, but (in the case of standard resolution) even specify the whole trade-off curve. We establish this by studying the following family of graphs (referred to in [LT82] as *bit reversal graphs*).

**Lemma 7.2 ([LT82]).** *There are explicitly constructible DAGs $G_n$ of size $\Theta(n)$ with a single sink and vertex indegree 2 having the following pebbling properties:*

1. *The black pebbling price of $G_n$ is $\mathsf{Peb}(G_n) = 3$.*

2. *Any black pebbling strategy $\mathcal{P}_n$ for $G_n$ that optimizes time given space constraints[9] $\mathrm{O}(n)$ exhibits a trade-off $\mathsf{time}(\mathcal{P}_n) = \Theta\bigl(n^2/\mathsf{space}(\mathcal{P}_n)\bigr)$.*

3. *Any black-white pebbling strategy $\mathcal{P}_n$ for $G_n$ that optimizes time given space constraints $\mathrm{O}(\sqrt{n})$ exhibits a trade-off $\mathsf{time}(\mathcal{P}_n) = \Theta\bigl((n/\mathsf{space}(\mathcal{P}_n))^2\bigr)$.*

What we would like to do now is to plug this theorem right into Theorem 6.5 and be done. Unfortunately, although this will prove to be a simple and successful strategy in general, it will not work for this particular family of formulas if we want to get tight trade-offs. This is so since there is a quadratic gap between the black-white and black-only pebbling trade-offs (see the discussion immediately after Theorem 6.2). However, analyzing the strcture of the proof of Lemma 7.2 a little bit closer, it turns out that in this particular case it is possible to simulate optimal black-white pebblings in resolution in a time- and space-preserving way. (This is a special case of a more general theorem proven in [Nor10b], and we refer to that paper for the details.)

**Lemma 7.3 ([Nor10b]).** *Let $F_n$ be pebbling substitution formulas over the graphs $G_n$ in Lemma 7.2. Then for any $s = \mathrm{O}(\sqrt{n})$ there are syntactic standard resolution refutations in length $\mathrm{O}((n/s)^2)$ and clause space $\mathrm{O}(s)$.*

Using both Lemma 7.2 and Lemma 7.3, we can establish the results stated next, which are tight for resolution.

**Theorem 7.4 (More detailed version of Theorem 2.6).** *For any fixed positive integer $K$ there are explicitly constructible CNF formulas $\{F_n\}_{n=1}^{\infty}$ of size $\Theta(n)$ and width $3(K+1)$ such that the following holds (where all multiplicative constants hidden in the asymptotic notation depend only on $K$):*

1. *The formulas $F_n$ are refutable in syntactic resolution in total space $\mathit{TotSp}_{\mathfrak{R}}(F_n \vdash 0) = \mathrm{O}(1)$.*

2. *For any $s(n) = \mathrm{O}(\sqrt{n})$ there are syntactic resolution refutations $\pi_n$ of $F_n$ in simultaneous length $L(\pi_n) = \mathrm{O}\bigl((n/s(n))^2\bigr)$ and total space $\mathit{TotSp}(\pi_n) = \mathrm{O}(s(n))$.*

3. *There are syntactic $\mathfrak{R}(K+1)$-refutations $\pi_n$ of $F_n$ in simultaneous length $L(\pi_n) = \mathrm{O}(n)$ and formula space $Sp(\pi_n) = \mathrm{O}(1)$.*

4. *For any semantic resolution refutation $\pi_n : F_n \vdash 0$ in clause space $Sp(\pi_n) \leq s(n)$ it holds that $L(\pi_n) = \Omega\bigl((n/s(n))^2\bigr)$.*

---

[9]The reason for including the upper bounds on space in the statement of the theorem is that no matter how much space is available, it is of course never possible to do better than linear time. Thus the trade-offs cannot hold when length dips below linear.

5. *For any $k \leq K$, any semantic $\Re(k)$-refutation $\pi_n : F_n \vdash 0$ in formula space $Sp(\pi_n) \leq s(n)$ must have length $L(\pi_n) = \Omega\left(\left(n/(s(n)^{1/(k+1)})\right)^2\right)$. In particular, any constant-space $\Re(k)$-refutation must also have quadratic length.*

¿From now on and for the rest of this section, we will assume unless stated otherwise that $d = K + 1$ and that $f = f_d(x_1, \ldots, x_d) = \bigoplus_{i=1}^{d} x_i$ is the exclusive or function over $d$ variables (although we will redundantly repeat this from time to time for increased clarity).

*Proof of Theorem 7.4.* Consider the pebbling formulas $Peb_{G_n}[f]$ defined over the bit reversal DAGs in Lemma 7.2. Combining Lemma 7.2 with Theorem 6.5, and improving the upper bounds by appealing to Lemma 7.3, the theorem follows.

Since this is our first trade-off proof, let us write it out in detail. Thus, the upper bound on total space in part 1 of Theorem 7.4 follows from the pebbling space upper bound in part 1 of Lemma 7.2 combined with the reduction from pebbling to resolution in part 2 of Theorem 6.5. As noted above, the upper bound on the length-space trade-off in Part 2 of Theorem 7.4 does *not* follow from Theorem 6.5, but is obtained by applying Lemma 7.3. The constant upper bound on $\Re(K+1)$-formula space in part 3 is part 3 of Theorem 6.5. The lower bound on the length-space trade-off for resolution in part 4 follows from the black-white pebbling trade-off in part 3 of Lemma 7.2 combined with the reduction from resolution to pebbling in part 5 of Theorem 6.5. Finally, the weaker bound on $\Re(k)$-tradeoffs in part 5 of Theorem 7.4 is what follows when we apply the weaker reduction from $k$-DNF resolution to pebbling in part 6 of Theorem 6.5 instead. □

## 7.4 Superpolynomial Trade-offs for any Non-constant Space

It is clear that we can never get superpolynomial trade-offs from DAGs pebblable in constant space, since such graphs must have constant-space pebbling strategies in polynomial time by a simple counting argument. However, perhaps somewhat surprisingly, as soon as we study *arbitrarily slowly* growing space, we can obtain superpolynomial trade-offs for formulas whose refutation space grows this slowly. This is a consequence of the following recent pebbling trade-off result from [Nor10b], extending a construction by Carlson and Savage [CS80, CS82].

**Lemma 7.5 ([Nor10b]).** *There is an explicitly constructible graph family $\Gamma(c, r)$, for $c, r \in \mathbb{N}^+$, with a unique sink and vertex indegree 2, having the following properties:*

1. *The graphs $\Gamma(c, r)$ are of size $|V(\Gamma(c, r))| = \Theta(cr^3 + c^3r^2)$.*

2. *$\Gamma(c, r)$ has black-white pebbling price $\mathsf{BW\text{-}Peb}(\Gamma(c, r)) = r + \mathrm{O}(1)$ and black pebbling price $\mathsf{Peb}(\Gamma(c, r)) = 2r + \mathrm{O}(1)$.[10]*

3. *There is a black-only pebbling of $\Gamma(c, r)$ in time linear in the graph size and in space $\mathrm{O}(c + r)$.*

4. *Suppose that $\mathcal{P}$ is a black-white pebbling of $\Gamma(c, r)$ with $\mathsf{space}(\mathcal{P}) \leq r + s$ for $0 < s \leq c/8$. Then the time required to perform $\mathcal{P}$ is lower-bounded by*

$$\mathsf{time}(\mathcal{P}) \geq \left(\frac{c - 2s}{4s + 4}\right)^r \cdot r! \ .$$

The graph family in Lemma 7.5 turns out to be surprisingly versatile and will occur several times below with different parameter settings. We now use it to prove the following theorem

---
[10]Note that item 2 says that the pebbling price grows linearly with $r$ but is independent of $c$. Thus, the parameter $s$ in item 4 can be thought of as the extra pebbling space "slack" governing how severe the time-space trade-off will be.

**Theorem 7.6 (More detailed version of Theorem 2.8).** *Let $g(n)$ be any arbitrarily slowly growing monotone function $\omega(1) = g(n) = \mathrm{O}(n^{1/7})$, and fix any $\epsilon > 0$ and positive integer $K$. Then there are explicitly constructible CNF formulas $\{F_n\}_{n=1}^{\infty}$ of size $\Theta(n)$ and width $3(K+1)$ such that the following holds:*

1. *The formulas $F_n$ are refutable in syntactic resolution in total space $TotSp_{\mathfrak{R}}(F_n \vdash 0) = \mathrm{O}(g(n))$.*

2. *There are syntactic resolution refutations $\pi_n$ of $F_n$ in simultaneous length $L(\pi_n) = \mathrm{O}(n)$ and total space $TotSp(\pi_n) = \mathrm{O}\left((n/g^2(n))^{1/3}\right)$.*

3. *There are syntactic $\mathfrak{R}(K+1)$-refutations $\pi_n$ of $F_n$ in simultaneous length $L(\pi_n) = \mathrm{O}(n)$ and formula space $Sp(\pi_n) = \mathrm{O}(1)$.*

4. *Any semantic resolution refutation of $F_n$ in clause space $\mathrm{O}\left((n/g^2(n))^{1/3-\epsilon}\right)$ must have superpolynomial length.*

5. *For any $k \leq K$, any semantic $\mathfrak{R}(k)$-refutation in formula space $\mathrm{O}\left((n/g^2(n))^{1/(3(k+1))-\epsilon}\right)$ must have superpolynomial length.*

*All multiplicative constants hidden in the asymptotic notation depend only on $K$, $\epsilon$ and $g$.*

We remark that the upper-bound condition $g(n) = \mathrm{O}(n^{1/7})$ is very mild and is there only for technical reasons in this theorem. If we allow the minimal space to grow as fast as $n^{\epsilon}$ for some $\epsilon > 0$, then there are other pebbling trade-off results that can give even stronger results for resolution than the one stated above. (see, in particular, Section 7.6). Thus, the interesting part is that $g(n)$ is allowed to grow arbitrarily slowly.

*Proof of Theorem 7.6.* Consider the graphs $\Gamma(c, r)$ in Lemma 7.5. We want to choose the parameters $c$ and $r$ in a suitable way so that we get a family of graphs in size $n = \Theta(cr^3 + c^3r^2)$. If we set

$$r = r(n) = g(n) \tag{7.1}$$

for $g(n) = \mathrm{O}(n^{1/7})$, this forces

$$c = c(n) = \Theta\left(\sqrt[3]{n/g^2(n)}\right) \ . \tag{7.2}$$

Consider the graph family $\{G_n\}_{n=1}^{\infty}$ defined by $G_n = \Gamma(c(n), r(n))$ as in (7.1) and (7.2), which is a family of size $\Theta(n)$. Consider the pebbling formulas $F_n = Peb_{G_n}[f]$, substitute a $k$-non-authoritarian Boolean function $f$ of arity $k+1$, say XOR over $k+1$ variables for concreteness, and appeal to the translation between pebbling and resolution in Theorem 6.5.

Part 2 of Lemma 7.5 yields that $TotSp_{\mathfrak{R}}(F_n \vdash 0) = \mathrm{O}(g(n))$. Also, the black pebbling of $G_n$ in part 3 yields a linear-time refutation $\pi_n : F_n \vdash 0$ with $TotSp(\pi_n) = \mathrm{O}(\sqrt[3]{n/g^2(n)})$.

Now set the parameter $s$ in part 4 of Lemma 7.5 to $s = c^{1-\epsilon'}$ for $\epsilon' = \kappa\epsilon$ where $\kappa$ is chosen large enough (depending on $K$). Then for large enough $n$ we have $s \leq c/8$ and the trade-off in part 4 applies. Combining the pebbling trade-off there with Theorem 6.5, we get that if the clause space is less than $(n/g^2(n))^{1/3-\epsilon}$, then the required length of the resolution refutation grows as $(\Omega(c\epsilon'))^r = (\Omega(n/g^2(n)))^{\epsilon g(n)}$ which is superpolynomial in $n$ for any $g(n) = \omega(1)$. The rest of the theorem follows in a similar fashion. In particular, to get Part 5 above use Part 6 of Theorem 6.5 which gives a bound that is a $(k+1)$-root of the bound we get for semantic resolution. □

## 7.5 Robust Superpolynomial Trade-offs

We now know that there are polynomial trade-offs in resolution for constant space, and that going ever so slightly above constant space we can get superpolynomial trade-offs. The next question we want to focus on is how robust trade-offs we can get. That is, over how large a range of space does the trade-off hold? Given minimal refutation space $s$, how much larger space is needed in order to obtain the linear length refutation that we know exists for any pebbling contradiction?

The answer is that we can get superpolynomial trade-offs that span almost the whole range between constant and linear space. We present two different results illustrating this.

**Theorem 7.7.** *For any fixed positive integer $K$, there are explicitly constructible CNF formulas $\{F_n\}_{n=1}^{\infty}$ of size $\Theta(n)$ and width $3(K+1)$ such that:*

1. *Every formula $F_n$ is refutable in syntactic resolution in total space $TotSp_{\mathfrak{R}}(F_n \vdash 0) = \mathrm{O}(\log n)$.*

2. *There is a syntactic resolution refutation $\pi_n : F_n \vdash 0$ in simultaneous length $L(\pi_n) = \mathrm{O}(n)$ and total space $TotSp(\pi_n) = \mathrm{O}\left(\sqrt[3]{n/\log^2 n}\right)$.*

3. *There are syntactic $\mathfrak{R}(K+1)$-refutations $\pi_n$ of $F_n$ in simultaneous length $L(\pi_n) = \mathrm{O}(n)$ and formula space $Sp(\pi_n) = \mathrm{O}(1)$.*

4. *There is a constant $\kappa > 0$ such that any semantic resolution refutation $\pi_n : F_n \vdash 0$ in clause space $Sp(\pi_n) \leq \kappa \sqrt[3]{n/\log^2 n}$ must have length $L(\pi_n) = n^{\Omega(\log \log n)}$.*

5. *For any $k \leq K$, any semantic $k$-DNF resolution refutation $\pi_n : F_n \vdash 0$ in formula space $Sp(\pi_n) = \mathrm{o}\left((n/\log^2 n)^{1/(3(k+1))}\right)$ must have length $L(\pi_n) = n^{\Omega(\log \log n)}$.*

*The constant $\kappa$ as well as the constants hidden in the asymptotic notation are independent of $n$.*

*Proof.* Consider the graphs $\Gamma(c, r)$ in Lemma 7.5 with parameters chosen so that $c = 2^r$. Then the size of $\Gamma(c, r)$ is $\Theta(r^2 2^{3r})$. Let $r(n) = \max\{r : r^2 2^{3r} \leq n\}$ and define the graph family $\{G_n\}_{n=1}^{\infty}$ by $G_n = \Gamma(2^r, r)$ for $r = r(n)$. Finally, consider the pebbling formulas $F_n = Peb_{G_n}[f]$ with the help of Theorem 6.5.

Translating from $G_n$ back to $\Gamma(c, r)$ we have parameters $r = \Theta(\log n)$ and $c = \Theta((n/\log^2 n)^{1/3})$, so Lemma 7.5 yields that $TotSp_{\mathfrak{R}}(F_n \vdash 0) = \mathrm{O}(\log n)$. Also, the black pebbling of $G_n$ in Lemma 7.5 yields a linear-time refutation $\pi_n : F_n \vdash 0$ with $TotSp(\pi_n) = \mathrm{O}((n/\log^2 n)^{1/3})$.

Setting $s = c/8$ in the trade-off in part 4 of Lemma 7.5 shows that there is a constant $\kappa$ such that if the clause space of a refutation $\pi_n : F_n \vdash 0$ drops below $\kappa \cdot (n/\log^2 n)^{1/3} \leq (r+2) + s$, then we must have

$$L(\pi_n) \geq \mathrm{O}(1)^r \cdot r! = n^{\Omega(\log \log n)} \tag{7.3}$$

(where we used that $r = \Theta(\log n)$ for the final equality). The rest of the theorem follows in an analogous fashion. In particular, to get Part 5 above use Part 6 of Theorem 6.5 which gives a bound that is a $(k+1)$-root of the bound we get for semantic resolution. $\square$

Sacrificing a square at the lower end of the interval, we can improve the upper end to $n/\log n$.

**Theorem 7.8 (More detailed version of Theorem 2.9).** *For any positive integer $K$, there are explicitly constructible CNF formulas $\{F_n\}_{n=1}^{\infty}$ of size $\Theta(n)$ and width $3(K+1)$ such that the following holds (where the hidden constants depend only on $K$):*

1. The formulas $F_n$ are refutable in syntactic resolution in total space $TotSp_{\mathfrak{R}}(F_n \vdash 0) = \mathrm{O}(\log^2 n)$.

2. There are syntactic resolution refutations $\pi_n$ of $F_n$ in simultaneous length $L(\pi_n) = \mathrm{O}(n)$ and total space $TotSp(\pi_n) = \mathrm{O}(n/\log n)$.

3. There are syntactic $\mathfrak{R}(K+1)$-refutations $\pi_n$ of $F_n$ in simultaneous length $L(\pi_n) = \mathrm{O}(n)$ and formula space $Sp(\pi_n) = \mathrm{O}(1)$.

4. Any semantic resolution refutation of $F_n$ in clause space $Sp(\pi_n) = \mathrm{o}(n/\log n)$ must have length $L(\pi_n) = n^{\Omega(\log \log n)}$.

5. For any $k \leq K$, any semantic $\mathfrak{R}(k)$-refutation in formula space $Sp(\pi_n) = \mathrm{o}\bigl((n/\log n)^{1/(k+1)}\bigr)$ must have length $L(\pi_n) = n^{\Omega(\log \log n)}$.

For this theorem we need the following graph family from [LT82], which is built as *stacks of superconcentrators*, i.e., by placing graphs with very good connectivity properties on top of one another.

**Lemma 7.9 ([LT82]).** *There is a family of explicitly constructible graphs $\Phi(m, r)$ with a unique sink and vertex indegree 2 such that the following holds:*

1. $\Phi(m, r)$ has size $\Theta(rm)$.

2. $\mathsf{Peb}\bigl(\Phi(m, r)\bigr) = \mathrm{O}(r \log m)$.

3. There is a linear-time black pebbling strategy $\mathcal{P}$ for $\Phi(m, r)$ with $\mathsf{space}(\mathcal{P}) = \mathrm{O}(m)$.

4. If $\mathcal{P}$ is a black-white pebbling strategy for $\Phi(m, r)$ in space $s \leq m/20$, then $\mathsf{time}(\mathcal{P}) \geq m \cdot \left(\frac{rm}{64s}\right)^r$.

*Proof of Theorem 7.8.* As usual, pick $f = f_d(x_1, \ldots, x_d) = \bigoplus_{i=1}^{d} x_i$ for $d = K+1$, and consider the pebbling formulas $Peb_{\Phi(m,r)}[f]$ defined over stacks of superconcentrators $\Phi(m, r)$ as in Lemma 7.9 with $m = 20T$ and $r = \lfloor n/T \rfloor$ for $T = \Theta(n/\log n)$. Theorem 7.8 now follows by combining Lemma 7.9 with Theorem 6.5. □

We remark that the results in Theorem 7.8 are perhaps slightly stronger than those in Theorem 7.7 since they span a much larger (although non-overlapping) space interval. However, they require a very much more involved graph construction with worse hidden constants than the very simple and clean construction underlying Theorem 7.7.

## 7.6 Exponential Trade-offs

Superpolynomial trade-offs are all fine and well, but can we get *exponential* trade-offs? We conclude this section by giving strong answers in the affirmative to this question.

The same counting argument that was mentioned in the beginning of Section 7.4 tells us that we can never expect to get exponential trade-offs from DAGs with polylogarithmic pebbling price. However, if we move to graphs with pebbling price $\Omega(n^\epsilon)$ for some constant $\epsilon > 0$, pebbling formulas over such graphs can exhibit exponential trade-offs.

We obtain our first such exponential trade-off, which also exhibits a certain robustness, by again studying the DAGs in Lemma 7.5.

**Theorem 7.10.** *For any positive integer $K$ set $\epsilon = \frac{1}{3K+8}$ and $\delta = (K+2)\epsilon$. Then there is a constant $0 < \gamma < 1$ (depending only on $K$) and explicitly constructible CNF formulas $\{F_n\}_{n=1}^{\infty}$ of size $\Theta(n)$ and width $3(K+1)$ such that the following holds:*

1. *The formulas $F_n$ are refutable in syntactic resolution in total space $TotSp_{\mathfrak{R}}(F_n \vdash 0) = \mathrm{O}(n^\epsilon)$.*

2. *There are syntactic resolution refutations $\pi_n$ of $F_n$ in simultaneous length $L(\pi_n) = \mathrm{O}(n)$ and total space $TotSp(\pi_n) = \mathrm{O}(n^\delta)$.*

3. *There are syntactic $\mathfrak{R}(K+1)$-refutations $\pi_n$ of $F_n$ in simultaneous length $L(\pi_n) = \mathrm{O}(n)$ and formula space $Sp(\pi_n) = \mathrm{O}(n^\delta)$).*

4. *Any semantic resolution refutation of $F_n$ in clause space $n^\delta/10$ must have length at least $n^\epsilon!$.*

5. *For any $k \leq K$, any semantic $\mathfrak{R}(k)$-refutation in formula space $\gamma n^{\delta/(k+1)}$ must have exponential length, i.e., length at least $n^\epsilon!$.*

*All multiplicative constants hidden in the asymptotic notation depend only on $K$.*

Note that since $\delta > (k+1)\epsilon$ the tradeoff we get for $k$-DNF resolution is nontrivial. By this we mean that although there exist refutations requiring space $O(n^\epsilon)$, as long as we use space that is somewhat smaller than $n^{\delta/(k+1)}$ the refutation length is exponential.

*Proof of Theorem 7.6.* Consider again the graphs $\Gamma(c, r)$ in Lemma 7.5. Set $r = r(n) = n^\epsilon$ and $c = c(n) = n^\delta$. The graph family $\{G_n\}_{n=1}^\infty$ defined by $G_n = \Gamma(c(n), r(n))$ as in (7.1) and (7.2), is a family of size $\Theta(n)$. Consider the pebbling formulas $F_n = Peb_{G_n}[f]$, where $f$ is a $k$-non-authoritarian Boolean function $f$ of arity $k+1$, say XOR over $k+1$ variables for concreteness, and appeal to the translation between pebbling and resolution in Theorem 6.5.

Part 2 of Lemma 7.5 yields that $TotSp_{\mathfrak{R}}(F_n \vdash 0) = \mathrm{O}(n^\epsilon)$. Also, the black pebbling of $G_n$ in part 3 yields a linear-time refutation $\pi_n : F_n \vdash 0$ with $TotSp(\pi_n) = \mathrm{O}(n^\delta)$.

Now if $s < n^\delta/10$ inspection of part 4 of Lemma 7.5 shows that for large enough $n$ we get an exponential lower bound on the pebbling time of $n^\epsilon!$. Combining this with Theorem 6.5 gives our time lower bound for semantic resolution. Finally, to get the lower bound for semantic $k$-DNF resolution we apply Part 6 of Theorem 6.5 which gives a bound that is a $(k+1)$-root of the bound we get for semantic resolution. □

We remark that there is nothing magic in our particular choice of parameters in the proof of Theorem 7.10. Other parameters could be plugged in instead and yield slightly different results.

Now that we know that there are robust exponential trade-offs for resolution and $k$-DNF resolution, we want to obtain exponential trade-offs for formulas with their minimal refutation space being as large as possible.

The higher the lower bound on space is, the more interesting the trade-off gets. It seems reasonable that to look at and analyze a CNF formula, a SAT solver will at some point use at least linear space. If so, it is not immediate to argue why the SAT solver would later work hard on optimizing lower order terms in the memory consumption and thus get stuck in a trade-off for relatively small space. Ideally, therefore, we would like to obtain trade-offs for linear or even superlinear space (if there are such trade-offs, that is). For such formulas, we would be more confident that the trade-off phenomena should also show up in practice.[11]

It is clear that pebbling contradictions can never yield any trade-off results in the superlinear regime, since they are always refutable in linear length and linear space simultaneously. Also, all trade-offs obtainable from the graphs in Lemma 7.5 will be for space far below linear. However, using the following two results we can get exponential trade-offs for space almost linear, or more precisely for space as large as $\Theta(n/\log n)$.

---

[11]Having said that, we also want to point out that the case can certainly be made that even sublinear space trade-offs might be very relevant for real life applications. Intriguingly enough, pebbling contradictions over so-called pyramid graphs ([Coo74, Kla85]) might in fact be an example of this. We know that these formulas have short, simple refutations, but in [SBK04] it was shown that state-of-the-art clause learning algorithms can have serious problems with even moderately large pebbling contradictions. (Their "grid pebbling formulas" are exactly our pebbling contradictions using substitution with binary, non-exclusive or.) We wonder whether the high lower bound on clause space can be part of the explanation behind this phenomenon.

**Lemma 7.11 ([LT82]).** *For every directed acyclic graph $G$ with $n$ vertices and bounded indegree $\ell$, and for every space parameter $s$ satisfying $(3\ell + 2)n/\log n \leq s \leq n$, there is a black pebbling strategy $\mathcal{P}$ for $G$ with $\mathsf{space}(\mathcal{P}) \leq s$ and $\mathsf{time}(\mathcal{P}) \leq s \cdot 2^{2^{\mathrm{O}(n/s)}}$.*

**Lemma 7.12 ([LT82]).** *There exist constants $\epsilon, \kappa > 0$ such that for all sufficiently large integers $n, s$ satisfying $\kappa n/\log n \leq s \leq n$, we can find an explicitly constructible single-sink DAG $G$ with indegree 2 and number of vertices at most $n$ such that any black-white pebbling strategy $\mathcal{P}$ for $G$ with $\mathsf{space}(\mathcal{P}) \leq s$ must have $\mathsf{time}(\mathcal{P}) \geq s \cdot 2^{2^{\epsilon n/s}}$.*

Note that the graph $G$ in Lemma 7.12 depends on the pebbling space parameter $s$. Lengauer and Tarjan conjecture that no single graph gives an exponential time-space tradeoff for the whole range of $s \in [n/\log n, n]$, but to the best of our knowledge this problem is still open.

**Theorem 2.10 (restated).** *Let $\kappa$ be any sufficiently large constant. Then there are $k$-CNF formulas $F_n$ of size $\mathrm{O}(n)$ and a constant $\kappa' \ll \kappa$ such that:*

1. *The formulas $F_n$ have syntactic resolution refutations in total space $\kappa' \cdot n/\log n$.*

2. *$F_n$ is also refutable in syntactic resolution in length $\mathrm{O}(n)$ and total space $\mathrm{O}(n)$ simultaneously.*

3. *However, any semantic refutation of $F_n$ in clause space at most $\kappa \cdot n/\log n$ has length $\exp(n^{\Omega(1)})$.*

*Proof.* Appeal to Lemmas 7.11 and 7.12 in combination with Theorem 6.5 in the same way as in previous proofs in this section. □

We remark that Lemma 7.12 in combination with Lemma 7.11 can be used to obtain DAGs (and thus CNF formulas) with other superpolynomial trade-offs as well for different space parameters in the range above $n/\log n$ up to $n/\log\log n$. For simplicity and conciseness, however, we only state the special case above.

As was discussed in Section 2.3, Theorem 2.10 does not yield any provably nontrivial trade-offs for $\mathfrak{R}(k)$, since the space range where the trade-off kicks in is so low that we do not know whether there actually exist any $\mathfrak{R}(k)$-refutation in such small space. We do get weaker, though exponential (and provably non-vacuous) trade-offs for $\mathfrak{R}(k)$ in Theorem 7.10.

## 8 Concluding Remarks

We end this paper by discussing some open questions related to our reported work.

**Resolution** For the length, width, and clause space measures in resolution, there are known upper and lower worst-case bounds that essentially match modulo constant factors. This is *not* the case for total space, however.

**Open Question 1.** *Are there polynomial-size CNF formulas of width $\mathrm{O}(1)$ which require total resolution refutation space $TotSp_{\mathfrak{R}}(F \vdash 0) = \Omega\big((\text{size of } F)^2\big)$?*

The answer has been conjectured by [ABRW02] to be "yes", but as far as we are aware, there are no stronger lower bounds on total space known than those that follow trivially from corresponding linear lower bounds on clause space. Thus, a first step would be to show superlinear lower bounds on total space.

One way of interpreting the results of the current paper is that time-space trade-offs in pebble games carry over more or less directly to the resolution proof system (modulo the technical restrictions discussed in Section 6). The resolution trade-off results obtainable by this method are inherently limited, however, in the sense that pebblings in small space can be seen never to take too much time by a simple counting argument. For resolution there are no such limitations, at least not a priori, since the corresponding counting argument does not apply. Thus, one can ask whether it is possible to demonstrate even more dramatic time-space trade-offs for resolution than those obtained via pebbling.

To be more specific, we are particularly interested in what trade-offs are possible at the extremal points of the space interval, where we can only get polynomial trade-offs for constant space and no trade-offs at all for linear space.

**Open Question 2.** *Are there superpolynomial trade-offs for formulas refutable in constant space?*

**Open Question 3.** *Are there formulas with trade-offs in the range space > formula size? Or can every refutation be carried out in at most linear space?*

We find Open Question 3 especially intriguing. Note that all bounds on clause space proven so far, inlcuding the trade-offs in the current paper, are in the regime where the space is less than formula size (which is quite natural, since by [ET01] we know the size of the formula is an upper bound on the minimal clause space needed). It is unclear to what extent such lower bounds on space are relevant to state-of-the-art SAT solvers, however, since such algorithms will presumably use at least a linear amount of memory to store the formula to begin with. For this reason, it seems to be a highly interesting problem to determine what can be said if we allow extra clause space above linear. Are there formulas exhibiting trade-offs in this superlinear regime, or is it always possible to carry out a minimal-length refutation in, say, at most a constant factor times the linear upper bound on the space required for any formula? As was noted above, pebbling formulas cannot help answer these two questions, since they are always refutable in linear time and linear space simultaneously by construction, and since constant pebbling space implies polynomial pebbling time.

A final problem related specifically to standard resolution is that it would be interesting to investigate the implications of our results for applied satisfiability algorithms.

**Open Question 4.** *Do the trade-off phenomena we have established in this paper show up "in real life" for state-of-the-art DPLL based SAT-solvers, when run on the appropriate pebbling contradictions (or variations of such pebbling contradictions)?*

**A stronger space separation for $k$-DNF resolution** We have proven a strict separation between $k$-DNF resolution and $(k+1)$-DNF resolution by exhibiting for every fixed $k$ a family of CNF formulas of size $n$ that require space $\Omega\bigl(\sqrt[k+1]{n/\log n}\bigr)$ for any $k$-DNF resolution refutation but can be refuted in constant space in $(k+1)$-DNF resolution. This shows that the family of $\mathfrak{R}(k)$ proof systems form a strict hierarchy with respect to space.

As has been said above, however, we have no reason to believe that the lower bound for $\mathfrak{R}(k)$ is tight. In fact, it seems reasonable that a tighter analysis should be able to improve the bound to at least $\Omega\bigl(\sqrt[k]{n/\log n}\bigr)$ and possibly even further. The only known *upper* bound on the space needed in $\mathfrak{R}(k)$ for these formulas is the $\mathrm{O}(n/\log n)$ bound that is easily obtained for standard resolution. Closing, or at least narrowing, the gap between $\Omega\bigl(\sqrt[k+1]{n/\log n}\bigr)$ and $\mathrm{O}(n/\log n)$ is hence an open question.

**Understanding minimally unsatisfiable $k$-DNF sets** It seems that the problem of getting better lower bounds on space for $k$-DNF resolution is related to the problem of better understanding the structure of minimally unsatisfiable sets of $k$-DNF formulas. Although the correspondence is more intuitive than formal, it would seem that progress on this latter problem would probably translate into sharper lower bounds for $\Re(k)$ as well. The reason for this hope is that the asymptotically optimal results for standard resolution in this paper can in some sense be seen to follow from (the proof technique used to obtain) the tight bound for CNF formulas in Theorem 2.13.

What we are able to prove in this paper is that any minimally unsatisfiable $k$-DNF set $\mathbb{D}$ (for $k$ a fixed constant) must have at least $\Omega\bigl(\sqrt[k+1]{|\mathbb{D}|}\bigr)$ variables (Theorem 2.14) but the only explicit constructions of such sets that we where able to obtain had $\mathrm{O}(|\mathbb{D}|)$ variables (Lemma 2.15). As has already been mention, the recent work [NR09] unexpectedly improved the lower bound to roughly $\mathrm{O}\bigl(\sqrt[k]{|\mathbb{D}|}\bigr)$. This appears to be a natural and interesting combinatorial problem in its own right, and it would be very nice to close the gap between the upper and lower bound.

We have the following conjecture, where for simplicity we fix $k$ to remove it from the asymptotic notation.

**Conjecture 5.** *Suppose that $\mathbb{D}$ is a minimally unsatisfiable $k$-DNF set for some arbitrary but fixed positive integer $k$. Then the number of variables in $\mathbb{D}$ is at most $\mathrm{O}(|\mathbb{D}|)^k$.*

Proving this conjecture would establish asymptotically tight bounds for minimally unsatisfiable $k$-DNF sets (ignoring factors involving the constant $k$).

**Generalizations to other proof systems** We have presented a "substitution space theorem" for resolution as a way of lifting lower bounds on the number of variables to lower bounds on (clause) space, and have then extended this result by lifting lower bounds on the number of variables *in resolution* to lower bounds on formula space in the *much stronger $k$-DNF resolution proof systems*. It is a natural question to ask whether our techniques can be extended to other proof systems as well.

We remark that the translations in Sections 4 and 5 of refutations of substitution formulas in some other proof system $\mathcal{P}$ via projection to resolution refutations of the original formula seem extremely generic and robust in that they do not at all depend on which derivation rules are used by $\mathcal{P}$ nor on the class of formulas with which $\mathcal{P}$ operates. The only place where the particulars of the proof system come into play is when we actually need to analyze the content of the proof blackboard. As described in the introduction, this happens at some critical point in time when we know that the blackboard of our translated (projected) resolution proof mentions a lot of variables, and want to argue that this implies that the blackboard of the $\mathcal{P}$-proof must contain a lot of formulas (or possibly some other resource that we want to lower-bound in $\mathcal{P}$). This part of the analysis is the (essentially tight) result for resolution in Lemma 4.7 and the (likely not tight) bound for $k$-DNF sets in Lemma 5.5 in this paper. Any corresponding result for some other proof system $\mathcal{P}$ would translate into lower bounds for $\mathcal{P}$ in terms of lower bounds on variable space in resolution.

UNDERSTANDING SPACE IN PROOF COMPLEXITY